\documentclass[a4paper,11pt]{article}
\usepackage{jheppub} 
\usepackage{lineno}

\usepackage{ytableau}
\usepackage[export]{adjustbox}
\usepackage{lscape}

\arxivnumber{2409.158641} 

\title{\boldmath The Seven-Point Two-Loop Full-Color All-Plus Helicity Yang-Mills Amplitude}

\author[]{Adam R. Dalgleish,  David C. Dunbar,
  Warren B. Perkins and Joseph M.W. Strong}

\affiliation[]{Department of Physics,\\
Faculty of Science and Engineering, \\
Swansea University,\\
Swansea, SA2 8PP, UK }

\emailAdd{d.c.dunbar@swansea.ac.uk}

\abstract{

The seven-gluon two-loop full-color Yang-Mills amplitude is presented in a compact analytic form where
we use the methods of four-dimensional unitarity cuts to obtain the polylogarithmic pieces and augmented recursion to obtain the rational pieces.  Furthermore, an $n$-point form for the full-color polylogarithmic piece is  presented.   
We use these results to  probe for potential relations amongst  the two-loop partial amplitudes.
}

\keywords{NNLO computations}

\def\spa#1.#2{\left\langle#1\,#2\right\rangle}
\def\spb#1.#2{\left[#1\,#2\right]}
\def\la{\langle}
\def\ra{\rangle}
\def\l{\ell}

\def\Trr{\rm T}
\def\eps{\epsilon}
\def\Ord{{\cal O}}

\def\tr{\mathop{\rm Tr}\nolimits}
\begin{document}
\maketitle
\flushbottom

 \section{Introduction}

Perturbative scattering amplitudes are 
a key component for comparing theory to experiment and 
with increasing experimental data  there is an insatiable demand for more accurate theoretical predictions~\citep{Amoroso:2020lgh}.   
Amplitudes are also
of interest both for comparing theory to experiment  and for viewing symmetries that may not be obvious from the perspective of the Lagrangian.  
Since the Standard Model of Particle Physics is a gauge theory,   amplitudes in gauge theories are of particular interest.  There has been great progress in the methodology of computing scattering amplitudes but they remain challenging.

In Yang-Mills theory,   we can expand the scattering amplitude of $n$-gluons in terms of the gauge coupling constant, $g$,
\begin{equation}\label{eq:loop_decomposition}
    \mathcal{A}_n = g^{n-2}  \sum_{\l \geq 0} a^\l \mathcal{A}_n^{(\l)}
\end{equation}
where $a = g^2 e^{-\epsilon \gamma_E} / (4\pi)^{2-\epsilon}$ (we use dimensional regularisation and work in $d=4-2\epsilon$ dimensions).
The $\mathcal{A}_n^{(\l)}$ are the $\l$-loop amplitudes.  These depend both upon kinematic information of the particles but also their gauge content.     It has proven useful to separate,   if possible,  the kinematic and color information by  expanding the amplitude in terms of color structures,
\begin{align} \label{eq:general colour decomposition}
    \mathcal{A}_n^{(\l)} = \sum_\lambda A^{(\l)}_{n:\lambda} C^\lambda   \; . 
\end{align}
We will use a color trace expansion where external states are labelled by matrices in the $SU(N_c)$ or $U(N_c)$ algebra.
The $C^\lambda$ then contain traces over the $SU(N_c)$ or $U(N_c)$ matrices of the 
color symmetry,  as well as factors of $N_c$~\citep{Cvitanovic:1980bu,Mangano:1990by,Mangano:1988kk}.
The  $A^{(\l)}_{n:\lambda}$ are known as partial amplitudes.
The  $\mathcal{A}_n$ are fully crossing symmetric but the  $A^{(\l)}_{n:\lambda}$
only  possess symmetries related to $C^\lambda$.  They can be arranged according to the accompanying power of $N_c$ with
the partial amplitude accompanying $N_c^\l$ is often referred to as  ``leading in color", with other partial amplitudes being  ``sub-leading in color",  ``sub-sub-leading in color" etc.

The form of an amplitude is also dependent on the helicities of the gluons involved.
Some helicity configurations,  such as all same (we take as positive),  lead to simpler amplitudes due to the increased symmetry.
The calculation of the two-loop all-plus amplitude is made easier because the tree-level all-plus amplitudes vanish and then consequentially the one-loop is purely rational.   This helicity amplitude was the first 
to be computed for all-$n$ at one loop~\cite{Bern:1993qk}.   The one-loop all-plus being related both to the $N=4$ supersymmetric MHV amplitude by dimension shifting~\cite{Bern:1996ja} and also is equal to that for self-dual Yang-Mills~\cite{Cangemi:1996rx,Chalmers:1996rq}.

There has been significant progress on calculating tree and one-loop gluon scattering amplitudes however 
for two-loop amplitudes,  work is ongoing with results only available for a small number of external legs and/or helicities.   The situation is much better in theories with enhanced symmetry such as maximal $N=4$ supersymmetric Yang--Mills.

First we summarise existing results.
The four gluon amplitude was the first  scattering amplitude to   be computed with full-color and for all helicity configurations. This was performed in various dimensional regularisation schemes~\citep{Glover:2001af,Bern:2002tk} and to all orders 
of the regulator~\citep{Ahmed:2019qtg}.
Beyond four points progress has been incremental and 
all  partial amplitudes of the five gluon amplitude have now been calculated: 
the all-plus leading in color partial amplitude was first calculated using a generalised unitarity procedure~\cite{Badger:2013gxa,Badger:2015lda} then presented more compactly in~\cite{Gehrmann:2015bfy}.
It was re-derived using a simpler method of four dimensional unitarity and augmented recursion in~\cite{Dunbar:2016aux}, which is the method we use.
Later, the remaining all-plus color structures were calculated in~\cite{Badger:2019djh,Dunbar:2019fcq}, completing the all-plus amplitude.   
The single-minus helicity leading in color partial amplitude was obtained in~\cite{Badger:2018enw} and the remaining leading in color helicity configurations were obtained in~\cite{Abreu:2019odu}, both using finite field numerical methods. 
Recently, the remaining color structures have been computed~\cite{DeLaurentis:2023nss,Agarwal:2023suw}.

For six gluons,  only the all-plus amplitude has been computed (leading in color~\cite{Dunbar:2016gjb}, then full-color~\cite{Dalgleish:2020mof}), using four dimensional unitarity and augmented recursion.
The seven gluon all-plus leading in color partial amplitude has also been calculated using this methodology~\cite{Dunbar:2017nfy}.
An $n$-point expression for the all-plus single color trace, $N_c^0$ partial amplitude was conjectured in~\cite{Dunbar:2020wdh} using various consistency conditions. and verified 
in~\cite{Kosower:2022bfv} up to eight points.  satisfying various consistency conditions.

This paper presents a compact analytic form of the two-loop seven gluon all-plus amplitude,  at full color, $\mathcal{A}_7^{(2)}(1^+,2^+,3^+,4^+,5^+,6^+,7^+)$.
This involves re-deriving the leading color piece and confirming the structure of the single trace $N_c$-independent piece,  as well as deriving the remaining partial amplitudes for the first time.
We use the method of four dimensional unitarity cuts to obtain the polylogarithmic parts of these results in a simple way.
Recursion is used to obtain the rational terms however
due to the presence of double poles in momenta,  augmented recursion~\cite{Alston:2012xd} must be used.

The results we present can be used as a test for possible linear relations among the partial amplitudes.   Any relation independent of helicity must hold for the all-plus although obviously the reverse is not true.  As such we can definitively rule out possible relations and leave remaining potential relation.    We use a decomposition of the partial amplitudes under the symmetric group which allows us to categorise all possible relations satisfied by the full seven point amplitude and also the polylogarithmic part of the eight and nine point amplitudes.  The nine-point triple trace partial amplitude satisfies relations beyond those obtained by decoupling although this is only verified for the all-plus amplitude and its polylogarithmic part.

We will structure the paper as follows:
The next section sets out conventions and background on full color amplitudes.
The following section describes the separation into pieces to be treated with differing techniques.
The method of four dimensional unitarity will be presented, followed by the polylogarithmic results.
Augmented recursion will be presented,  followed by rational results.   (with details of the new currents in an appendix). 
We then use the  full results for $n=7$ and the polylogarithmic parts of $n=8,9$ as a probe to search for linear relations amongst the partial amplitudes.

\section{Full Color Amplitudes}

\subsection{Color Decomposition}

 We first review the color decomposition of the $n$-gluon amplitude at tree and one-loop level. 
At tree-level,  the color decomposition is
\begin{align}
    \mathcal{A}_n^{(0)}(1,2,\cdots,n) = \sum_{S_n/\mathcal{P}_{n:1}} \Trr[T^{a_1}\cdots  T^{a_n}] A_{n:1}^{(0)}(a_1,\cdots,a_n)
\end{align}
in a color trace basis~\cite{Cvitanovic:1980bu,Mangano:1987xk,Mangano:1988kk,Mangano:1990by}.  While we are generally interested in amplitudes with gluons in a pure $SU(N_c)$ theory, it can be useful to calculate 
those structures that occur in a $U(N_c)$ theory,  so we allow for either.  The $T^{a}$ are generators in the fundamental representation of $SU(N_c)$ or $U(N_c)$.
The partial amplitudes $A_{n:1}^{(0)}(a_1,\cdots,a_n)$ are gauge invariant and cyclicly symmetric and hence the $S_n/\mathcal{P}_{n:1}$ sum is over the $(n-1)!$ permutations of $(1,\cdots,n)$ up to this cyclic symmetry.

At one-loop level,   the decomposition is~\cite{Bern:1990ux}
\begin{align}
    \mathcal{A}_n^{(1)}(1,2,\cdots,n) =&  N_c \sum_{S_n/\mathcal{P}_{n:1}} \Trr[T^{a_1}\cdots  T^{a_n}] A_{n:1}^{(1)}(a_1,\cdots,a_n)
    \notag\\
    + \sum_{r=2}^{\lfloor n/2 \rfloor +1}  \sum_{S_n/\mathcal{P}_{n:r}}   \Trr[T^{a_1} &\cdots  T^{a_{r-1}}] \Trr[T^{b_r}\cdots  T^{b_n}] A_{n:r}^{(1)}(a_1,\cdots,a_{r-1} ; b_r,\cdots,b_n)
\end{align}
where the 
$A_{n:r}^{(1)}(a_1,\cdots,a_{r-1} ; b_r,\cdots,b_n)$ partial amplitudes have cyclic symmetry to match their accompanying traces.
Where the two traces have the same length, the partial amplitude also has an additional $Z_2$ symmetry upon interchanging these sets of 
legs. The $S_n/\mathcal{P}_{n:r}$ sum is over  permutations of external legs up to these symmetries, so that every possible color trace structure occurs exactly once in the overall sum.

At two-loop level~\cite{Dunbar:2019fcq},
\begin{align}
    \mathcal{A}_n^{(2)}(1,2, & \cdots,n) = N_c^2 \sum_{S_n/\mathcal{P}_{n:1}} \Trr[T^{a_1}\cdots  T^{a_n}] A_{n:1}^{(2)}(a_1,\cdots,a_n)
    \notag\\
    +& N_c \sum_{r=2}^{\lfloor n/2 \rfloor +1} \sum_{S_n/\mathcal{P}_{n:r}} \Trr[T^{a_1}\cdots  T^{a_{r-1}}] \Trr[T^{b_r}\cdots  T^{b_n}] A_{n:r}^{(2)}(a_1,\cdots,a_{r-1} ; b_r,\cdots,b_n)
    \notag\\
    +& \sum_{s=1}^{\lfloor n/3 \rfloor} \sum_{t=s}^{\lfloor (n-s)/2 \rfloor} \sum_{S_n/\mathcal{P}_{n:s,t}} \Trr[T^{a_1}\cdots  T^{a_{s}}] \Trr[T^{b_{s+1}}\cdots  T^{b_{s+t}}] \Trr[T^{c_{s+t+1}}\cdots  T^{c_{n}}]
    \notag\\&\hskip4.0truecm
    \times A_{n:s,t}^{(2)}(a_1,\cdots,a_{s} ; b_{s+1},\cdots,b_{s+t} ; c_{s+t+1},\cdots,c_n)
    \notag\\
    + &\sum_{S_n/\mathcal{P}_{n:1}} \Trr[T^{a_1}\cdots  T^{a_n}] A_{n:1B}^{(2)}(a_1,\cdots,a_n)
\end{align}
with three-trace partial amplitudes and a new single-trace $N_c$-independent amplitude $A_{n:1B}^{(2)}$  appearing.
The symmetry factors $\mathcal{P}_{n:1}$, $\mathcal{P}_{n:r}$ and $\mathcal{P}_{n:s,t}$ in each case describe the symmetries of cycling the arguments of the trace structures, or interchanging two or three trace structures when they are of equal length.
The partial amplitudes themselves are invariant under the relevant $\mathcal{P}_{n:1}$, $\mathcal{P}_{n:r}$ or $\mathcal{P}_{n:s,t}$.
For example, $A^{(2)}_{7:2,2}(1,2;3,4;5,6,7)$ is invariant under 
$\mathcal{P}_{7;2,2} = Z_2(1,2)\times Z_2(3,4)\times Z_3(5,6,7) \times Z_2(\{1,2\},\{3,4\})$.
The sum is then over permutations of the legs up to this symmetry, $S_n/\mathcal{P}_n$.
Additionally,   amplitudes are flip-invariant under reflection,  where the order within each trace is reversed together with a $(-1)^n$.   That is, 
\begin{align}
A_{n:s,t}^{(2)}(a_1,\cdots,a_{s} ; & b_{s+1},\cdots,b_{s+t} ; c_{s+t+1},\cdots,c_n)
\notag\\
&=(-1)^n A_{n:s,t}^{(2)}(a_s,\cdots,a_{1} ; b_{s+t},\cdots,b_{s+1} ; c_n, \cdots , c_{s+t+1})
\; . 
\end{align} 

In the $SU(N_c)$ theory,  factors of $\Trr[T^a]$ vanish, so  terms $A_{n:2}^{(2)}$ and $A_{n:1,t}^{(2)}$ would not appear in the above expansions.

For tree and one-loop amplitudes other color decompositions exist.  Specifically the decomposition~\cite{DelDuca:1999rs} exists for tree amplitudes which reduces the number of independent amplitudes to $(n-2)!$.   This has implications for relations amongst the amplitudes which we will look at later.  Although, this decomposition also extends to one-loop  it is less well developed beyond and so we will work in the color trace formalism  and accept the redundancy as a usefull consistency check.

\subsection{Decoupling Identities}

The partial amplitudes solely associated with a $U(N_c)$ symmetry, $A^{(2)}_{n:2}$ and $A^{(2)}_{n:1,t}$ still play a  role in our calculations.  
Specifically, there are {\it  decoupling identities} which  allow us to express the $A^{(2)}_{n:2,2}$ and $A^{(2)}_{n:2,3}$ in terms of the $A^{(2)}_{n:1,t}$. 
Further decoupling identities then relate these to the sub-leading partial amplitudes  $A^{(2)}_{n:2,2}$ $r> 2$.    The first triple trace amplitude which is {\it not} obtainable via decoupling identities is the nine point $A^{(2)}_{9:3,3}$. 

Letting a gluon lie in the $U(1)$ part of $U(N_c)$, we would expect the amplitude to vanish.
Choosing,  say,  $T^1 \rightarrow T^1_{U(1)}=N_c^{-1/2} I_n$,   we see 
$\Trr[T^1_{U(1)}]=N_c^{+1/2}$ and $\Trr[T^1_{U(1)}T^2 \cdots  \cdots T^r] = N_c^{-1/2}\Trr[T^2 \cdots T^r]$.
Examining the coefficient of a given trace structure,  and equating to zero we find relations between partial amplitudes known as decoupling identities.
These allow the $U(N_c)$ partial amplitudes to be related to $SU(N_c)$ 
partial amplitudes and include relations among the $SU(N_c)$ partial amplitudes.

The two-loop seven-point amplitude possesses five $SU(N_c)$ partial amplitudes:
$A^{(2)}_{7:1}$, $A^{(2)}_{7:3}$, $A^{(2)}_{7:4}$, $A^{(2)}_{7:2,2}$ and $A^{(2)}_{7:1B}$,
together with four purely $U(N_c)$ partial amplitudes:
$A^{(2)}_{7:2}$, $A^{(2)}_{7:1,1}$, $A^{(2)}_{7:1,2}$ and $A^{(2)}_{7:1,3}$.
These are related by the decoupling identities:

\def\Rdec{A}
\begin{align}
    \Rdec_{7:1}^{(2)}(1, 2, 3, 4, 5, 6, 7) + \Rdec_{7:1}^{(2)}(1, 3, 4, 5, 6, 7, 2) +
    \Rdec_{7:1}^{(2)}(1, 4, 5, 6, 7, 2, 3) &\notag\\ 
    + \Rdec_{7:1}^{(2)}(1, 5, 6, 7, 2, 3, 4) +  \Rdec_{7:1}^{(2)}(1, 6, 7, 2, 3, 4, 5) + \Rdec_{7:1}^{(2)}(1, 7, 2, 3, 4, 5, 6) &\notag\\
    + \Rdec_{7:2}^{(2)}({1}; {2, 3, 4, 5, 6, 7}) & = 0
\\
    \Rdec_{7:1,1}^{(2)}({1}; {2}; {3, 4, 5, 6, 7}) + \Rdec_{7:2}^{(2)}({2}; {1, 3, 4, 5, 6, 7}) + \Rdec_{7:2}^{(2)}({2}; {1, 4, 5, 6, 7, 3}) &\notag\\
    + \Rdec_{7:2}^{(2)}({2}; {1, 5, 6, 7, 3, 4}) + \Rdec_{7:2}^{(2)}({2}; {1, 6, 7, 3, 4, 5}) + \Rdec_{7:2}^{(2)}({2}; {1, 7, 3, 4, 5, 6}) &\notag\\
    + \Rdec_{7:3}^{(2)}({1, 2}; {3, 4, 5, 6, 7}) &=0
\end{align}
\begin{align}
    \Rdec_{7:1,2}^{(2)}({1; 2, 3; 4, 5, 6, 7}) + \Rdec_{7:3}^{(2)}({2, 3; 1, 4, 5, 6, 7}) + \Rdec_{7:3}^{(2)}({2, 3; 1, 5, 6, 7, 4}) &\notag\\
    + \Rdec_{7:3}^{(2)}({2, 3; 1, 6, 7, 4, 5}) + \Rdec_{7:3}^{(2)}({2, 3; 1, 7, 4, 5, 6}) + \Rdec_{7:4}^{(2)}({1, 2, 3; 4, 5, 6, 7}) &\notag\\
    + \Rdec_{7:4}^{(2)}({1, 3, 2; 4, 5, 6, 7}) &=0
\\
    \Rdec_{7:1,3}^{(2)}({1; 2, 3, 4; 5, 6, 7}) + \Rdec_{7:4}^{(2)}({2, 3, 4; 1, 5, 6, 7}) + \Rdec_{7:4}^{(2)}({2, 3, 4; 1, 6, 7, 5}) &\notag\\
    + \Rdec_{7:4}^{(2)}({2, 3, 4; 1, 7, 5, 6}) + \Rdec_{7:4}^{(2)}({5, 6, 7; 1, 2, 3, 4}) + \Rdec_{7:4}^{(2)}({5, 6, 7; 1, 3, 4, 2}) &\notag\\
    + \Rdec_{7:4}^{(2)}({5, 6, 7; 1, 4, 2, 3})&=0
\\
    \Rdec_{7:1,1}^{(2)}({2; 3; 1, 4, 5, 6, 7}) + \Rdec_{7:1,1}^{(2)}({2; 3; 1, 5, 6, 7, 4}) + \Rdec_{7:1,1}^{(2)}({2; 3; 1, 6, 7, 4, 5}) &\notag\\
    + \Rdec_{7:1,1}^{(2)}({2; 3; 1, 7, 4, 5, 6}) + \Rdec_{7:1,2}^{(2)}({2; 1, 3; 4, 5, 6, 7}) + \Rdec_{7:1,2}^{(2)}({3; 1, 2; 4, 5, 6, 7}) &=0
\\
    \Rdec_{7:1,2}^{(2)}({2; 3, 4; 1, 5, 6, 7}) + \Rdec_{7:1,2}^{(2)}({2; 3, 4; 1, 6, 7, 5}) + \Rdec_{7:1,2}^{(2)}({2; 3, 4; 1, 7, 5, 6}) &\notag\\
    + \Rdec_{7:1,3}^{(2)}({2; 1, 3, 4; 5, 6, 7}) + \Rdec_{7:1,3}^{(2)}({2; 1, 4, 3; 5, 6, 7}) + \Rdec_{7:2,2}^{(2)}({1, 2; 3, 4; 5, 6, 7})&=0
\\
    \Rdec_{7:2,2}^{(2)}({2, 3; 4, 5; 1, 6, 7}) + \Rdec_{7:2,2}^{(2)}({2, 3; 4, 5; 1, 7, 6}) + \Rdec_{7:2,2}^{(2)}({2, 3; 6, 7; 1, 4, 5}) &\notag\\
    + \Rdec_{7:2,2}^{(2)}({2, 3; 6, 7; 1, 5, 4}) + \Rdec_{7:2,2}^{(2)}({4, 5; 6, 7; 1, 2, 3}) + \Rdec_{7:2,2}^{(2)}({4, 5; 6, 7; 1, 3, 2})&=0
\\
    \Rdec_{7:1B}^{(2)}(1, 2, 3, 4, 5, 6, 7) + \Rdec_{7:1B}^{(2)}(1, 3, 4, 5, 6, 7, 2) + \Rdec_{7:1B}^{(2)}(1, 4, 5, 6, 7, 2, 3) &\notag\\
    + \Rdec_{7:1B}^{(2)}(1, 5, 6, 7, 2, 3, 4) + \Rdec_{7:1B}^{(2)}(1, 6, 7, 2, 3, 4, 5) + \Rdec_{7:1B}^{(2)}(1, 7, 2, 3, 4, 5, 6)&=0
\; .  
\end{align}

Our calculational strategy has been to independently calculate all nine $U(N_c)$ partial amplitudes, then use the above decoupling identities as a consistency check on the results.   The decoupling identities would allow us to express $ \Rdec_{7:2,2}^{(2)}$ in terms of the
$\Rdec_{7:1,2}^{(2)}$ and $\Rdec_{7:1,1}^{(2)}$ which then can be expressed in terms of the sub-leading $\Rdec_{7:3}^{(2)}$ and 
 $\Rdec_{7:4}^{(2)}$.  This works for any  $\Rdec_{n:2,r}^{(2)}$ but not for  $\Rdec_{n:r,s}^{(2)}$ where $r > 2$. 

The decoupling identities do not fully exhaust the relations between structures in this color decomposition.
Further relations have been derived explicitly at four-, five-, and six-point~\cite{Edison:2011ta,Edison:2012fn,Dunbar:2023ayw}.  
We will use our results to explore possible identities later.

\section{Functional Structure of the Amplitude}

\subsection{IR Singular Pieces}

We present our result as an unrenormalised amplitude, calculated in the four-dimensional helicity scheme~\cite{Bern:1991aq}.
As the singular structures of these partial amplitudes are known in general~\cite{Catani:1998bh}, we subdivide the amplitude into terms that contain divergences as $\epsilon\to 0$, $U^{(2)}_{n:\lambda}$, and those that are finite, $F^{(2)}_{n:\lambda}$,

\begin{align}
    A^{(2)}_{n:\lambda} = U^{(2)}_{n:\lambda} + F^{(2)}_{n:\lambda} + \Ord(\eps).
\end{align}
where $\lambda$ labels the various color structures. 
In a general two-loop amplitudes we would expect to see UV divergences, soft IR divergences and collinear IR divergences appearing in $U^{(2)}_{n:\lambda}$, with contributions up to $1/\eps^4$~\cite{Catani:1998bh}.
However, due to the all-plus tree amplitude vanishing, the UV and collinear IR divergences cancel, leaving only soft IR singularities~\cite{Kunszt:1994np}, which have at most $1/\eps^2$.
The vanishing of the tree amplitude also leads to regularisation scheme-independence in the two-loop all-plus amplitude~\cite{Catani:1998bh}. 

The form of the IR singular structure for the  all-plus two-loop amplitude was presented in a color trace basis in~\cite{Dunbar:2019fcq}, which we reproduce here (with modified notation) for completeness.   First, defining
\begin{align}
    I_{i,j} \equiv - \frac{(s_{ij})^{-\eps}}{\eps^2} \;,
\end{align}
and for  lists $S_1 = \{ a_1, a_2, \cdots, a_s \}$ and
$S_2 = \{ b_1, b_2, \cdots, b_t \}$
\begin{align}
    I_r[S_1] &= I_r[\{ a_1, a_2, \cdots, a_s \}] \equiv \sum_{i=1}^s I_{a_i,a_{i+1}}\;,
    \notag\\
          I_j[S_1,S_2] &= I_j[ \{ a_1, a_2, \cdots, a_s \} , \{ b_1, b_2, \cdots, b_t \} ] \equiv (I_{a_1,a_s} + I_{b_1,b_t} - I_{a_1,b_1} - I_{a_s,b_t})
    \notag\\
    I_k[S_1,S_2] &= I_k[ \{ a_1, a_2, \cdots, a_s \} , \{ b_1, b_2, \cdots, b_t \} ] \equiv (I_{a_1,b_t} + I_{b_1,a_s} - I_{a_1,b_1} - I_{a_s,b_t})
    \notag \\ 
\end{align}
where $I_{a_s,a_{s+1}} \equiv I_{a_s,a_{1}}$.   
 With these definitions there is the identity 
\begin{align}
    I_r[S_1 \oplus S_2] =  I_r[S_1] + I_r[S_2] + I_k[S_1,S_2] - I_j[S_1,S_2]\;,
\end{align}
where $\{ a_1, a_2, \cdots, a_s \} \oplus \{ b_1, b_2, \cdots, b_t \} = \{ a_1, a_2, \cdots, a_s, b_1, b_2, \cdots, b_t \}$.

With these definitions,  the IR singular pieces of the two-loop all-plus partial amplitudes are~\cite{Dunbar:2019fcq}:
\begin{align}
    U_{n:1}^{(2)}(S) =& A_{n:1}^{(1)}(S) \times I_r[S]
    \notag\\
    U_{n:r}^{(2)}(S_1;S_2) =& A_{n:r}^{(1)}(S_1;S_2) \times (I_r[S_1]+I_r[S_2])
    \notag\\
    &+ \sum_{S_1' \in C(S_1)} \sum_{S_2' \in C(S_2)} A_{n:1}^{(1)}(S_1' \oplus S_2') \times I_j[S_1', S_2']
    \notag\\
    U_{n:s,t}^{(2)}(S_1;S_2;S_3) =& \sum_{S_2' \in C(S_2)} \sum_{S_3' \in C(S_3)} A_{n:r}^{(1)}(S_1'; S_2' \oplus S_3') \times I_j[S_2', S_3']
    \notag\\
    &+ \sum_{S_1' \in C(S_1)} \sum_{S_3' \in C(S_3)} A_{n:r}^{(1)}(S_2'; S_1' \oplus S_3') \times I_j[S_1', S_3']
    \notag\\
    &+ \sum_{S_1' \in C(S_1)} \sum_{S_2' \in C(S_2)} A_{n:r}^{(1)}(S_3'; S_1' \oplus S_2') \times I_j[S_1', S_2']
    \notag\\
    U_{n:1B}^{(2)}(S) =& \sum_{U(S)} A_{n:r}^{(1)}(S_1';S_2') \times I_k[S_1',S_2']
\label{IRbits}
\end{align}
where $C(S)$ is the set of cyclic permutations of $S$ and
$U(S)$ is the set of all distinct pairs of lists $(S_1',S_2')$ such that $S_1' \oplus S_2' = S$ and the size of $S_i'$ is greater than one.
For example, at seven-point:
\begin{align}
    U(\{1,2,3,4,5,6,7\}) = \Big\{  (\{1,2\},\{3,4,5,6,7\}), \; \; &(\{1,2,3\},\{4,5,6,7\}),
    \notag\\
    (\{2,3\},\{4,5,6,7,1\}),\;  (\{2,3,4\},\{5,6,7,1\}), \;
  & 
    (\{3,4\},\{5,6,7,1,2\}), \; (\{3,4,5\},\{6,7,1,2\}), 
    \notag\\
     (\{4,5\},\{6,7,1,2,3\}), \; (\{4,5,6\},\{7,1,2,3\}),\;
    &
     (\{5,6\},\{7,1,2,3,4\}),\; (\{5,6,7\},\{1,2,3,4\}),\;
    \notag\\
     (\{6,7\},\{1,2,3,4,5\}), \;(\{6,7,1\},\{2,3,4,5\}),\;
    & (\{7,1\},\{2,3,4,5,6\}), \;(\{7,1,2\},\{3,4,5,6\})
    \Big\}\;.
\end{align}

\subsection{Polylogarithmic and Rational Pieces}

We further  separate the finite pieces into polylogarithmic terms, $P^{(2)}_{n:\lambda}$, and rational terms, $R^{(2)}_{n:\lambda}$:

\begin{align}
    F^{(2)}_{n:\lambda} = P^{(2)}_{n:\lambda} + R^{(2)}_{n:\lambda}.
\end{align}
The polylogarithmic pieces are determined using four-dimensional unitarity and the rational pieces using augmented recursion.
In many approaches these two parts of the amplitude are closely tied together although in  a Dimensional reduction methodology
where Unitarity is used in multiple integer dimensions and integrands are reconstructed as a polynomial in $(D-2)$ there can also be a separation. In~ \cite{Badger:2016ozq} is was shown the 
rational terms in a five and six point computation of the all-plus arise  solely from the leading
$(D-2)^2$ term.   In~\cite{Kosower:2022bfv} dimensional constructions was used to obtain the rational term for the all-plus amplitude for the six point and the leading in color for the 
seven-point together with the sub-sub-leading for seven and eight points.   These different methods have given results in agreement with those computed using our methodology.

\section{Polylogarithmic Terms}

\label{section:four}

\begin{figure}[h]
    \centering
    \includegraphics[width=7cm]{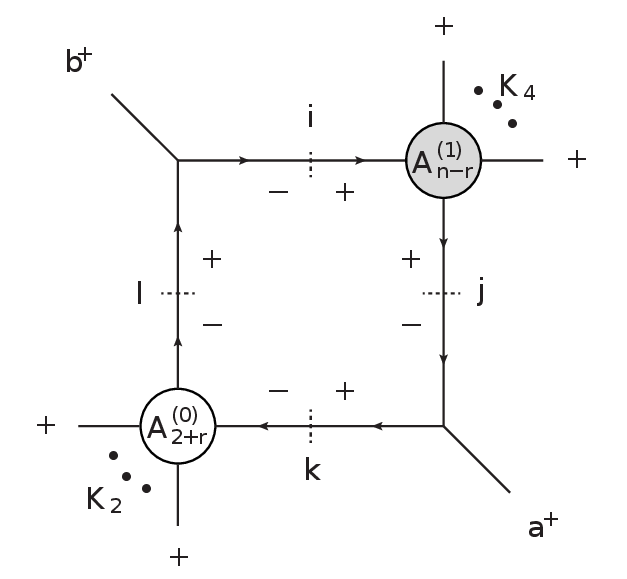}
    \caption{The quadruple cut of the two-loop all-plus amplitude involving an all-plus
one-loop vertex (shown in grey). $K_2$ may be null but $K_4$ must contain at least two
external legs. $k_a$ and $k_b$ are single legs and all external legs are positive helicity.
    }
    \label{fig:quad_cut_box}
\end{figure}

We use unitarity techniques~\cite{Bern:1994zx,Bern:1994cg,Britto:2004nc}
to determine the polylogarithmic terms. In particular, we use four-dimensional unitarity~\cite{Dunbar:2009ax}, where the cut legs are taken to be four-dimensional. Cutting three propagators can yield a product of two tree amplitudes. In four dimensions all such triple cuts of an all-plus amplitude vanish, as there are insufficient negative helicity legs to form two MHV tree amplitudes.  Cuts that yield the product of a one-loop amplitude and tree amplitudes can be obtained by considering diagrams where the one-loop all-plus amplitudes is inserted as a vertex~\cite{Dunbar:2016cxp,Dunbar:2017nfy}. Figure~\ref{fig:quad_cut_box} illustrates this process with the insertion of the one-loop amplitude into a box configuration. Such 'two mass easy' box configurations have non-vanishing quadruple cuts and contribute polylogarithms to the amplitude.
Both box and triangle configurations contribute to $U_{n:\lambda}^{(2)}$, while
bubble configurations were shown to have zero coefficient in~\cite{Dunbar:2009ax}.

To compute the Unitary cuts we need the one-loop partial amplitudes in closed form.    These are~\cite{Bern:1993qk,Dunbar:2019fcq},
\begin{eqnarray}
A_{n:1}^{(1)}(1^+,2^+,3^+,\cdots, n^+)  &=& -\frac{1}{3}{ \sum_{i<j<k<l} 
 \tr_-(ijkl)  
\over \spa1.2\spa2.3\cdots \spa{n}.1}
\notag\\
A_{n:2}^{(1)}(1^+ ;2^+,3^+,\cdots, n^+) &=& -{ \sum_{i<j}  [1|ij|1]  \over \spa2.3\spa3.4 \cdots \spa{n}.2} 
\\
A_{n:3}^{(1)}(1^+,2^+ ; 3^+,\cdots, n^+) &=& 2{ \spb1.2^2 \over \spa3.4 \spa4.5  \cdots \spa{n}.3} 
\notag \\
A_{n:r}^{(1)}(1^+,2^+,\cdots, r-1^+ ;  r^+, \cdots , n^+) &=& -2 {  (P_{1\ldots r-1}^2)^2   \over 
(\spa1.2\spa2.3 \cdots \spa{(r-1)}.1)  ( \spa{r}.{(r+1)}  \cdots \spa{n}.{r}  } \; , 
\notag
\end{eqnarray}
where we are using a spinor helicity formalism as defined in appendix~\ref{app:spin}.

\subsection{Full Color Results at $n$-point}

We express the results for the partial amplitudes in terms of ``box functions'' containing the polylogarithms  together with rational coefficients.  
The full box integral function contains both IR singular terms and finite terms which we separate, 
\begin{align}
    I_4^{2me}(S,T,K_2^2,K_4^2) = I_4^{2me} \Bigg\vert_{IR} - \frac{2}{ ST-K_2^2 K_4^2 } F^{2m}(S,T,K_2^2,K_4^2)
\end{align}
where $F^{2m}$ is a dimensionless function of polylogarithms.
Combining the IR singular terms from the box and triangle configurations yields  the required IR singular term, but truncated to $\Ord(\eps^0)$~\cite{Dunbar:2016cxp,Dunbar:2019fcq}.
Promoting this to its all-$\eps$ form yields the full singularity structure~\eqref{IRbits}.
The IR finite piece of the two mass easy box integral is
\begin{align}
    F^{2m}(S,T,K_2^2,K_4^2) =& \text{Li}_2 \left(1-\frac{K_2^2}{S}\right) + \text{Li}_2 \left(1-\frac{K_2^2}{T}\right) + \text{Li}_2 \left(1-\frac{K_4^2}{S}\right) + \text{Li}_2 \left(1-\frac{K_4^2}{T}\right) 
    \notag\\
    &- \text{Li}_2 \left( 1 - \frac{K_2^2 K_4^2}{ST} \right) + \frac{1}{2} \text{ln}^2 \left( \frac{S}{T}  \right)
\end{align}
and when $K_2^2=0$,
\begin{align}
    F^{2m}(S,T,0,K_4^2) =& \text{Li}_2 \left(1-\frac{K_4^2}{S}\right) + \text{Li}_2 \left(1-\frac{K_4^2}{T}\right) + \frac{1}{2} \text{ln}^2 \left( \frac{S}{T}  \right) + \frac{\pi^2}{6} \; , 
\end{align}
\def\Pbit{P}
\def\plF{{\rm F}}
with $S=(K_2+k_a)^2=(K_4+k_b)^2$ and $T=(K_2+k_b)^2=(K_4+k_a)^2$  as in fig. ~\ref{fig:quad_cut_box}.

It is convenient to move to a notation which focusses on the legs attached to each corner of the box. If the massive corners are described by the sets of external legs $A$ and $B$ 
and the null corners by the single legs $a$ and $b$ we define,
\begin{align}
    F(a,b;A;B)&= F^{2m}[K_{a,A}^2,K_{A,b}^2,K_{A}^2,K_{B}^2],
    \notag\\
    \hbox{ and }
    F(a,b;\{\},B) &= F(a,b;A,\{\}) = 0,
\end{align}
where $K^2_{a,A}=(k_{a}+k_{A_1}+k_{A_2}+\cdots+k_{A_s})^2$, etc.

The coefficients of the integral functions are obtained using quadruple cuts~\cite{Britto:2004nc} and are the products of on-shell amplitudes evaluated on the cut.   The different,  partial amplitudes give rise to a small set of coefficient functions given in the following.    For a box with the one-loop corner having $n_T$ external legs and the opposite corner whose tree amplitude has $n_S$ external legs,  firstly a term when the one-loop amplitude is the 
leading in color $A_{n_T+2:1}^{(1)}$, 
\def\mtr{n_S+2}
\def\mol{n_T+2}
\begin{align}
   & C_{1}(a,b,S_1,S_2,T_1,T_2)
    \notag\\
    &=A_{\mtr}^{(0)}(k^-,S_1,l^-,S_2)
    A_3^{(0)}(a^+,k^+,j^-)A_3^{(0)}(l^+,b^+,i^-)
    A
    ^{(1)}_{\mol:1}(i^+,T_1,j^+,T_2)
    \notag\\
    &=\frac{1}3\spa{a}.b^2C_{PT}(a,S_1,b,S_2)C_{PT}(b,T_1,a,T_2)
    \notag\\
    &\times\Bigg(
    \frac{\la b|T_1T_2|b\ra \la a|T_1T_2|a\ra}{\spa{a}.b^2}
    +\hspace{-0.5cm}\sum_{u<v<w<x\in K_4}\hspace{-0.5cm}\mathrm{tr}_-[uvwx]
    +\sum_{u<v\in T_1}
    \frac{K_4^2\la b|uv|a\ra+\la a|T_2uvK_4|b\ra}{\spa{b}.a}
    \notag\\
    &\hspace{1cm}+\sum_{u<v\in T_2}
    \frac{\la b|K_4uvT_2|a\ra }{\spa{b}.a}
    +\sum_{u<v<w\in K_4}\frac{\la b|uvwK_4|a\ra}{\spa{a}.b}
    +\sum_{u<v<w\in T_1}\frac{\la b|K_4uvw|a\ra}{\spa{a}.b}
    \notag\\
    &\hspace{1cm}
    +\sum_{u<v<w\in T_2}\frac{\la a|uvwK_4|b\ra}{\spa{b}.a}
   \Bigg)
   \times F(a,b;K_2=S_1\oplus S_2; K_4=T_1\oplus T_2),
    \label{eq:leadpolco}
\end{align}
where $|S_1\oplus S_2|=n_S$ and $|T_1\oplus T_2|=n_T$ and  the ``Parke-Taylor'' factor $C_{PT}$ is
\begin{equation}
C_{PT} (a_1,a_2,\cdots a_r) = \frac{1}{\spa{a_1}.{a_2}\spa{a_2}.{a_3} \cdots \spa{a_{r-1}}.{a_r} \spa{a_r}.{a_1}}  \; . 
\end{equation}
For $A_{n:r}^{(1)}$ with $r>2$ there are two cases,
\begin{align}
   & C_{2}(a,b,S_1,S_2,T_1,T_2,T_3)
    \notag\\
    &=A_{\mtr}^{(0)}(k^-,S_1,l^-,S_2)
    A_3^{(0)}(a^+,k^+,j^-)A_3^{(0)}(l^+,b^+,i^-)
    A
    ^{(1)}_{\mol:r}(i^+,T_1,j^+,T_2;T_3)
    \notag\\
    &=2\spa{a}.b^2C_{PT}(b,T_1,a,T_2)\,C_{PT}(T_3)C_{PT}(a,S_1,b,S_2)\times \Big(K_{T_3}^2\Big)^2
     \notag\\
   &\times F(a,b;S_1\oplus S_2; T_1\oplus T_2\oplus T_3)
   \end{align}
   where $|T_3|=r-1$ and
\begin{align}
&C_{3}(a,b,S_1,S_2,T_1,T_2)
 \notag\\
    &\equiv A_{\mtr}^{(0)}(k^-,S_1,l^-,S_2)
    A_3^{(0)}(a^+,k^+,j^-)A_3^{(0)}(l^+,b^+,i^-)
    A
    ^{(1)}_{\mol:r}(i^+,T_1; j^+,T_2)
    \notag\\
&=2\la a|K_{T_2}K_{T_1}|b\ra^2 C_{PT}(a,S_1,b,S_2)C_{PT}(b,T_1)C_{PT}(T_2,a).
 \notag\\
   &\times F(a,b;S_1\oplus S_2; T_1\oplus T_2)
\label{eq:coefnr}
\end{align}
where $|T_2|=r-2$.  For the case where we have a $A_{n:2}^{(1)}$  corner there is 
\begin{align}
&C_{4}(a,b,S_1,S_2,T_1,T_2,t_3)
 \notag\\
    &\equiv A_{\mtr}^{(0)}(k^-,S_1,l^-,S_2)
    A_3^{(0)}(a^+,k^+,j^-)A_3^{(0)}(l^+,b^+,i^-)
    A
    ^{(1)}_{\mol:2}(t_3\,;i^+,T_1, j^+,T_2)
    \notag\\
&=\spa{a}.b^2 C_{PT}(a,S_1,b,S_2)C_{PT}(b,T_1,a,T_2)
\notag\\
&\times \Bigg(
\frac{[t_3|K_4|a\ra[t_3|(K_{T_1}-K_{T_2})|b\ra}{\spa{a}.b}
+2[t_3|T_2T_1|t_3]
+\sum_{v<w\in K_4}[t_3|vw|t_3]
\Bigg)
 \notag\\
   &\times F(a,b;S_1\oplus S_2; T_1\oplus T_2\oplus t_3)
\end{align}
where $t_3$ is a single leg within $K_4$ so $K_4=K_{T_1}+K_{T_2}+k_3=-i-j$.  
\def\EXTRA{ and
\begin{align}
   & C_{5}(a,b,S_1,S_2,T_1)
    \notag\\
    &=A_{\mtr}^{(0)}(k^-,S_1,l^-,S_2)
    A_3^{(0)}(a^+,k^+,j^-)A_3^{(0)}(l^+,b^+,i^-)
    A
    ^{(1)}_{\mol:2}(i^+;j^+,T_1)
    \notag\\
    &=-\,C_{PT}(a,T_1)\,C_{PT}(a,S_1,b,S_2)
    \times \Big(
     K_4^2\la a|T_1K_4|a\ra
     + \sum_{u<v\in T_1}
   \la a|K_4uvK_4|a\ra
   \Big)
    \notag\\
   &\times F(a,b;S_1\oplus S_2; T_1).
   \end{align}
This fifth coefficient will never contribute to the boxes when fully color dressed due to the ever present color contribution
\begin{align}
    (\mathrm{Tr}[bil]-\mathrm{Tr}[bli])\mathrm{Tr}[i]=\mathrm{Tr}[bl]-\mathrm{Tr}[bl]=0,
\end{align}
where $b,\,i$ and $l$ are the momentum around the tree corner indicated in figure~\ref{fig:quad_cut_box}.
}  

The coefficient of a  box function will be sums of these kinematic terms which we now obtain by color dressing the two-mass boxes

\subsection{Leading in Color Term}
Defining the sets 
\begin{align}
    U_{ab}=\{a+1,a+2,\cdots,b-1\}\; \text{and} \; V_{ab}=\{b+1,\cdots,a-1\}  \; . 
\end{align}
That is split the list $\{1,2,\cdots,n\}$ into $\{a,U_{ab}\,,b,V_{ab}\}$ where $a$ is cycled to the front we have
\begin{align}
   \Pbit_{n:1}^{(2)}(1^+,2^+,\cdots,n^+)
    &=
    \sum_{1\leq a<b\leq n}
    \Bigg(
    C_{1}(a,b,U_{ab}\,,0,V_{ab}\,,0)
    +C_{1}(a,b,0,V_{ab}\,,0,U_{ab})
    \Bigg)  \; . 
\end{align}
This is alternate form of the previous result~\cite{Dunbar:2017nfy}.

\subsection{$\mathbf{SU(N_c)}$ Double Trace Terms}

Considering terms Tr$[X]$Tr$[Y]=$Tr$[x_1,x_2,\cdots,x_{r-1}]$Tr$[y_1,y_2,\cdots,y_{n+1-r}]$, if $a$ and $b$ are within the same trace we define $U_{ab}$ and $V_{ab}$ as before with respect to the elements of this trace, and define new lists $X_i,Y_{j}=X-\{i\}$ and $Y-\{j\}$ respectively. The ordering of these sets matters and  $X_i$ is defined to start with the $(i+1)$th element,
\begin{align}
    X=\{1,2,\cdots,r\}\to \{i,i+1,\cdots,i-1\}\to X_i= \{i+1,i+2,\cdots,i-1\}.
\end{align}
We also need to define $Spl_2$ as the set of splits of a list into two lists maintaining list order. So if $U=\{u_1,u_2,\cdots,u_r\}$
\begin{align}
    Spl_2(U)=\{U^i\}=(\{u_1,u_2,\cdots,u_i\},\{u_{i+1},\cdots,u_r\}).
\end{align}
This includes splits involving the empty set $(\{\},U)$ and $(U,\{\})$ and counts them separately. For later convenience we will define the sum over $Spl_2(U_a)$ as the sum over the sets $\{(A_1^i,A_2^i)\}\in Spl_2(U_a)$ and similarly for leg $b$, $\{(B_1^i,B_2^i)\}\in Spl_2(U_b)$, where legs $a$ and $b$ are the legs on the massless corners of the two-mass box.

Finally it will also be useful to define the following double sum 
\begin{align}
    \sum_{CSpl_2(U)}\equiv\sum_{V\in Z(U)}\sum_{Spl_2(V)},
\end{align}
which is a sum over the splits of all cycles of the set $U$. 
We can now write the amplitude as
\begin{align}
   &\Pbit_{n:r}^{(2)}(1^+,2^+,\cdots (r-1)^+; r^+,\cdots, n^+)
   = \Pbit_{n:r}^{(2)}(X;Y)
   \notag\\
   =&-\sum_{a\in X}\sum_{b\in Y}
   \sum_{Spl_2(X_a)}\sum_{Spl_2(Y_b)}
   \Big[
   C_{1}(a,b,B_2^j,A_2^i,B_1^j,A_1^i)+C_{1}(a,b,A_1^i,B_1^j,A_2^i,B_2^j)
   \Big]
   \notag\\
   &+\sum_{Z_2(X,Y)}\sum_{a<b\in X}
   \Bigg(
   \Big[ 
   C_{2}(a,b,0,V_{ab},0,U_{ab},Y)+
   C_{2}(a,b,U_{ab},0,V_{ab},0,Y)
   \Big]
   \notag\\
   &+
   \sum_{(A_1^i,A_2^i)\in CSpl_2(Y)}
   \Big[ 
   C_{1}(a,b,U_{ab},A_1^i,V_{ab},A_2^i)
   +C_{1}(a,b,A_1^i,V_{ab},A_2^i,U_{ab})
   \Big] 
   \Bigg),
\end{align}
where we have suppressed notation such as the sum over $Z_2$ meaning swapping $X$ and $Y$ within that sum and $a<b\in X$ being in terms of the ordering of $X.$
 This expression works for $r=2$ with the suitable $U(1)$ modification.  When $|Y|=1$ we replace $C_{2}(a,b,0,V_{ab},0,U_{ab},Y)$ with $C_{4}(a,b,0,V_{ab},0,U_{ab},Y)$ and many of the above sums become trivial.

\subsection{{$\mathbf{SU(N_c)}$}{} Triple Trace Terms}
We now consider terms Tr$[X]$Tr$[Y]$Tr$[Z]$ using obvious generalisations of previously defined sets.

\begin{align}
    &\Pbit_{n:s,t}^{(2)}(1^+,\cdots,s^+;(s+1)^+,\cdots,(s+t)^+;(s+t+1)^+,\cdots,n^+)
    =\Pbit_{n:s,t}^{(2)}(X;Y;Z)
    \notag\\
    &=\sum_{Z_{3}(X,Y,Z)}
    \Big(\sum_{Z_2(X,Y)}\sum_{a<b\in X}\sum_{(A_1^i,A_2^i)\in CSpl_2(Y)}
    \Big[
    C_{2}(a,b,U_{ab},A_1^i,V_{ab},A_2^i,Z)+
    C_{2}(a,b,A_1^i,V_{ab},A_2^i,U_{ab},Z)
    \Big]
    \notag\\
    &-\sum_{a\in X}\sum_{b\in Y}
    \sum_{Spl_2(X_a)}\sum_{Spl_2(Y_b)}
    \Big[
    C_{2}(a,b,B_2^j,A_2^i,B_1^j,A_1^i,Z)+C_{2}(a,b,A_1^i,B_1^j,A_2^i,B_2^j,Z)
    \Big]
    \Big).
\end{align}

Again if $s=1$ or $s=t=1$ we simply replace $C_{2}$ with $C_{4}$ in the sum where appropriate.

\subsection{{$N_c$}{}-independent Single Trace Term}

Finally, completing the $A_{n:1B}^{(2)}$ all-plus amplitude, we have 
\begin{align}
&\Pbit_{n:1B}^{(2)}(1^+,2^+,3^+,\cdots,n^+)=
\notag\\
& \sum_{a<b}\Bigg(
-\sum_{(U^i_1:U^i_2)\in Spl_2(U_{ab})}\sum_{(V^i_1:V^i_2)\in Spl_2(V_{ab})}
\Bigl[C_{3}(a,b,U_2^i,V_2^i,V_1^i,U_1^i)+C_{3}(a,b,U_1^i,V_1^i,U_2^i,V_2^i)\Bigr]
\notag\\
&+\sum_{(V_1^i,V_2^i,V_3^i)\in Spl_3(V_{ab})}C_{3}(a,b,U_{ab},V_2^i,V_1^i,V_3^i)
+\sum_{(U_1^i,U_2^i,U_3^i)\in Spl_3(U_{ab})}C_{3}(a,b,U_2^i,V_{ab},U_3^i,U_1^i)
\Bigg).
\label{eq:poly1b}
\end{align}
\subsection{Checks}
Although not manifest,  these expressions have the correct cyclic and flip symmetries. Additionally, the  following decoupling identities have been tested up to $10$-points:
\begin{equation}
\Pbit_{n:2}^{(2)}(1 ; 2,3,\cdots,n)  +\sum_{\sigma_1} \Pbit_{n:1}^{(2)}(\sigma_1)   =0   \; , 
\end{equation}
\begin{equation}
\Pbit^{(2)}_{n:3}(1,2;3,4,\ldots,n)+\Pbit^{(2)}_{n:1,1}(1;2;3,4,\ldots,n)
- \hspace{-0.8cm} \sum_{\sigma\in OP\{2,1\}\{3,\ldots,n\}} \Pbit_{n;1}^{(2)}(\sigma)=0  \; , 
\label{eq_COP}
\end{equation}

\begin{equation}
\Pbit^{(2)}_{n:1,s}(1;2,3,\ldots,s+1;s+2,\ldots,n)
+\sum_{\sigma_2}\Pbit^{(2)}_{n:s+2}(\sigma_2;s+2,\ldots,n)
+\sum_{\sigma_3}\Pbit^{(2)}_{n:s+1}(2,\ldots,s+1;\sigma_3)
=0
\end{equation}
where the sums over $\sigma_{1,2,3}$ are the sums over the different ways of inserting $1$ 
into $\{2,3,\ldots,n\},$ $\{2,3,\ldots,s+1\}$ and $\{s+2,\ldots,n\}$ respectively.

\section{Rational Terms}

The remaining rational pieces of the partial amplitudes, $R^{(2)}_{7:\lambda}$, are calculated using 
augmented recursion~\cite{Alston:2012xd,Alston:2015gea,Dunbar:2017nfy}.

Tree amplitudes are fully rational  and Britto, Cachazo, Feng and Witten showed how these can be obtained recursively from lower-point amplitudes by treating the amplitude as a function of complex momenta and investigating its pole structure~\cite{Britto:2005fq}.
For BCFW recursion, a complex shift is applied to two of the gluon momenta, $p_1$ and $p_2$, shifting the spinors as
\begin{align}
    \bar{\lambda}_1 \rightarrow \bar{\lambda}_{\hat{1}}(z) = \bar{\lambda}_1 - z \bar{\lambda}_2
\;\; , \; 
    \lambda_2 \rightarrow \lambda_{\hat{2}}(z) = \lambda_2 + z \lambda_1
    \label{BCFW_shift}
\end{align}
where $z$ is a new complex variable.
The momenta $\hat{p}_1(z)$ and $\hat{p}_2(z)$ remain on-shell and overall momentum conservation is preserved.
The rational amplitude can now be considered a complex function, $R(z)$.

Applying Cauchy's theorem to $R(z)/z$ over a contour at infinity, and assuming that $R(z)$ vanishes at large $|z|$, gives
\begin{align} 
    R(0) = -\sum_{z_{ij} \neq 0} \text{Res} \Big[ \frac{R(z)}{z} \Big] \Big\vert_{z_{ij}}
\label{eq:cauchy}\end{align}
where $z_{ij}$ are the positions of poles in $R(z)$.
$R(0)$ is the original  function that we wish to find.
For tree amplitudes a Feynman diagram decomposition shows that only simple poles arise and that these appear as a result of propagators going on shell:
\begin{align}
    \frac{1}{ \hat{p}^2_{ij}(z) } \equiv \frac{1}{ (p_i + \cdots + \hat{p}_1(z) + \cdots + p_j)^2 }
    \notag\\
    = \frac{1}{p_{ij}^2 - z \la 2|p_{ij}|1]}
    = \frac{1}{z-z_{ij}} \frac{(-1)}{ \la 2|p_{ij}|1] }
\end{align}
where 
\begin{align}
    z_{ij} = \frac{p_{ij}^2}{ \la 2|p_{ij}|1]}.
\end{align}
When a particular propagator goes on-shell, the structures on either side can be written as lower-point amplitudes,

\begin{align}
    \lim_{z \rightarrow z_{ij}} 
    A^{(tree)}(z) =  \frac{(-1)}{ \la 2|p_{ij}|1] } \sum_{h=\pm 1} A^{(tree):h}_L(z_{ij}) \frac{1}{z-z_{ij}} A^{(tree):-h}_R(z_{ij})
\end{align}
where $A^{(tree)}_L$ and $A^{(tree)}_R$ are tree amplitudes and the superscript $\pm h$ is shorthand for the helicity on the $p_{ij}$ leg entering the propagator.
This specifies the residues in these amplitudes and eq.~(\ref{eq:cauchy}) becomes
\begin{align}
    A^{(tree)} = \sum_{z_{ij} \neq 0} \sum_{h=\pm 1} A^{(tree):h}_L(z_{ij}) \frac{1}{p_{ij}^2} A^{(tree):-h}_R(z_{ij})\;,
\end{align}
which determines the tree amplitude on the left entirely in terms of lower point amplitudes.

There are alternatives to the  BCFW shift \eqref{BCFW_shift} that also introduce a complex parameter whilst maintaining momentum conservation. Different shifts can generate different behaviours as $|z|$ becomes large. A particularly useful example is the 
``Risager'' shift~\cite{Risager:2005vk}:
\begin{align}
    \lambda_1 \rightarrow \lambda_{\hat{1}}(z) &= \lambda_1 + z [23] \lambda_\eta  \; , 
    \notag\\
    \lambda_2 \rightarrow \lambda_{\hat{2}}(z) &= \lambda_2 + z [31] \lambda_\eta \; , 
    \notag\\
    \lambda_3 \rightarrow \lambda_{\hat{3}}(z) &= \lambda_3 + z [12] \lambda_\eta
\end{align}
where $\lambda_\eta$ is some spinor chosen such that $\spa{i}.{\eta} \neq 0$ for $i \in \{1,2,\cdots,n\}$.
That the final result should be independent of $\lambda_\eta$ is a powerful consistency check.

For loop amplitudes, the situation is more complicated.
Eq. ~(\ref{eq:cauchy}) can still be applied to the rational part of the amplitude,  but
there is now the possibility of double poles occurring. The complex analysis remains straightforward: given a Laurent expansion of the rational piece,

\begin{align}
    R^{(2)}_{n:\lambda}(z) = \sum_{z_{ij}} \Bigg( \frac{a^{(ij)}_{-2}}{(z-z_{ij})^2} + \frac{a^{(ij)}_{-1}}{(z-z_{ij})} + \Ord( (z-z_{ij})^0 ) \Bigg) \;,
\end{align}
the residues in eq.~(\ref{eq:cauchy}) are

\begin{align}
    \text{Res} \Big[ \frac{R(z)}{z} \Big] \Big\vert_{z_{ij}} = - \frac{a_{-2}}{z^2_{ij}} + \frac{a_{-1}}{z_{ij}}\;.
\end{align}
However, only the residue of the leading pole can be found by  straightforward factorisation.
The sub-leading pole receives non-factorising contributions, meaning we must find these through other means.
The method we follow is that of ``augmented recursion"~\cite{Dunbar:2016aux}.

For the all-plus amplitude the BCFW shift does not lead to $R(z)$ vanishing as $|z|$ becomes large~\cite{Dunbar:2016aux},
therefore we make use of the Risager shift.
Applying the Risager shift to $R^{(2)}_{7:\lambda}$ excites three types of pole structure:

\begin{itemize}
    \item Tree to two-loop factorisations
    \item One-loop to one-loop factorisations with at least three external momenta on either side
    \item One-loop to one-loop factorisations with two external legs on one side
\end{itemize}

\begin{figure}[h]
    \centering
    \hspace*{-0.5truecm}
    \includegraphics{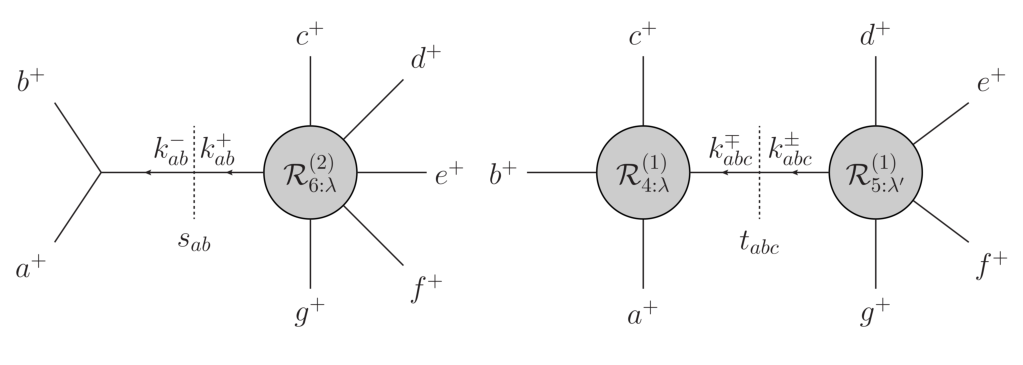}
    \caption{The tree to two-loop and one-loop to one-loop factorisations. These give rise to  simple poles in $s_{ab}$ and $t_{abc}$ respectively [general momentum labels $\{a,b,c,d,e,f,g\}$ are used to avoid confusion with the shifted set $\{\hat{1},\hat{2},\hat{3},4,5,6,7\}$].}
    \label{fig:nice_facts}
\end{figure}

The first two cases as shown in fig.~\ref{fig:nice_facts}. These contain only simple poles 
and so can be treated with simple recursion.

\begin{figure}[h]
    \centering
    \includegraphics{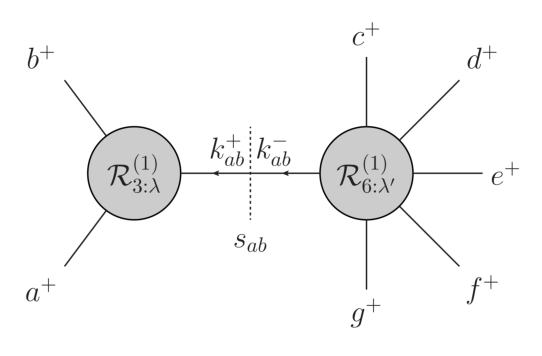}
    \caption{The one-loop to one-loop factorisation that gives rise to double poles in $\spa{a}.b$.}
    \label{fig:double_pole_factorisation}
\end{figure}

The third case introduces double poles.
Naively, we may wish to consider the one-loop to one-loop factorisation with the propagator $1/s_{ab}$ as 
shown in fig.~\ref{fig:double_pole_factorisation}. 
However, if the  momenta are complex, the one-loop three-point all-plus vertex itself contains a factor of $1/\spa{a}.b$,  giving a double pole overall. Only the leading term can be obtained from the factorisation and, as discussed above, both the leading and sub-leading poles are needed to evaluate the 
residue in eq. ~(\ref{eq:cauchy}).

\begin{figure}[h]
    \centering
    \includegraphics{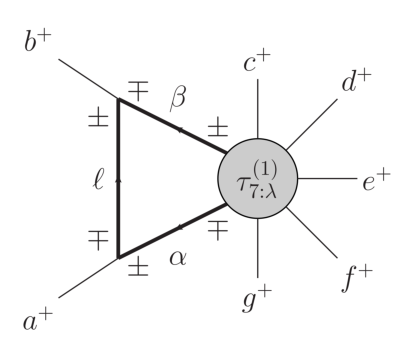}
    \caption{Augmented recursion diagram containing the poles in $s_{ab}$.
    Thick lines indicate off-shell propagators, over which the loop momentum integration 
    is performed.}
    \label{fig:augmented_recursion_diagram}
\end{figure}

We adopt the augmented recursion 
procedure~\cite{Alston:2012xd,Alston:2015gea,Dunbar:2016aux,Dunbar:2017nfy}  
and focus on the loop integral that yields the three-point one-loop all-plus vertex. The diagram of
interest is shown in fig.~\ref{fig:augmented_recursion_diagram}. The loop momentum integration  
is to be performed, so the propogators $\ell$, $\alpha$ and $\beta$ can be off-shell and  
$\tau_{7:\lambda}^{(1)}$ is a current with $\alpha$ and $\beta$ off-shell. As we are only interested in terms with poles as $s_{ab}\to 0$, $\tau_{7:\lambda}^{(1)}$ does not need to be exact the arbitrary $\alpha^2$ and $\beta^2$, instead it only needs to match certain limits~\cite{Alston:2015gea,Dunbar:2016aux}. This allows the $\tau_{7:\lambda}^{(1)}$ to be systematically 
generated from the known $A_{7:\lambda}^{(1)}$. 

The color dressed current contains multiple configurations of the external legs, including new currents that have not been required previously.
The leading in color rational piece $R^{(2)}_{7:1}$ required only $\tau^{(1)}_{7:1}(\alpha^-,\beta^+,c^+,d^+,e^+,f^+,g^+)$ for its derivation~\cite{Dunbar:2017nfy}.
Exploiting symmetries and decoupling identities allows us to reduce the number of distinct currents appearing the full color problem, leaving two new currents, $\tau^{(1)}_{7:1}(\alpha^-,c^+,\beta^+,d^+,e^+,f^+,g^+)$ and $\tau^{(1)}_{7:1}(\alpha^-,c^+,d^+,\beta^+,e^+,f^+,g^+)$, to consider. Performing the loop integration then yields the leading and sub-leading poles as $\spa{a}.b\to 0$. 

The currents we use together with details of their construction are given in appendix~\ref{app:cur}.

\subsection{Results and consistency checks}

Summing all of the recursive contributions for each color trace structure gives the rational pieces of the partial amplitudes, $R^{(2)}_{7:\lambda}$.  The augmented recursion procedure involves combining a large number of contributions at the seven-point level,  leading to results with many terms,  many of which depend on $\eta$.
Nonetheless,  the results are independent of the choice of Risager shift  spinor 
$\lambda_\eta$, which is strong evidence that the procedure was successful and that an appropriate shift was chosen: if $R(z)$ did not vanish as $|z|$ became large, 
as eq.~(\ref{eq:cauchy}) requires it to, then a $\lambda_\eta$ dependence would likely remain~\cite{Dunbar:2016aux}.
The $R^{(2)}_{7:\lambda}$ also have the correct cyclic symmetries $\mathcal{P}_{7:\lambda}$ in their arguments as well as satisfying all decoupling identities.

Compact, explicitly $\eta$ independent expressions for the $R^{(2)}_{7:\lambda}$ are obtained by fitting to the raw recursion result guided by the factorisation structure of the amplitude.
For some of the color structures it is convenient to base the analytic result on a decoupling identity, i.e. some results are quoted as the combination of other color structures appearing in a decoupling identity. We emphasis that all the color structures were calculated independently and all of the decoupling identities are satisfied by our raw forms, with none of the decoupling identities being assumed to hold.

\subsection{R$_{7:1}$}

We confirm the result of~\cite{Dunbar:2017nfy} and reproduce that form here for completeness,
\begin{align}
    R_{7:1}(a,b,c,d,e,f,g) =& \frac{i}{9} \sum_{\mathcal{P}_{7:1}} \frac{G_{7:1}^1 + G_{7:1}^2 + G_{7:1}^3 + G_{7:1}^4 + G_{7:1}^5 + G_{7:1}^6 + G_{7:1}^7}{ \spa{a}.{b}\spa{b}.{c}\spa{c}.{d}\spa{d}.{e}\spa{e}.{f}\spa{f}.{g}\spa{g}.{a} } ,
\end{align}
where
\begin{align}
    G^1_{7:1} =
    \frac{\spa{g}.{a}}{t_{abc}t_{efg}} \Bigg( &
    \frac{ \spa{c}.{d} \spb{e}.{g} [d|k_{abc}|e\ra [a|k_{abc}|e\ra [c|k_{abc}|f\ra }{ \spa{e}.{f} }
    -\frac{ \spa{d}.{e} \spb{c}.{a} [d|k_{efg}|c\ra [g|k_{efg}|c\ra [e|k_{efg}|b\ra }{\spa{b}.{c}}
    \notag\\
     +&\frac{ \spa{e}.{f} \spa{c}.{d} \spb{c}.{a} \spb{f}.{g} [e|k_{efg}|a\ra [d|k_{efg}|b\ra }{\spa{a}.{b}}
    -\frac{ \spa{b}.{c} \spa{d}.{e} \spb{e}.{g} \spb{a}.{b} [c|k_{abc}|g\ra [d|k_{abc}|f\ra }{\spa{f}.{g}}
    \Bigg),
\end{align}
\begin{align}
    G^2_{7:1} = \frac{1}{t_{abc}t_{efg}} s_{cd} s_{de} \spa{g}.{a} [g|k_{efg}k_{abc}|a],
\end{align}
\begin{align}
    G^3_{7:1} = \frac{1}{t_{cde}} \Bigg( &
    s_{ce} \bigg( \frac{s_{ef} \la c|k_{ab}k_{fga}|d\ra }{\spa{c}.{d}} - \frac{ s_{bc} \la e|k_{fg}k_{gab}|d\ra }{\spa{d}.{e}} \bigg)
    + \frac{ \spa{e}.{f}\spa{b}.{c}\spb{f}.{b} [c|k_{cde}|g\ra [e|k_{cde}|a\ra }{\spa{g}.{a}}
    \notag\\
    & + \frac{ \spa{b}.{c} [c|k_{cde}|b\ra [e|k_{cde}|a\ra [b|k_{fg}|e\ra }{\spa{a}.{b}}
  + \frac{ \spa{e}.{f} [e|k_{cde}|f\ra [c|k_{cde}|g\ra [f|k_{ab}|c\ra }{\spa{f}.{g}}
    \Bigg),
\end{align}
\begin{align}
    G^4_{7:1} = \frac{\spb{g}.{a}}{\spa{g}.{a}} \spa{g}.{e} \spa{a}.{e}
    \left( \frac{\spb{d}.{e}}{\spa{d}.{e}} \spa{d}.{g} \spa{d}.{a} +  \frac{\spb{e}.{f}}{\spa{e}.{f}} \spa{f}.{g} \spa{f}.{a} \right),
\end{align}
\begin{align}
    G^5_{7:1} = \frac{1}{t_{cde}} \Big( &
    \spb{c}.{e} ( \spa{e}.{f} \spb{d}.{f} \la c|k_{ab}k_{fga}|d\ra + 
    \spa{b}.{c} \spb{d}.{b} \la e|k_{fg}k_{gab}|d\ra )
    \notag\\
    & + \spa{b}.{c}\spa{e}.{f} (2 \spa{g}.{a} \spb{c}.{e}\spb{f}.{g}\spb{a}.{b} 
    + \spb{b}.{f} [e|k_{ab}k_{fg}|c] )
    \Big),
\end{align}
\begin{align}
    G^6_{7:1} = \frac{1}{\spa{g}.{a}} ( \la g|fk_{bc}|a\ra t_{efg} - \la a|bk_{ef}|g\ra t_{abc} )
\end{align}
and
\begin{align}
    G^7_{7:1} =&
    s_{bf}^2 - 2s_{ga}^2 - 3s_{db}s_{df} + 4s_{da}s_{dg} 
    - 6s_{ac}s_{eg} + 7(s_{eb}s_{fc}+s_{ea}s_{gc}) + s_{ab}s_{fg}
    \notag\\
    & + 3s_{fa}s_{gb} + s_{ce} \big( s_{cf} + s_{eb} - 4(s_{ab}+s_{fg}+s_{ga}) + 5(s_{dg}+s_{ad}) \big)
    \notag\\
    & + 4[e|bcf|e\ra - 2[f|gab|f\ra + 3[g|baf|g\ra + 2[g|cea|g\ra.
\end{align}

\subsection{R$_{7:2}$}

Expressed in terms of the leading in color partial amplitude, via a decoupling identity, we have 
\begin{align}
    R_{7:2}(a;b,c,d,e,f,g) =& - R_{7:1}(a, b, c, d, e, f, g) - R_{7:1}(a, c, d, e, f, g, b)  \notag\\ 
    &- R_{7:1}(a, d, e, f, g, b, c) - R_{7:1}(a, e, f, g, b, c, d) \notag\\ 
    &- R_{7:1}(a, f, g, b, c, d, e) - R_{7:1}(a, g, b, c, d, e, f) \notag\\
    =& - \sum_{Z_6(bcdefg)} R_{7:1}(a,b,c,d,e,f,g).
\end{align}

\subsection{R$_{7:3}$}

The first new $SU(N_c)$ rational piece to be calculated is $R_{7:3}$.
Using the decoupling identity, we can express it in terms of the previously defined partial amplitude and the new $R_{7:1,1}$,
\begin{align}
    R_{7:3}(a,b;c,d,e,f,g) =& -R_{7:1,1}(a;b;c,d,e,f,g) -R_{7:2}(b;a,c,d,e,f,g) \notag\\
    &- R_{7:2}(b;a, d, e, f, g, c) - R_{7:2}(b;a, e, f, g, c, d) \notag\\
    &- R_{7:2}(b;a, f, g, c, d, e) - R_{7:2}(b;a, g, c, d, e, f) \notag\\
    =& -R_{7:1,1}(a;b;c,d,e,f,g) -\sum_{Z_5(cdefg)} R_{7:2}(b;a,c,d,e,f,g).
\end{align}
We make this choice because $R_{7:1,1}$ contains only simple poles whereas $R_{7:3}$ also contains double poles, so the former can be stated more compactly.

\subsection{R$_{7:4}$} \label{sec:R74-rational-compact-results}

The second new $SU(N_c)$ partial amplitude to be calculated is $R_{7:4}$.
The result obtained from augmented recursion is analytic, however the manifestly symmetric form presented here required some additional work to obtain in such a compact form.
\begin{equation}
    R_{7:4}(a,b,c;d,e,f,g) = \sum_{\mathcal{P}_{7:4}} B_{7:4}(a,b,c,d,e,f,g),
\end{equation}
where the basis
\begin{align}
    B_{7:4}(a,b,c,d,e,f,g) =& B_{7:4}^{(ts)} + B_{7:4}^{(t1)} + B_{7:4}^{(t2)} + B_{7:4}^{(t3)} 
    \notag\\
    & + B_{7:4}^{(t4)} + B_{7:4}^{(s1)} + B_{7:4}^{(s2)}
\end{align}
can be divided into various denominator structures:
\begin{align}
    B_{7:4}^{(ts)} = &\frac{2i}{9} \frac{s_{df}}{\spa{f}.{g}^2 t_{fgd}} \frac{\spb{d}.{e}^2}{\spa{a}.{b} \spa{b}.{c} \spa{c}.{a}}
    \notag\\
    & - \frac{2i}{9} \frac{s_{ef}}{\spa{f}.{g}^2 t_{efg}} \frac{\spb{e}.{d}^2}{\spa{a}.{b} \spa{b}.{c} \spa{c}.{a}},
\end{align}
\begin{align}
    B_{7:4}^{(t1)} = - \frac{1}{9} \frac{i}{t_{abc}} & \frac{1}{\spa{b}.{c}  \spa{c}.{a}} \frac{1}{ \spa{d}.{e}  \spa{e}.{f}  \spa{f}.{g}  \spa{g}.{d} } 
    \notag\\
    ( & [a|k_{abc}|e\ra  [b|k_{abc}|d\ra  \spb{d}.{e} +
    [a|k_{abc}|e\ra  s_{fg}  \spb{b}.{e}
    \notag\\
    & + [b|k_{fg}|d\ra  s_{fg}  \spb{a}.{d} +
    s_{fg}  \spa{f}.{g}  \spb{a}.{g}  \spb{b}.{f} 
    \notag\\
    & + \spa{e}.{f}  \spa{f}.{g}  \spb{a}.{f}  \spb{b}.{f}  \spb{e}.{g}
    ),
\end{align}
\begin{align}
    B_{7:4}^{(t2)} = \frac{1}{9} & \frac{i}{t_{abc}} \frac{\spb{a}.{b}}{ \spa{a}.{b}  \spa{b}.{c}  \spa{c}.{a} } \frac{1}{ \spa{d}.{e}  \spa{d}.{g}  \spa{e}.{f}  \spa{f}.{g} }
    \notag\\
    & ( [f|k_{abc}|b\ra  \spa{a}.{g}  \spa{e}.{f}  \spb{e}.{g} - 
    [f|k_{abc}|b\ra  \spa{a}.{d}  \spa{e}.{f}  \spb{d}.{e}
    \notag\\
    & + [g|k_{abc}|b\ra  s_{ef}  \spa{a}.{g} -
    [d|k_{abc}|a\ra  [g|k_{abc}|b\ra  \spa{d}.{g}
    ),
\end{align}
\begin{align}
    B_{7:4}^{(t3)} = \frac{1}{9} \frac{i}{t_{efg}} & \frac{1}{\spa{e}.{f}  \spa{f}.{g}} 
    \notag\\
    \Bigg( & \frac{ ( [e|k_{efg}|b\ra  [g|k_{efg}|c\ra  \spb{b}.{c} + [e|k_{efg}|a\ra  [g|k_{efg}|d\ra  \spb{d}.{a} ) }{\spa{a}.{b}  \spa{b}.{c}  \spa{c}.{d}  \spa{d}.{a}} 
    \notag\\
    & + \frac{([e|k_{efg}|a\ra  [g|k_{efg}|c\ra  \spb{c}.{a}  + [e|k_{efg}|b\ra  [g|k_{efg}|d\ra  \spb{d}.{b})}{\spa{a}.{d}  \spa{b}.{c}  \spa{c}.{a}  \spa{d}.{b}} 
    \notag\\
    & + \frac{ ([e|k_{efg}|a\ra  [g|k_{efg}|b\ra  \spb{a}.{b}  + [e|k_{efg}|c\ra  [g|k_{efg}|d\ra  \spb{d}.{c}) }{\spa{a}.{b}  \spa{b}.{d}  \spa{c}.{a}  \spa{d}.{c}}
    \Bigg),
\end{align}
\begin{align}
    B_{7:4}^{(t4)} = \frac{2i}{9} & \frac{1}{t_{efg}} \frac{1}{\spa{a}.{b} \spa{b}.{c} \spa{c}.{a} } \notag\\
    & \Bigg( \frac{\spb{f}.{g} \spb{d}.{e}^2 }{ \spa{f}.{g} } 
    + 2 \frac{ \spb{e}.{g} \spb{d}.{g} \spb{d}.{e} \spa{e}.{g} }{ \spa{e}.{f} \spa{f}.{g} }
    \Bigg),
\end{align}
\begin{align}
    B_{7:4}^{(s1)} = - \frac{i}{3} & \frac{ \spb{a}.{b} }{ \spa{a}.{b}^2 } \frac{1}{ \spa{c}.{d} \spa{d}.{e}  \spa{e}.{f}  \spa{f}.{g}  \spa{g}.{c} }
    \notag\\
    &( - \la b|c|g|a\ra + \la b|d|e|a\ra + \la b|d|f|a\ra + \la b|e|f|a\ra )
\end{align}
and
\begin{align}
    B_{7:4}^{(s2)} = & - \frac{2i}{3} \frac{1}{\spa{f}.{g}^2} \frac{G_{7:4}^{1}}{ \spa{a}.{b}\spa{b}.{c}\spa{c}.{a} } \frac{1}{ \spa{d}.{e}\spa{e}.{f}\spa{g}.{d} } \frac{1}{\spa{a}.{d}} 
    \notag\\
    & - \frac{i}{72} \frac{1}{\spa{f}.{g}^2} \frac{ G_{7:4}^{2} + G_{7:4}^{3} + G_{7:4}^{4} }{ \spa{a}.{b}\spa{b}.{c}\spa{c}.{a} } \frac{1}{ \spa{d}.{e}\spa{e}.{f}\spa{g}.{d} } \frac{1}{\spa{a}.{e}}.
    \label{eq:denom}
\end{align}
The numerators in the latter piece can be written
\begin{align}
    G_{7:4}^{1} =  &
    \spa{a}.{f}  \spa{b}.{d}  \spa{b}.{f}  \spa{d}.{g}  \spb{b}.{d}  \spb{b}.{f} 
    + 
    \spa{a}.{f}  \spa{b}.{d}  \spa{c}.{g}  \spa{d}.{f}  \spb{b}.{f}  \spb{c}.{d}
    \notag\\&+
    \spa{a}.{f}  \spa{b}.{f}  \spa{c}.{d}  \spa{d}.{g}  \spb{b}.{d}  \spb{c}.{f}
    +
    \spa{a}.{f}  \spa{c}.{d}  \spa{c}.{f}  \spa{d}.{g}  \spb{c}.{d}  \spb{c}.{f}
    \notag\\& +
    2 \spa{a}.{f}  \spa{b}.{f}  \spa{d}.{e}  \spa{d}.{g}  \spb{b}.{f} \spb{d}.{e}
    -
    \spa{a}.{f}  \spa{b}.{e}  \spa{d}.{f}  \spa{d}.{g}  \spb{b}.{f} \spb{d}.{e}
    \notag\\& +
    2 \spa{a}.{e}  \spa{c}.{g}  \spa{d}.{f} ^2 \spb{c}.{f}  \spb{d}.{e}
    -
    \spa{a}.{f}  \spa{c}.{e}  \spa{d}.{f}  \spa{d}.{g}  \spb{c}.{f} \spb{d}.{e}
    \notag\\& -
    \spa{a}.{f}  \spa{b}.{f}  \spa{d}.{e}  \spa{d}.{g}  \spb{b}.{e} \spb{d}.{f}
    -
    \spa{a}.{e}  \spa{c}.{g}  \spa{d}.{f} ^2 \spb{c}.{e}  \spb{d}.{f}
    \notag\\& +
    \spa{a}.{e}  \spa{d}.{f}  \spa{d}.{g}  \spa{e}.{f}  \spb{d}.{e} \spb{e}.{f},
\end{align}
\begin{align}
    G_{7:4}^{2} =  &
    -12 \spa{a}.{g}  \spa{b}.{d}  \spa{b}.{f}  \spa{c}.{e}  \spb{b}.{c}  \spb{b}.{d}  + 
    44 \spa{a}.{g}  \spa{b}.{d}  \spa{b}.{f}  \spa{e}.{d}  \spb{b}.{d} ^2 
    \notag\\& -
    32 \spa{a}.{f}  \spa{b}.{d} ^2 \spa{e}.{g}  \spb{b}.{d} ^2 + 
    138 \spa{a}.{g}  \spa{b}.{e}  \spa{b}.{f}  \spa{c}.{e}  \spb{b}.{c}  \spb{b}.{e}
    \notag\\&  - 
    138 \spa{a}.{g}  \spa{b}.{e}  \spa{b}.{f}  \spa{e}.{d}  \spb{b}.{d}  \spb{b}.{e}  - 
    740 \spa{a}.{f}  \spa{b}.{d}  \spa{b}.{e}  \spa{e}.{g}  \spb{b}.{d}  \spb{b}.{e} 
    \notag\\& - 
    70 \spa{a}.{f}  \spa{b}.{e} ^2 \spa{e}.{g}  \spb{b}.{e} ^2 + 
    12 \spa{a}.{f}  \spa{b}.{d}  \spa{b}.{f}  \spa{e}.{g}  \spb{b}.{d}  \spb{b}.{f} 
    \notag\\& +  
    646 \spa{a}.{f}  \spa{b}.{e}  \spa{b}.{f}  \spa{e}.{g}  \spb{b}.{e}  \spb{b}.{f}  + 
    8 \spa{a}.{g}  \spa{b}.{f}  \spa{b}.{g}  \spa{e}.{d}  \spb{b}.{d}  \spb{b}.{g}
    \notag\\&  -  
    32 \spa{a}.{g}  \spa{b}.{d}  \spa{b}.{g}  \spa{e}.{f}  \spb{b}.{d}  \spb{b}.{g}  + 
    36 \spa{a}.{g}  \spa{b}.{d}  \spa{b}.{f}  \spa{e}.{g}  \spb{b}.{d}  \spb{b}.{g} 
    \notag\\& -  
    138 \spa{a}.{g}  \spa{b}.{e}  \spa{b}.{f}  \spa{e}.{g}  \spb{b}.{e}  \spb{b}.{g}  + 
    8 \spa{a}.{g}  \spa{b}.{d}  \spa{c}.{f}  \spa{e}.{d}  \spb{b}.{d}  \spb{c}.{d} 
    \notag\\& -  
    12 \spa{a}.{f}  \spa{b}.{d}  \spa{c}.{g}  \spa{e}.{d}  \spb{b}.{d}  \spb{c}.{d}  + 
    36 \spa{a}.{d}  \spa{b}.{d}  \spa{c}.{g}  \spa{e}.{f}  \spb{b}.{d}  \spb{c}.{d} 
    \notag\\& -  
    32 \spa{a}.{f}  \spa{b}.{d}  \spa{c}.{d}  \spa{e}.{g}  \spb{b}.{d}  \spb{c}.{d}  - 
    54 \spa{a}.{f}  \spa{b}.{d}  \spa{c}.{e}  \spa{e}.{g}  \spb{b}.{e}  \spb{c}.{d} 
    \notag\\& -  
    8 \spa{a}.{f}  \spa{b}.{f}  \spa{c}.{g}  \spa{e}.{d}  \spb{b}.{f}  \spb{c}.{d}  - 
    24 \spa{a}.{g}  \spa{b}.{d}  \spa{c}.{f}  \spa{e}.{f}  \spb{b}.{f}  \spb{c}.{d} 
    \notag\\& +  
    32 \spa{a}.{f}  \spa{b}.{d}  \spa{c}.{g}  \spa{e}.{f}  \spb{b}.{f}  \spb{c}.{d}  - 
    40 \spa{a}.{f}  \spa{c}.{d}  \spa{c}.{g}  \spa{e}.{d}  \spb{c}.{d} ^2 
    \notag\\&+  
    40 \spa{a}.{d}  \spa{c}.{d}  \spa{c}.{g}  \spa{e}.{f}  \spb{c}.{d} ^2 + 
    24 \spa{a}.{g}  \spa{b}.{f}  \spa{c}.{e} ^2 \spb{b}.{c}  \spb{c}.{e}  
    \notag\\&-  
    48 \spa{a}.{g}  \spa{b}.{e}  \spa{c}.{f}  \spa{e}.{d}  \spb{b}.{d}  \spb{c}.{e}  - 
    106 \spa{a}.{f}  \spa{b}.{e}  \spa{c}.{g}  \spa{e}.{d}  \spb{b}.{d}  \spb{c}.{e}
    \notag\\&  +  
    690 \spa{a}.{d}  \spa{b}.{e}  \spa{c}.{g}  \spa{e}.{f}  \spb{b}.{d}  \spb{c}.{e}  - 
    710 \spa{a}.{f}  \spa{b}.{d}  \spa{c}.{e}  \spa{e}.{g}  \spb{b}.{d}  \spb{c}.{e} 
    \notag\\& +  
    40 \spa{a}.{e}  \spa{b}.{e}  \spa{c}.{g}  \spa{e}.{f}  \spb{b}.{e}  \spb{c}.{e}  - 
    70 \spa{a}.{f}  \spa{b}.{e}  \spa{c}.{e}  \spa{e}.{g}  \spb{b}.{e}  \spb{c}.{e}
    \notag\\&  -  
    40 \spa{a}.{g}  \spa{b}.{e}  \spa{c}.{f}  \spa{e}.{f}  \spb{b}.{f}  \spb{c}.{e}  + 
    560 \spa{a}.{f}  \spa{b}.{e}  \spa{c}.{g}  \spa{e}.{f}  \spb{b}.{f}  \spb{c}.{e} ,
    \intertext{}
    G_{7:4}^{3} =  &
    54 \spa{a}.{f}  \spa{b}.{f}  \spa{c}.{e}  \spa{e}.{g}  \spb{b}.{f}  \spb{c}.{e}  - 
    24 \spa{a}.{g}  \spa{b}.{g}  \spa{c}.{e}  \spa{e}.{f}  \spb{b}.{g}  \spb{c}.{e} 
    \notag\\& -  
    108 \spa{a}.{f}  \spa{b}.{e}  \spa{c}.{g}  \spa{e}.{g}  \spb{b}.{g}  \spb{c}.{e}  - 
    104 \spa{a}.{f}  \spa{c}.{e}  \spa{c}.{g}  \spa{e}.{d}  \spb{c}.{d}  \spb{c}.{e} 
    \notag\\& + 
    108 \spa{a}.{e}  \spa{c}.{f}  \spa{c}.{g}  \spa{e}.{d}  \spb{c}.{d}  \spb{c}.{e}  + 
    664 \spa{a}.{d}  \spa{c}.{e}  \spa{c}.{g}  \spa{e}.{f}  \spb{c}.{d}  \spb{c}.{e} 
    \notag\\& - 
    24 \spa{a}.{f}  \spa{c}.{d}  \spa{c}.{e}  \spa{e}.{g}  \spb{c}.{d}  \spb{c}.{e}  - 
    14 \spa{a}.{e}  \spa{c}.{e}  \spa{c}.{g}  \spa{e}.{f}  \spb{c}.{e} ^2 
    \notag\\&+ 
    706 \spa{a}.{f}  \spa{b}.{f}  \spa{c}.{e}  \spa{e}.{g}  \spb{b}.{e}  \spb{c}.{f}  + 
    44 \spa{a}.{f}  \spa{c}.{f}  \spa{c}.{g}  \spa{e}.{d}  \spb{c}.{d}  \spb{c}.{f} 
    \notag\\& + 
    8 \spa{a}.{f}  \spa{c}.{d}  \spa{c}.{g}  \spa{e}.{f}  \spb{c}.{d}  \spb{c}.{f}  - 
    52 \spa{a}.{d}  \spa{c}.{f}  \spa{c}.{g}  \spa{e}.{f}  \spb{c}.{d}  \spb{c}.{f} 
    \notag\\& + 
    536 \spa{a}.{f}  \spa{c}.{e}  \spa{c}.{g}  \spa{e}.{f}  \spb{c}.{e}  \spb{c}.{f}  - 
    652 \spa{a}.{e}  \spa{c}.{f}  \spa{c}.{g}  \spa{e}.{f}  \spb{c}.{e}  \spb{c}.{f} 
    \notag\\& - 
    16 \spa{a}.{f}  \spa{c}.{e}  \spa{c}.{f}  \spa{e}.{g}  \spb{c}.{e}  \spb{c}.{f}  + 
    8 \spa{a}.{g}  \spa{b}.{g}  \spa{c}.{f}  \spa{e}.{d}  \spb{b}.{d}  \spb{c}.{g} 
    \notag\\& - 
    8 \spa{a}.{g}  \spa{b}.{g}  \spa{c}.{d}  \spa{e}.{f}  \spb{b}.{d}  \spb{c}.{g}  - 
    4 \spa{a}.{f}  \spa{c}.{g} ^2 \spa{e}.{d}  \spb{c}.{d}  \spb{c}.{g} 
    \notag\\&  +
    4 \spa{a}.{d}  \spa{c}.{g} ^2 \spa{e}.{f}  \spb{c}.{d}  \spb{c}.{g}  + 
    54 \spa{a}.{e}  \spa{c}.{g} ^2 \spa{e}.{f}  \spb{c}.{e}  \spb{c}.{g}  
    \notag\\&- 
    108 \spa{a}.{f}  \spa{c}.{e}  \spa{c}.{g}  \spa{e}.{g}  \spb{c}.{e}  \spb{c}.{g}  - 
    32 \spa{a}.{f}  \spa{b}.{f}  \spa{d}.{g}  \spa{e}.{d}  \spb{b}.{d}  \spb{d}.{f} 
    \notag\\& - 
    32 \spa{a}.{f}  \spa{b}.{f}  \spa{d}.{g}  \spa{e}.{f}  \spb{b}.{f}  \spb{d}.{f}  + 
    32 \spa{a}.{f}  \spa{b}.{f}  \spa{d}.{f}  \spa{e}.{g}  \spb{b}.{f}  \spb{d}.{f} 
    \notag\\& - 
    8 \spa{a}.{f}  \spa{b}.{f}  \spa{d}.{g}  \spa{e}.{g}  \spb{b}.{g}  \spb{d}.{f}  - 
    231 \spa{a}.{f}  \spa{c}.{g}  \spa{d}.{f}  \spa{e}.{d}  \spb{c}.{d}  \spb{d}.{f}
    \notag\\&  + 
    8 \spa{a}.{f}  \spa{c}.{f}  \spa{d}.{g}  \spa{e}.{d}  \spb{c}.{d}  \spb{d}.{f}  + 
    183 \spa{a}.{d}  \spa{c}.{g}  \spa{d}.{f}  \spa{e}.{f}  \spb{c}.{d}  \spb{d}.{f} 
    \notag\\& + 
    32 \spa{a}.{f}  \spa{c}.{d}  \spa{d}.{f}  \spa{e}.{g}  \spb{c}.{d}  \spb{d}.{f}  - 
    24 \spa{a}.{f}  \spa{c}.{g}  \spa{d}.{f}  \spa{e}.{f}  \spb{c}.{f}  \spb{d}.{f} 
    \notag\\& - 
    16 \spa{a}.{f}  \spa{c}.{f}  \spa{d}.{g}  \spa{e}.{f}  \spb{c}.{f}  \spb{d}.{f}  + 
    40 \spa{a}.{f}  \spa{c}.{f}  \spa{d}.{f}  \spa{e}.{g}  \spb{c}.{f}  \spb{d}.{f} 
    \notag\\& +
    16 \spa{a}.{f}  \spa{c}.{g}  \spa{d}.{g}  \spa{e}.{f}  \spb{c}.{g}  \spb{d}.{f}  - 
    16 \spa{a}.{f}  \spa{c}.{g}  \spa{d}.{f}  \spa{e}.{g}  \spb{c}.{g}  \spb{d}.{f} 
    \notag\\& - 
    8 \spa{a}.{f}  \spa{c}.{f}  \spa{d}.{g}  \spa{e}.{g}  \spb{c}.{g}  \spb{d}.{f}  + 
    32 \spa{a}.{f}  \spa{d}.{f} ^2 \spa{e}.{g}  \spb{d}.{f} ^2 
    \notag\\&+ 
    8 \spa{a}.{g}  \spa{b}.{f}  \spa{d}.{g}  \spa{e}.{d}  \spb{b}.{d}  \spb{d}.{g}  - 
    32 \spa{a}.{f}  \spa{b}.{d}  \spa{d}.{g}  \spa{e}.{g}  \spb{b}.{d}  \spb{d}.{g} 
    \notag\\& + 
    32 \spa{a}.{g}  \spa{b}.{g}  \spa{d}.{f}  \spa{e}.{f}  \spb{b}.{f}  \spb{d}.{g}  - 
    24 \spa{a}.{f}  \spa{b}.{f}  \spa{d}.{g}  \spa{e}.{g}  \spb{b}.{f}  \spb{d}.{g} 
    \notag\\& - 
    24 \spa{a}.{f}  \spa{c}.{g}  \spa{d}.{g}  \spa{e}.{f}  \spb{c}.{f}  \spb{d}.{g}  + 
    32 \spa{a}.{f}  \spa{c}.{g}  \spa{d}.{f}  \spa{e}.{g}  \spb{c}.{f}  \spb{d}.{g}
    \notag\\& +
    32 \spa{a}.{f}  \spa{d}.{f}  \spa{d}.{g}  \spa{e}.{g}  \spb{d}.{f}  \spb{d}.{g}  - 
    506 \spa{a}.{f}  \spa{b}.{d}  \spa{e}.{d}  \spa{e}.{g}  \spb{b}.{d}  \spb{e}.{d} 
    \notag\\& - 
    180 \spa{a}.{d}  \spa{b}.{d}  \spa{e}.{f}  \spa{e}.{g}  \spb{b}.{d}  \spb{e}.{d}  - 
    491 \spa{a}.{f}  \spa{b}.{e}  \spa{e}.{d}  \spa{e}.{g}  \spb{b}.{e}  \spb{e}.{d} 
    \notag\\& + 
    334 \spa{a}.{d}  \spa{b}.{e}  \spa{e}.{f}  \spa{e}.{g}  \spb{b}.{e}  \spb{e}.{d}  + 
    832 \spa{a}.{f}  \spa{b}.{f}  \spa{e}.{d}  \spa{e}.{g}  \spb{b}.{f}  \spb{e}.{d}    
    \intertext{and}
    G_{7:4}^{4} = &
    6 \spa{a}.{f}  \spa{b}.{d}  \spa{e}.{f}  \spa{e}.{g}  \spb{b}.{f}  \spb{e}.{d}  + 
    54 \spa{a}.{f}  \spa{b}.{d}  \spa{e}.{g} ^2 \spb{b}.{g}  \spb{e}.{d}  
    \notag\\&- 
    203 \spa{a}.{e}  \spa{c}.{g}  \spa{e}.{d}  \spa{e}.{f}  \spb{c}.{e}  \spb{e}.{d}  - 
    87 \spa{a}.{d}  \spa{c}.{e}  \spa{e}.{f}  \spa{e}.{g}  \spb{c}.{e}  \spb{e}.{d} 
    \notag\\& - 
    199 \spa{a}.{f}  \spa{c}.{g}  \spa{e}.{d}  \spa{e}.{f}  \spb{c}.{f}  \spb{e}.{d}  + 
    167 \spa{a}.{d}  \spa{c}.{g}  \spa{e}.{f} ^2 \spb{c}.{f}  \spb{e}.{d} 
    \notag\\& + 
    40 \spa{a}.{f}  \spa{c}.{f}  \spa{e}.{d}  \spa{e}.{g}  \spb{c}.{f}  \spb{e}.{d}  - 
    24 \spa{a}.{f}  \spa{c}.{d}  \spa{e}.{f}  \spa{e}.{g}  \spb{c}.{f}  \spb{e}.{d} 
    \notag\\& - 
    668 \spa{a}.{d}  \spa{c}.{g}  \spa{e}.{f}  \spa{e}.{g}  \spb{c}.{g}  \spb{e}.{d}  + 
    24 \spa{a}.{f}  \spa{c}.{d}  \spa{e}.{g} ^2 \spb{c}.{g}  \spb{e}.{d}  
    \notag\\&+ 
    48 \spa{a}.{d}  \spa{d}.{f}  \spa{e}.{f}  \spa{e}.{g}  \spb{d}.{f}  \spb{e}.{d}  + 
    48 \spa{a}.{f}  \spa{d}.{g}  \spa{e}.{d}  \spa{e}.{g}  \spb{d}.{g}  \spb{e}.{d} 
    \notag\\& - 
    48 \spa{a}.{d}  \spa{d}.{f}  \spa{e}.{g} ^2 \spb{d}.{g}  \spb{e}.{d}  + 
    16 \spa{a}.{g}  \spa{e}.{d} ^2 \spa{e}.{f}  \spb{e}.{d} ^2
    \notag\\& - 
    192 \spa{a}.{f}  \spa{e}.{d} ^2 \spa{e}.{g}  \spb{e}.{d} ^2 + 
    89 \spa{a}.{d}  \spa{e}.{d}  \spa{e}.{f}  \spa{e}.{g}  \spb{e}.{d} ^2 
    \notag\\&- 
    682 \spa{a}.{f}  \spa{b}.{f}  \spa{e}.{d}  \spa{e}.{g}  \spb{b}.{d}  \spb{e}.{f}  - 
    764 \spa{a}.{f}  \spa{b}.{d}  \spa{e}.{f}  \spa{e}.{g}  \spb{b}.{d}  \spb{e}.{f} 
    \notag\\& - 
    70 \spa{a}.{f}  \spa{b}.{e}  \spa{e}.{f}  \spa{e}.{g}  \spb{b}.{e}  \spb{e}.{f}  + 
    78 \spa{a}.{f}  \spa{b}.{f}  \spa{e}.{f}  \spa{e}.{g}  \spb{b}.{f}  \spb{e}.{f} 
    \notag\\& - 
    658 \spa{a}.{f}  \spa{b}.{f}  \spa{e}.{g} ^2 \spb{b}.{g}  \spb{e}.{f}  + 
    32 \spa{a}.{d}  \spa{c}.{g}  \spa{e}.{f} ^2 \spb{c}.{d}  \spb{e}.{f}  
    \notag\\&+ 
    32 \spa{a}.{f}  \spa{c}.{f}  \spa{e}.{d}  \spa{e}.{g}  \spb{c}.{d}  \spb{e}.{f}  - 
    48 \spa{a}.{f}  \spa{c}.{d}  \spa{e}.{f}  \spa{e}.{g}  \spb{c}.{d}  \spb{e}.{f}
    \notag\\&  + 
    195 \spa{a}.{e}  \spa{c}.{g}  \spa{e}.{f} ^2 \spb{c}.{e}  \spb{e}.{f}  - 
    740 \spa{a}.{f}  \spa{c}.{g}  \spa{e}.{f}  \spa{e}.{g}  \spb{c}.{g}  \spb{e}.{f} 
    \notag\\& + 
    160 \spa{a}.{f}  \spa{c}.{f}  \spa{e}.{g} ^2 \spb{c}.{g}  \spb{e}.{f}  - 
    24 \spa{a}.{f}  \spa{d}.{g}  \spa{e}.{f} ^2 \spb{d}.{f}  \spb{e}.{f}  
    \notag\\&- 
    24 \spa{a}.{f}  \spa{d}.{g}  \spa{e}.{f}  \spa{e}.{g}  \spb{d}.{g}  \spb{e}.{f}  - 
    87 \spa{a}.{d}  \spa{e}.{f} ^2 \spa{e}.{g}  \spb{e}.{d}  \spb{e}.{f}  
    \notag\\&- 
    740 \spa{a}.{f}  \spa{b}.{d}  \spa{e}.{g} ^2 \spb{b}.{d}  \spb{e}.{g}  - 
    70 \spa{a}.{f}  \spa{b}.{e}  \spa{e}.{g} ^2 \spb{b}.{e}  \spb{e}.{g}  
    \notag\\&+ 
    736 \spa{a}.{f}  \spa{b}.{f}  \spa{e}.{g} ^2 \spb{b}.{f}  \spb{e}.{g}  - 
    601 \spa{a}.{f}  \spa{c}.{g}  \spa{e}.{d}  \spa{e}.{g}  \spb{c}.{d}  \spb{e}.{g} 
    \notag\\& + 
    1269 \spa{a}.{d}  \spa{c}.{g}  \spa{e}.{f}  \spa{e}.{g}  \spb{c}.{d}  \spb{e}.{g}  - 
    24 \spa{a}.{f}  \spa{c}.{d}  \spa{e}.{g} ^2 \spb{c}.{d}  \spb{e}.{g}  
    \notag\\&+ 
    1175 \spa{a}.{e}  \spa{c}.{g}  \spa{e}.{f}  \spa{e}.{g}  \spb{c}.{e}  \spb{e}.{g}  + 
    692 \spa{a}.{f}  \spa{c}.{g}  \spa{e}.{f}  \spa{e}.{g}  \spb{c}.{f}  \spb{e}.{g} 
    \notag\\& - 
    112 \spa{a}.{f}  \spa{c}.{f}  \spa{e}.{g} ^2 \spb{c}.{f}  \spb{e}.{g}  - 
    87 \spa{a}.{d}  \spa{e}.{f}  \spa{e}.{g} ^2 \spb{e}.{d}  \spb{e}.{g}  
    \notag\\&+ 
    32 \spa{a}.{f}  \spa{d}.{f}  \spa{e}.{g}  \spa{f}.{g}  \spb{d}.{f}  \spb{f}.{g}  - 
    48 \spa{a}.{f}  \spa{e}.{g} ^2 \spa{f}.{g}  \spb{e}.{g}  \spb{f}.{g}.
\end{align}

\subsection{R$_{7:1,1}$} \label{sec:R711-rational-compact-results}

The first new exclusively $U(N_c)$ partial amplitude to be calculated is $R_{7:1,1}$.
Although it is a structure not present in the $SU(N_c)$ theory, we chose to reference it in the statement of the $R_{7:3}$ rational piece because it is the simpler of the two structures.
As with the other partial amplitudes, the result produced by augmented recursion was analytic, and after simplifying manipulations takes the form:
\begin{equation}
    R_{7:1,1}(a;b;c,d,e,f,g) = \sum_{\mathcal{P}_{7:1,1}} B_{7:1,1}(a,b,c,d,e,f,g) ,
\end{equation}
with a basis function
\begin{align}
    B_{7:1,1}(a,b,c,d,e,f,g) =& B_{7:1,1}^{(tt)}(a,b,c,d,e,f,g) + B_{7:1,1}^{(ts)}(a,b,c,d,e,f,g) \notag\\
    & + B_{7:1,1}^{(s)}(a,b,c,d,e,f,g)
\end{align}
containing the structures
\begin{align}
    B_{7:1,1}^{(tt)}(a,b,c,d,e,f,g) =& \frac{i}{t_{aef}  t_{bcd}} \frac{ G_{7:1,1}^{1} }{ \spa{a}.{e}  \spa{a}.{f}  \spa{a}.{g}  \spa{b}.{c}  \spa{c}.{d} \spa{e}.{f} } 
    \notag\\
    & + \frac{i}{t_{aef}  t_{bcd}} \frac{ [a|k_{ef}|b \ra  [g|k_{bcd}|d \ra  \spb{a}.{e}  \spb{b}.{d} }{ \spa{b}.{c}  \spa{b}.{d}  \spa{c}.{d}  \spa{e}.{f}  \spa{f}.{g} },
\end{align}
\begin{equation}
    B_{7:1,1}^{(ts)}(a,b,c,d,e,f,g) = \frac{i}{t_{aef}} \frac{G_{7:1,1}^{2}}{ \spa{a}.{e} \spa{a}.{f}  \spa{a}.{g} \spa{b}.{c} \spa{b}.{d} \spa{b}.{g} } \frac{1}{ \spa{c}.{d}  \spa{c}.{g}  \spa{e}.{f}  \spa{f}.{g} }
\end{equation}
and
\begin{equation}
    B_{7:1,1}^{(s)}(a,b,c,d,e,f,g) = \frac{i}{3} \frac{G_{7:1,1}^{3}+G_{7:1,1}^{4}+G_{7:1,1}^{5}}{ \spa{c}.{d}\spa{d}.{e}\spa{e}.{f}\spa{f}.{g}\spa{g}.{c} } \frac{1}{ \spa{b}.{c}\spa{b}.{d} } \frac{1}{ \spa{a}.{e}\spa{a}.{f} }.
\end{equation}
The numerators are
\begin{align}
    G_{7:1,1}^{1} = -& [b|k_{bcd}|a\ra  [d|k_{bcd}|a\ra  [g|k_{bcd}|e\ra  \spb{a}.{e} 
    \notag\\
    & - [e|k_{bcd}|g\ra  s_{ag}  \spa{a}.{e}  \spb{b}.{g}  \spb{d}.{g},
\end{align}
\begin{align}
    G_{7:1,1}^{2} = -& [a|k_{aef}|a\ra  [b|k_{bcd}|a\ra  [d|k_{bcd}|c\ra  \spa{b}.{d}  \spa{b}.{g}  \spa{f}.{g}
    \notag\\
    & +[b|k_{bcd}|a\ra  [d|k_{aef}|a\ra \spa{b}.{d}  \spa{b}.{g}  \spa{c}.{e}  \spa{f}.{g}  \spb{a}.{e}
    \notag\\
    &  +[a|k_{aef}|b\ra  [f|k_{aef}|d\ra  \spa{a}.{f}  \spa{a}.{g} \spa{b}.{g}  \spa{c}.{f}  \spb{b}.{d} 
    \notag\\
    & +[b|k_{bcd}|a\ra  \spa{a}.{f} \spa{b}.{c}  \spa{b}.{d}  \spa{b}.{g}  \spa{f}.{g}  \spb{a}.{f} \spb{b}.{d} 
    \notag\\
    & +[a|k_{aef}|b\ra  [f|k_{aef}|g\ra  \spa{a}.{f} \spa{a}.{g}  \spa{b}.{c}  \spa{f}.{g}  \spb{b}.{g} 
    \notag\\
    & +[d|k_{aef}|a\ra \spa{a}.{g}  \spa{b}.{d}  \spa{b}.{g}  \spa{c}.{g}  \spa{f}.{g}  \spb{a}.{g}  \spb{b}.{g} 
    \notag\\
    & +\spa{a}.{e}  \spa{a}.{g}  \spa{b}.{d}  \spa{b}.{g}  \spa{c}.{g}  \spa{f}.{g}  \spb{a}.{g}  \spb{b}.{g}  \spb{d}.{e} 
    \notag\\
    & +[b|k_{aef}|c\ra  \spa{a}.{f}  \spa{a}.{g}  \spa{b}.{d} \spa{b}.{g}  \spa{f}.{g}  \spb{a}.{g}  \spb{d}.{f}
    \notag\\
    &  +[a|k_{aef}|c\ra  [f|k_{aef}|b\ra  \spa{a}.{f}  \spa{a}.{g}  \spa{d}.{g}  \spa{f}.{g}  \spb{d}.{g}
    \notag\\
    &  -[a|k_{aef}|b\ra  [f|k_{aef}|c\ra \spa{a}.{f}  \spa{a}.{g}  \spa{d}.{g}  \spa{f}.{g}  \spb{d}.{g}
    \notag\\
    &  -[b|k_{bcd}|a\ra  \spa{a}.{e}  \spa{b}.{d}  \spa{b}.{g}  \spa{c}.{g}  \spa{f}.{g}  \spb{a}.{e}  \spb{d}.{g} 
    \notag\\
    & +\spa{a}.{f}  \spa{a}.{g}  \spa{b}.{d}  \spa{b}.{g}  \spa{c}.{d}  \spa{f}.{g}  \spb{a}.{g}  \spb{b}.{d}  \spb{f}.{d} 
    \notag\\
    & +\spa{a}.{f}  \spa{a}.{g}  \spa{b}.{d}  \spa{b}.{g}  \spa{c}.{d}  \spa{f}.{g}  \spb{a}.{d}  \spb{b}.{d}  \spb{f}.{g} 
    \notag\\
    & +[a|k_{aef}|d\ra  [f|k_{aef}|c\ra  \spa{a}.{f}  \spa{a}.{g}  \spa{b}.{g}  \spa{f}.{g}  \spb{g}.{d} 
    \notag\\
    & -[a|k_{aef}|c\ra [f|k_{aef}|d\ra  \spa{a}.{f}  \spa{a}.{g}  \spa{b}.{g}  \spa{f}.{g}  \spb{g}.{d} 
    \notag\\
    & +[f|k_{aef}|c\ra  \spa{a}.{f}  \spa{a}.{g}  \spa{b}.{d} \spa{b}.{g}  \spa{f}.{g}  \spb{a}.{b}  \spb{g}.{d},
\end{align}
\begin{align}
G_{7:1,1}^{3} =& 
\frac{123}{10} \spa{a}.{c} \spa{a}.{d} \spa{b}.{e} \spa{b}.{f} \spb{a}.{b}^2 +
\frac{27}{10} \spa{a}.{b} \spa{a}.{e}  \spa{b}.{c} \spa{d}.{f}  \spb{a}.{b} ^2
\notag\\& -
\frac{11}{5} \spa{a}.{b}  \spa{a}.{d} \spa{b}.{c}  \spa{e}.{f}  \spb{a}.{b} ^2 -
\frac{3}{2} \spa{a}.{c}^2 \spa{b}.{e}  \spa{d}.{f}  \spb{a}.{b}  \spb{a}.{c} 
\notag\\& -
\frac{3}{2} \spa{a}.{c} \spa{a}.{d} \spa{b}.{c} \spa{e}.{f} \spb{a}.{b} \spb{a}.{c} -
\frac{17}{2} \spa{a}.{c}^2 \spa{b}.{d} \spa{e}.{f} \spb{a}.{b} \spb{a}.{c}
\notag\\&  +
\frac{29}{2} \spa{a}.{c} \spa{a}.{d}  \spa{b}.{e} \spa{d}.{f} \spb{a}.{b} \spb{a}.{d} -
\frac{7}{2} \spa{a}.{c} \spa{a}.{d} \spa{b}.{d} \spa{e}.{f} \spb{a}.{b} \spb{a}.{d}
\notag\\& +
\frac{29}{2} \spa{a}.{c} \spa{a}.{d} \spa{b}.{e} \spa{e}.{f} \spb{a}.{b} \spb{a}.{e} -
\frac{9}{2} \spa{a}.{c}  \spa{a}.{e}  \spa{b}.{f} \spa{d}.{f}  \spb{a}.{b}  \spb{a}.{f} 
\notag\\& -
\frac{84}{5} \spa{a}.{b}  \spa{b}.{d}  \spa{c}.{e}  \spa{c}.{f}  \spb{a}.{b}  \spb{b}.{c}  +
\frac{9}{5} \spa{a}.{b} \spa{b}.{c} \spa{c}.{f} \spa{d}.{e} \spb{a}.{b} \spb{b}.{c}
\notag\\& +
\frac{21}{2} \spa{a}.{b} \spa{b}.{c}  \spa{c}.{e}  \spa{d}.{f}  \spb{a}.{b}  \spb{b}.{c}  -
\frac{29}{2} \spa{a}.{b}  \spa{c}.{d}  \spa{c}.{e}  \spa{c}.{f} \spb{a}.{c}  \spb{b}.{c}
\notag\\& -
\frac{15}{2} \spa{a}.{b} \spa{c}.{d}  \spa{c}.{e}  \spa{e}.{f}  \spb{a}.{e}  \spb{b}.{c} +
\frac{51}{2} \spa{a}.{b}  \spa{b}.{d}  \spa{c}.{e}  \spa{d}.{f}  \spb{a}.{b}  \spb{b}.{d} 
\notag\\& -
\frac{33}{2} \spa{a}.{b}  \spa{b}.{c}  \spa{d}.{e} \spa{d}.{f}  \spb{a}.{b}  \spb{b}.{d}  + 
\frac{33}{2} \spa{a}.{b}  \spa{c}.{d}  \spa{d}.{e} \spa{d}.{f}  \spb{a}.{d}  \spb{b}.{d} 
\notag\\& -
\frac{39}{2} \spa{a}.{b}  \spa{c}.{d}  \spa{d}.{f}  \spa{e}.{f}  \spb{a}.{f}  \spb{b}.{d}  -
\frac{33}{2} \spa{a}.{b}  \spa{c}.{d}  \spa{d}.{f} \spa{e}.{g}  \spb{a}.{g}  \spb{b}.{d} 
\notag\\& -
\frac{11}{5} \spa{a}.{b} \spa{b}.{e}  \spa{c}.{e}  \spa{d}.{f}  \spb{a}.{b}  \spb{b}.{e}  + 
\frac{36}{5} \spa{a}.{b}  \spa{b}.{d}  \spa{c}.{e}  \spa{e}.{f} \spb{a}.{b}  \spb{b}.{e} 
\notag\\& + 
\frac{19}{5} \spa{a}.{b}  \spa{b}.{c} \spa{d}.{e}  \spa{e}.{f}  \spb{a}.{b}  \spb{b}.{e}  -
\frac{9}{2} \spa{a}.{c}  \spa{b}.{d}  \spa{c}.{e}  \spa{e}.{f}  \spb{a}.{c}  \spb{b}.{e} 
\notag\\& -
\frac{9}{2} \spa{a}.{c}  \spa{b}.{d}  \spa{d}.{e} \spa{e}.{f}  \spb{a}.{d}  \spb{b}.{e}  +
\frac{47}{2} \spa{a}.{b} \spa{c}.{e}  \spa{d}.{e}  \spa{f}.{g}  \spb{a}.{g}  \spb{b}.{e} 
\notag\\& + 
\frac{19}{5} \spa{a}.{b}  \spa{b}.{e}  \spa{c}.{f}  \spa{d}.{f} \spb{a}.{b}  \spb{b}.{f}  -
\frac{9}{2} \spa{a}.{c}  \spa{b}.{d}  \spa{d}.{f}  \spa{e}.{f}  \spb{a}.{d}  \spb{b}.{f} 
\notag\\& +
\frac{29}{2} \spa{a}.{b} \spa{c}.{e}  \spa{d}.{f}  \spa{f}.{g}  \spb{a}.{g}  \spb{b}.{f}  - 
\frac{11}{5} \spa{a}.{b}  \spa{b}.{e}  \spa{c}.{f}  \spa{d}.{g} \spb{a}.{b}  \spb{b}.{g} 
\notag\\& +
\frac{29}{2} \spa{a}.{b}  \spa{c}.{e} \spa{d}.{g}  \spa{f}.{g}  \spb{a}.{g}  \spb{b}.{g}  +
\frac{3}{2} \spa{a}.{c}  \spa{b}.{d}  \spa{c}.{f}  \spa{d}.{e}  \spb{a}.{b}  \spb{c}.{d} 
\notag\\& -
\frac{21}{2} \spa{a}.{c}  \spa{b}.{c} \spa{d}.{e}  \spa{d}.{f}  \spb{a}.{b}  \spb{c}.{d}  - 
\frac{3}{2} \spa{a}.{c}  \spa{b}.{d}  \spa{c}.{f} \spa{e}.{f}  \spb{a}.{b}  \spb{c}.{f}  ,
\end{align}
\begin{align}
G_{7:1,1}^{4} =& 
16 \spa{a}.{c} \spa{a}.{d}  \spa{b}.{e}  \spa{c}.{f}  \spb{a}.{b}  \spb{a}.{c} -
6 \spa{a}.{d} ^2 \spa{b}.{c}  \spa{e}.{f}  \spb{a}.{b} \spb{a}.{d}
\notag\\&  +
3 \spa{a}.{d} ^2 \spa{c}.{e} \spa{c}.{f}  \spb{a}.{c}  \spb{a}.{d}  -
3 \spa{a}.{c}  \spa{a}.{d} \spa{c}.{e}  \spa{d}.{f}  \spb{a}.{c}  \spb{a}.{d}
\notag\\&  +
6 \spa{a}.{c}^2 \spa{d}.{e}  \spa{d}.{f}  \spb{a}.{c}  \spb{a}.{d}  +
3 \spa{a}.{c}  \spa{a}.{d}  \spa{c}.{d}  \spa{e}.{f}  \spb{a}.{c} \spb{a}.{d} 
\notag\\& -
6 \spa{a}.{d} ^2 \spa{c}.{e}  \spa{d}.{f}  \spb{a}.{d}^2 +
6 \spa{a}.{c} \spa{a}.{d}  \spa{d}.{e}  \spa{d}.{f} \spb{a}.{d}^2
\notag\\& +
6 \spa{a}.{d} ^2 \spa{c}.{d}  \spa{e}.{f} \spb{a}.{d} ^2 -
9 \spa{a}.{c}  \spa{a}.{e}  \spa{b}.{e}  \spa{d}.{f} \spb{a}.{b}  \spb{a}.{e}  
\notag\\&  +
3 \spa{a}.{d} \spa{a}.{e}  \spa{c}.{e}  \spa{c}.{f}  \spb{a}.{c}  \spb{a}.{e}  -
3 \spa{a}.{c}  \spa{a}.{e}  \spa{c}.{e}  \spa{d}.{f}  \spb{a}.{c} \spb{a}.{e}
\notag\\&  +
3 \spa{a}.{c}  \spa{a}.{d}  \spa{c}.{e}  \spa{e}.{f} \spb{a}.{c}  \spb{a}.{e}  -
3 \spa{a}.{c} ^2 \spa{d}.{e}  \spa{e}.{f} \spb{a}.{c}  \spb{a}.{e} 
\notag\\& -
6 \spa{a}.{d}  \spa{a}.{e}  \spa{c}.{e} \spa{d}.{f}  \spb{a}.{d}  \spb{a}.{e}  +
6 \spa{a}.{c}  \spa{a}.{e} \spa{d}.{e}  \spa{d}.{f}  \spb{a}.{d}  \spb{a}.{e} 
\notag\\& +
6 \spa{a}.{d}^2 \spa{c}.{e}  \spa{e}.{f}  \spb{a}.{d}  \spb{a}.{e}  -
6 \spa{a}.{c}  \spa{a}.{d}  \spa{d}.{e}  \spa{e}.{f}  \spb{a}.{d} \spb{a}.{e} 
\notag\\& +
3 \spa{a}.{d}  \spa{a}.{e} \spa{c}.{f} ^2 \spb{a}.{c}  \spb{a}.{f}  -
3 \spa{a}.{c}  \spa{a}.{e} \spa{c}.{f}  \spa{d}.{f}  \spb{a}.{c}  \spb{a}.{f} 
\notag\\& -
4 \spa{a}.{b}  \spa{c}.{d}  \spa{c}.{f} \spa{d}.{e}  \spb{a}.{d}  \spb{b}.{c}  +
6 \spa{a}.{b}  \spa{c}.{e}  \spa{c}.{f}  \spa{d}.{f}  \spb{a}.{f} \spb{b}.{c} 
\notag\\& +
4 \spa{a}.{b}  \spa{c}.{d}  \spa{c}.{f}  \spa{e}.{f} \spb{a}.{f}  \spb{b}.{c}  +
4 \spa{a}.{b}  \spa{c}.{d}  \spa{c}.{g} \spa{e}.{f}  \spb{a}.{g}  \spb{b}.{c} 
\notag\\& - 
18 \spa{a}.{b} \spa{b}.{d} \spa{c}.{f}  \spa{d}.{e}  \spb{a}.{b}  \spb{b}.{d}  +
6 \spa{a}.{c} \spa{b}.{d} \spa{c}.{f}  \spa{d}.{e}  \spb{a}.{c}  \spb{b}.{d} 
\notag\\& -
3 \spa{a}.{c} \spa{b}.{d}  \spa{c}.{e}  \spa{d}.{f}  \spb{a}.{c}  \spb{b}.{d}  -
3 \spa{a}.{c} \spa{b}.{d}  \spa{d}.{e}  \spa{d}.{f}  \spb{a}.{d} \spb{b}.{d} 
\notag\\& + 
6 \spa{a}.{b} \spa{c}.{d} \spa{d}.{e}  \spa{e}.{f}  \spb{a}.{e}  \spb{b}.{d}  + 
3 \spa{a}.{b} \spa{c}.{e}  \spa{d}.{f} ^2 \spb{a}.{f}  \spb{b}.{d} 
\notag\\& -
3 \spa{a}.{c} \spa{b}.{e} \spa{d}.{e}  \spa{e}.{f}  \spb{a}.{e}  \spb{b}.{e}  - 
6 \spa{a}.{b} \spa{c}.{e}  \spa{d}.{e}  \spa{e}.{f}  \spb{a}.{e}  \spb{b}.{e} 
\notag\\& - 
9 \spa{a}.{b} \spa{c}.{e}  \spa{d}.{f}  \spa{e}.{f}  \spb{a}.{f} \spb{b}.{e}  + 
6 \spa{a}.{b} \spa{c}.{d}  \spa{e}.{f} ^2 \spb{a}.{f} \spb{b}.{e} 
 \notag\\& - 
9 \spa{a}.{b}  \spa{c}.{e}  \spa{d}.{f}  \spa{e}.{g} \spb{a}.{g}  \spb{b}.{e}  +
6 \spa{a}.{b}  \spa{c}.{d}  \spa{e}.{f} \spa{e}.{g}  \spb{a}.{g}  \spb{b}.{e} 
\intertext{and}
G_{7:1,1}^{5} =& 
7 \spa{a}.{b}  \spa{b}.{d}  \spa{c}.{f} \spa{e}.{f}  \spb{a}.{b}  \spb{b}.{f}  -
6 \spa{a}.{c}  \spa{b}.{d} \spa{c}.{f}  \spa{e}.{f}  \spb{a}.{c}  \spb{b}.{f}
\notag\\&  - 
6 \spa{a}.{b}  \spa{c}.{f}  \spa{d}.{f}  \spa{e}.{f} \spb{a}.{f}  \spb{b}.{f}  -
6 \spa{a}.{b}  \spa{c}.{f}  \spa{d}.{f} \spa{e}.{g}  \spb{a}.{g}  \spb{b}.{f} 
\notag\\& + 
12 \spa{a}.{c}  \spa{b}.{d}  \spa{c}.{e}  \spa{d}.{f} \spb{a}.{b}  \spb{c}.{d}  +
6 \spa{a}.{d} \spa{c}.{f} ^2 \spa{d}.{e}  \spb{a}.{f}  \spb{c}.{d} 
\notag\\& -
6 \spa{a}.{c} \spa{c}.{g}  \spa{d}.{e}  \spa{d}.{f}  \spb{a}.{g}  \spb{c}.{d}  - 
3 \spa{a}.{c}  \spa{b}.{c}  \spa{d}.{e}  \spa{e}.{f}  \spb{a}.{b} \spb{c}.{e}
\notag\\&  - 
3 \spa{a}.{d}  \spa{c}.{e}  \spa{c}.{f}  \spa{d}.{e} \spb{a}.{d}  \spb{c}.{e}  + 
3 \spa{a}.{c}  \spa{c}.{f}  \spa{d}.{e}^2 \spb{a}.{d}  \spb{c}.{e} 
\notag\\& +
3 \spa{a}.{d}  \spa{c}.{e}^2 \spa{d}.{f}  \spb{a}.{d}  \spb{c}.{e}  - 
3 \spa{a}.{c}  \spa{c}.{e} \spa{d}.{e}  \spa{d}.{f}  \spb{a}.{d}  \spb{c}.{e} 
\notag\\& + 
3 \spa{a}.{d} \spa{c}.{e} ^2 \spa{e}.{f}  \spb{a}.{e}  \spb{c}.{e}  - 
3 \spa{a}.{c} \spa{c}.{e}  \spa{d}.{e}  \spa{e}.{f}  \spb{a}.{e}  \spb{c}.{e} 
 \notag\\& + 
6 \spa{a}.{c}  \spa{b}.{e}  \spa{c}.{f}  \spa{d}.{f}  \spb{a}.{b} \spb{c}.{f}  - 
6 \spa{a}.{c}  \spa{c}.{f} \spa{d}.{e}  \spa{d}.{f}  \spb{a}.{d}  \spb{c}.{f} 
\notag\\& + 
3 \spa{a}.{d} \spa{c}.{e}  \spa{c}.{f}  \spa{e}.{f}  \spb{a}.{e}  \spb{c}.{f}  - 
3 \spa{a}.{c}  \spa{c}.{f}  \spa{d}.{e}  \spa{e}.{f}  \spb{a}.{e} \spb{c}.{f} 
\notag\\& - 
3 \spa{a}.{d}  \spa{c}.{e}  \spa{c}.{f}  \spa{c}.{g} \spb{a}.{c}  \spb{c}.{g}  + 
3 \spa{a}.{c}  \spa{c}.{f}  \spa{c}.{g} \spa{d}.{e}  \spb{a}.{c}  \spb{c}.{g} 
\notag\\& + 
6 \spa{a}.{d}  \spa{c}.{f} \spa{c}.{g}  \spa{d}.{e}  \spb{a}.{d}  \spb{c}.{g}  + 
9 \spa{a}.{c} \spa{b}.{d}  \spa{d}.{e}  \spa{e}.{f}  \spb{a}.{b}  \spb{d}.{e} 
\notag\\& + 
3 \spa{a}.{d}  \spa{c}.{e}  \spa{c}.{f}  \spa{d}.{e}  \spb{a}.{c} \spb{d}.{e}  - 
3 \spa{a}.{c}  \spa{c}.{f}  \spa{d}.{e} ^2 \spb{a}.{c} \spb{d}.{e} 
\notag\\& + 
6 \spa{a}.{c}  \spa{c}.{e}  \spa{d}.{e}  \spa{d}.{f} \spb{a}.{c}  \spb{d}.{e}  - 
6 \spa{a}.{d}  \spa{c}.{f}  \spa{d}.{e}^2 \spb{a}.{d}  \spb{d}.{e} 
\notag\\& - 
6 \spa{a}.{e}  \spa{c}.{d} \spa{d}.{e}  \spa{d}.{f}  \spb{a}.{d}  \spb{d}.{e}  + 
6 \spa{a}.{d} \spa{c}.{e}  \spa{d}.{e}  \spa{d}.{f}  \spb{a}.{d}  \spb{d}.{e}
\notag\\&  + 
6 \spa{a}.{c}  \spa{b}.{e}  \spa{d}.{f} ^2 \spb{a}.{b}  \spb{d}.{f}  +
3 \spa{a}.{d}  \spa{c}.{e}  \spa{c}.{f}  \spa{d}.{f}  \spb{a}.{c} \spb{d}.{f}
\notag\\&  + 
9 \spa{a}.{c}  \spa{c}.{f}  \spa{d}.{e}  \spa{d}.{f} \spb{a}.{c}  \spb{d}.{f}  - 
6 \spa{a}.{d}  \spa{c}.{f}  \spa{d}.{e} \spa{d}.{f}  \spb{a}.{d}  \spb{d}.{f} 
\notag\\& - 
6 \spa{a}.{e}  \spa{c}.{d} \spa{d}.{f} ^2 \spb{a}.{d}  \spb{d}.{f}  + 
6 \spa{a}.{d}  \spa{c}.{e} \spa{d}.{f} ^2 \spb{a}.{d}  \spb{d}.{f} 
\notag\\& + 
6 \spa{a}.{e} \spa{c}.{f}  \spa{d}.{e}  \spa{d}.{f}  \spb{a}.{e}  \spb{d}.{f}  - 
6 \spa{a}.{e}  \spa{c}.{e}  \spa{d}.{f} ^2 \spb{a}.{e}  \spb{d}.{f} 
\notag\\& -
6 \spa{a}.{d}  \spa{c}.{f}  \spa{d}.{e}  \spa{e}.{f}  \spb{a}.{e} \spb{d}.{f}  + 
6 \spa{a}.{d}  \spa{c}.{e}  \spa{d}.{f}  \spa{e}.{f} \spb{a}.{e}  \spb{d}.{f}.
\end{align}

\subsection{R$_{7:1,2}$}

The second new exclusively $U(N_c)$ rational piece can be expressed in terms of the new $SU(N_c)$ structures via a decoupling identity,
\begin{align}
    R_{7:1,2}(a;b,c;d,e,f,g) =& -R_{7:3}(b,c;a,d,e,f,g) -R_{7:3}(b,c;a, e, f, g, d) \notag\\
    &- R_{7:3}(b,c;a,f,g,d,e) -R_{7:3}(b,c;a,g,d,e,f) \notag\\
    &- R_{7:4}(a, b, c;d, e, f, g) - R_{7:4}(a, c,b;d, e, f, g) \notag\\
    \intertext{\begin{equation}
        = -\sum_{Z_4(defg)} R_{7:3}(b,c;a,d,e,f,g) - \sum_{Z_2(bc)} R_{7:4}(a,b,c;d,e,f,g).
    \end{equation}}
\end{align}

\subsection{R$_{7:1,3}$}

From the decoupling identity, the new $U(N_c)$ rational piece $R_{7:1,3}$ can be expressed solely in terms of $R_{7:4}$,
\begin{align}
    R_{7:1,3}(a;b,c,d;e,f,g) =& -R_{7:4}(b,c,d;a,e,f,g) - R_{7:4}(b,c,d;a,f,g,e) \notag\\
    &- R_{7:4}(b,c,d;a,g,e,f) - R_{7:4}(e, f, g;a, b, c, d) \notag\\
    &- R_{7:4}(e, f, g;a, c, d, b) - R_{7:4}(e, f, g;a, d, b, c) \notag\\
    \intertext{\begin{equation}
        = -\sum_{Z_3(efg)} R_{7:4}(b,c,d;a,e,f,g) -\sum_{Z_3(bcd)} R_{7:4}(e,f,g;a,b,c,d).
    \end{equation}}
\end{align}

\subsection{R$_{7:2,2}$}

The final new $SU(N_c)$ rational piece can be expressed in terms of the previous two $U(N_c)$ structures,
\begin{align}
    R_{7:2,2}(a,b;c,d;e,f,g) =& -R_{7:1,2}(b;c, d;a, e, f, g) - R_{7:1,2}(b;c, d;a, f, g, e) \notag\\
    &-R_{7:1,2}(b;c, d;a, g, e, f) - R_{7:1,3}(b;a, c, d;e, f, g) \notag\\
    &-R_{7:1,3}(b;a, d, c;e, f, g) \notag\\
    \intertext{
    \begin{equation}
        = -\sum_{Z_3(efg)} R_{7:1,2}(b;c,d;a,e,f,g) -\sum_{Z_2(cd)} R_{7:1,3}(b;a,c,d;e,f,g).
    \end{equation}
    }
\end{align}

\subsection{R$_{7:1B}$}

Lastly is $R_{7:1B}$, the $SU(N_c)$ partial amplitude that appears with a single color trace and no factors of $N_c$ in the color decomposition.
(This differs from $R_{7:1}$, which appears multiplied by $N_c^2$ and a single trace.)
The augmented recursion result calculated here finds agreement with the $n$-point form postulated in~\cite{Dunbar:2020wdh}.
We reconstruct a version matching that form.
The function has cyclic symmetry in the momenta, but unlike other color structures does not appear with an explicit $\mathcal{P}_{7:\lambda}$ sum.

For compactness, the Parke--Taylor denominator is defined with
\begin{align}
    C_{PT}(a,b,c,d,e,f,g) = \frac{1}{ \spa{a}.{b}\spa{b}.{c}\spa{c}.{d}\spa{d}.{e}\spa{e}.{f}\spa{f}.{g}\spa{g}.{a} }.
\end{align}
Also useful is the epsilon function
\begin{align}
    \epsilon(a,b,c,d) =& [a|b|c|d|a\ra - \la a|b|c|d|a]
    \notag\\
    = & \spb{a}.{b}\spa{b}.{c}\spb{c}.{d}\spa{d}.{a} - \spa{a}.{b}\spb{b}.{c}\spa{c}.{d}\spb{d}.{a},
\end{align}
with a further compact notation
\begin{align}
    & \epsilon(\{a_1,...,a_r\},b,c,\{d_1,...,d_s\}) = \sum_{i=1}^r \sum_{j=1}^s \epsilon(a_i,b,c,d_j).
\end{align}

With these identifications, the partial amplitude can be written in two pieces,
\begin{align}
    R_{7:1B}(a,b,c,d,e,f,g) = R^A_{7:1B}(a,b,c,d,e,f,g) + R^B_{7:1B}(a,b,c,d,e,f,g) ,
\end{align}
where
\begin{align}
    R^A_{7:1B}(a,b,c,d,e,f,g) =&  -2i C_{PT}  ({a, b, c, d, e, f, g}) \times
    \notag\\
    \Big( \epsilon({\{a, b, c\}, d, f, {g}}) & +
    \epsilon({\{a, b\}, c, e, \{f, g\}}) 
    +
    \epsilon({\{a, b\}, c, f, {g}})+
     \notag\\& 
        \epsilon({{a}, b, d, \{e, f, g\}}) 
    +
    \epsilon({{a}, b, e, \{f, g\}}) +
    \epsilon({{a}, b, f, {g}}) \Big) 
\end{align}
and
\begin{align}
    R^B_{7:1B} & (a,b,c,d,e,f,g) = 
    \notag\\
    +4i \Big( &
    C_{PT}({a, b, c, e, f, d, g}) \epsilon({\{a, b, c\}, e, d, {g}}) - 
    C_{PT}({a, b, c, f, e, d, g}) \epsilon({\{a, b, c\}, f, d, {g}}) \notag\\&+
    C_{PT}({a, b, c, f, d, e, g}) \epsilon({\{a, b, c\}, f, e, {g}}) + 
    C_{PT}({a, b, d, e, c, f, g}) \epsilon({\{a, b\}, d, c, \{f, g\}}) \notag\\& +
    C_{PT}({a, b, d, e, f, c, g}) \epsilon({\{a, b\}, d, c, {g}}) - 
    C_{PT}({a, b, e, d, c, f, g}) \epsilon({\{a, b\}, e, c, \{f, g\}}) \notag\\& -
    C_{PT}({a, b, e, d, f, c, g}) \epsilon({\{a, b\}, e, c, {g}}) -
    C_{PT}({a, b, e, f, d, c, g}) \epsilon({\{a, b\}, e, c, {g}}) \notag\\& +
    C_{PT}({a, b, e, c, d, f, g}) \epsilon({\{a, b\}, e, d, \{f, g\}}) +
    C_{PT}({a, b, e, c, f, d, g}) \epsilon({\{a, b\}, e, d, {g}})  \notag\\&+
    C_{PT}({a, b, e, f, c, d, g}) \epsilon({\{a, b\}, e, d, {g}}) + 
    C_{PT}({a, b, f, e, d, c, g}) \epsilon({\{a, b\}, f, c, {g}}) \notag\\& -
    C_{PT}({a, b, f, c, e, d, g}) \epsilon({\{a, b\}, f, d, {g}}) - 
    C_{PT}({a, b, f, e, c, d, g}) \epsilon({\{a, b\}, f, d, {g}})  \notag\\&+
    C_{PT}({a, b, f, c, d, e, g}) \epsilon({\{a, b\}, f, e, {g}}) +
    C_{PT}({a, c, d, b, e, f, g}) \epsilon({{a}, c, b, \{e, f, g\}}) \notag\\& +
    C_{PT}({a, c, d, e, b, f, g}) \epsilon({{a}, c, b, \{f, g\}}) +
    C_{PT}({a, c, d, e, f, b, g}) \epsilon({{a}, c, b, {g}}) 
    \notag\\&- 
    C_{PT}({a, d, c, b, e, f, g}) \epsilon({{a}, d, b, \{e, f, g\}}) - 
    C_{PT}({a, d, c, e, b, f, g}) \epsilon({{a}, d, b, \{f, g\}}) \notag\\& -
    C_{PT}({a, d, e, c, b, f, g}) \epsilon({{a}, d, b, \{f, g\}}) - 
    C_{PT}({a, d, c, e, f, b, g}) \epsilon({{a}, d, b, {g}}) 
    \notag\\&- 
    C_{PT}({a, d, e, c, f, b, g}) \epsilon({{a}, d, b, {g}}) - 
    C_{PT}({a, d, e, f, c, b, g}) \epsilon({{a}, d, b, {g}})
    \notag\\& + 
    C_{PT}({a, d, b, c, e, f, g}) \epsilon({{a}, d, c, \{e, f, g\}}) +
    C_{PT}({a, d, b, e, c, f, g}) \epsilon({{a}, d, c, \{f, g\}}) \notag\\& +
    C_{PT}({a, d, e, b, c, f, g}) \epsilon({{a}, d, c, \{f, g\}}) + 
    C_{PT}({a, d, b, e, f, c, g}) \epsilon({{a}, d, c, {g}})
    \notag\\& + 
    C_{PT}({a, d, e, b, f, c, g}) \epsilon({{a}, d, c, {g}}) + 
    C_{PT}({a, d, e, f, b, c, g}) \epsilon({{a}, d, c, {g}})
    \notag\\& + 
    C_{PT}({a, e, d, c, b, f, g}) \epsilon({{a}, e, b, \{f, g\}}) +
    C_{PT}({a, e, d, c, f, b, g}) \epsilon({{a}, e, b, {g}})
    \notag\\& + 
    C_{PT}({a, e, d, f, c, b, g}) \epsilon({{a}, e, b, {g}}) + 
    C_{PT}({a, e, f, d, c, b, g}) \epsilon({{a}, e, b, {g}})
    \notag\\& - 
    C_{PT}({a, e, b, d, c, f, g}) \epsilon({{a}, e, c, \{f, g\}}) -
    C_{PT}({a, e, d, b, c, f, g}) \epsilon({{a}, e, c, \{f, g\}}) \notag\\& -
    C_{PT}({a, e, b, d, f, c, g}) \epsilon({{a}, e, c, {g}}) - 
    C_{PT}({a, e, b, f, d, c, g}) \epsilon({{a}, e, c, {g}})
    \notag\\& - 
    C_{PT}({a, e, d, b, f, c, g}) \epsilon({{a}, e, c, {g}}) - 
    C_{PT}({a, e, d, f, b, c, g}) \epsilon({{a}, e, c, {g}})
    \notag\\& - 
    C_{PT}({a, e, f, b, d, c, g}) \epsilon({{a}, e, c, {g}}) - 
    C_{PT}({a, e, f, d, b, c, g}) \epsilon({{a}, e, c, {g}})
    \notag\\& + 
    C_{PT}({a, e, b, c, d, f, g}) \epsilon({{a}, e, d, \{f, g\}}) +
    C_{PT}({a, e, b, c, f, d, g}) \epsilon({{a}, e, d, {g}})
    \notag\\& + 
    C_{PT}({a, e, b, f, c, d, g}) \epsilon({{a}, e, d, {g}}) + 
    C_{PT}({a, e, f, b, c, d, g}) \epsilon({{a}, e, d, {g}})
    \notag\\& - 
    C_{PT}({a, f, e, d, c, b, g}) \epsilon({{a}, f, b, {g}}) + 
    C_{PT}({a, f, b, e, d, c, g}) \epsilon({{a}, f, c, {g}})
    \notag\\& + 
    C_{PT}({a, f, e, b, d, c, g}) \epsilon({{a}, f, c, {g}}) + 
    C_{PT}({a, f, e, d, b, c, g}) \epsilon({{a}, f, c, {g}})
    \notag\\& - 
    C_{PT}({a, f, b, c, e, d, g}) \epsilon({{a}, f, d, {g}}) - 
    C_{PT}({a, f, b, e, c, d, g}) \epsilon({{a}, f, d, {g}})
    \notag\\& - 
    C_{PT}({a, f, e, b, c, d, g}) \epsilon({{a}, f, d, {g}}) + 
    C_{PT}({a, f, b, c, d, e, g}) \epsilon({{a}, f, e, {g}})
    \Big).
\end{align}

The rational terms are available in analytic form in the linked Mathematica file {\it R7terms.m}. 

\newpage

\section{Relations Between the Partial Amplitudes}

\label{section:group}

\def\YZ{$I=0$}
\def\YY{$I$}

In this section we will look at possible linear relations amongst the partial amplitudes.   We will examine these for specific  six and seven point all-plus amplitudes.    Many of these are contained within the decoupling identities however,  as shown 
in refs.~\cite{Edison:2011ta,Edison:2012fn} there are relations beyond those the decoupling identities.  For two loop amplitudes it is these relationships which are probably the most interesting. 

To do so, following~\cite{Naculich:2011ep,Edison:2011ta,Edison:2012fn}  we find 
the content of the six and seven point partial amplitudes in terms of the irreducible representations of the symmetric groups $S_6$ and $S_7$ and look for relations amongst
these combinations.  This approach has also been used to look at higher loop relations for four and five point amplitudes~\cite{Naculich:2011ep,Huang:2016iqf,Huang:2017ont,Naculich:2024fiy}.  
The six-point is a review.   These are of course,  only guaranteed to be  satisfied for the all plus amplitude but are probes for the possible relationships satisfied by all amplitudes.   There can be no further possible linear relations valid for all helicities beyond these.  
To be clear, we are looking for relations of the form
\begin{equation}
\sum  c_{x,i}  A_{7:x}^{(2)}  ( \sigma_i (1,2,3,4,5,6,7))
\end{equation}
where the $c_{x,i}$ are constant coefficients independent of kinematics and helicity.   At tree level,  these are well understood. 
At tree level there are relations beyond linear where the coefficients are functions of the kinematic variable~\cite{Bjerrum-Bohr:2009ulz}.  These are a consequence of the color-kinematic duality~\cite{Bern:2008qj}. Our study will focus however upon linear relations.

The irreducible representations of $S_n$ can be described by Young diagrams containing exactly $n$ boxes.      
There are 11 such diagrams for $S_6$ and 15 for $S_7$.   The partial amplitudes of Yang-Mills form a representation of $S_n$ which satisfy cyclic symmetry and are for $n=6$
 flip symmetric and   flip antisymmetric for $n=7$.    As such they decompose into a limited set of irreducible representations of $S_n$.   
For example, the single trace representation 
$A_{7:1}^{(2)}$ has irreducible representation content

\ytableausetup{centertableaux, boxsize=0.5em}
$3\times $ \begin{ytableau}  
\null  & \null   &\null  & \null 
&  \null 
\\
\null \\
\null
\end{ytableau}
$+$  
\begin{ytableau}  
\null  & \null   &\null  & \null \\
&  \null &\null &
\null
\end{ytableau}
$+2\times $  
\begin{ytableau}  
\null  & \null   &\null  & \null \\
&  \null &\null \\
\null
\end{ytableau}
$+$  
\begin{ytableau}  
\null  & \null   &\null  & \null \\
&  \null \\ \null \\
\null
\end{ytableau}
$+3\times $  
\begin{ytableau}  
\null  & \null   &\null  \\ 
\null &  \null &  \null \\
\null
\end{ytableau}
$+3\times $  
\begin{ytableau}  
\null  & \null   &\null  \\ 
\null &  \null \\  \null \\
\null
\end{ytableau}
$+ $ 
\begin{ytableau}  
\null  & \null   \\
\null  & \null \\  \null &  \null \\
\null
\end{ytableau}
$+2\times $ 
\begin{ytableau}  
\null  & \null   \\
\null  & \null \\  \null \\ \null \\
\null
\end{ytableau}
$+ $ 
\begin{ytableau}  
\null  \\ \null   \\
\null \\ \null \\  \null \\ \null \\
\null
\end{ytableau}
This is the  group theory decomposition based solely on the cyclic and flip properties.  As such it is also the decomposition of the tree amplitude 
$A^{(0)}_7$ the one-loop $A^{(1)}_{7:1}$ and both the two-loop partial amplitudes $A^{(2)}_{7:1}$ and  $A^{(2)}_{7:1B}$.  Although these have the same decomposition,  the kinematic structures are very different.
The decomposition for the various partial amplitudes is  indicated in 
 tables~ \ref{table:sixpt} and \ref{table:sevenpt}.     In terms of notation we label the representation in terms of the number of elements in each row so $3,2,1^2$ is the representation, 
\begin{ytableau}  
\null  & \null   &\null  \\
 \null 
&  \null \\
\null \\
\null
\end{ytableau}

Before presenting our results, we provide an overview of relations which appear at tree and one-loop level which provide motivation to look beyond decoupling identities.    At tree level and one-loop, the identities contained in decoupling identities are maximal for low numbers of external legs but for higher numbers of legs there exist more identities which can reduce the number of independent amplitudes. Specifically,    at tree level there are the  Kleiss-Kuijf relations \cite{Kleiss:1988ne},
\begin{equation}
A_{n}^{(0)} (1,\{\alpha\}, n, \{\beta\}) =  
(-1)^{|\beta|} \sum_{\sigma \in OP(\alpha ,\beta^T)}  
A_{n}^{(0)}( 1 , \{\sigma\} , n)
\label{eq:KK}
\end{equation}
where $\{ \alpha\}$ and $\{\beta\}$ are  sets of the remaining indices i.e. 
$\{\alpha\} =\{ a_2 , \cdots a_p\}$ and $\{\beta\} =\{a_{p+1},\cdots a_{n-1}\}$.   
The summation is over the order permutations of $\alpha$ and $\beta^T$. That is permutations of the union of the sets where the ordering of 
$\alpha$ and $\beta^T$ are preserved.   This identity reduces the 
$(n-1)!/2$ independent amplitudes to $(n-2)!$.  For $n\leq 6$,   the decoupling identities may be used to show (\ref{eq:KK}) however for 
$n\geq 7$ they require further information.  For $n=7$,   the rank of the decoupling system is 239 whereas the rank of the Kleiss-Kuijf relations is just larger at 240: a single extra relation exists.   

For the one-loop amplitude,  the double trace terms are not independent but
can be expressed in terms of the leading~\cite{Bern:1994zx}
\begin{equation}
A_{n:r}^{(1)} (a_1,a_2,\cdots, a_{r-1} ; a_{r}, \cdots , a_n)=(-1)^{r} \sum_{\sigma\in COP\{\alpha\}\{\beta^T\}} 
A_{n:1}^{(1)}  (\sigma )
\label{eq:OneLoopSubLeading}
\end{equation}
where $\{\alpha\}=\{ a_1 ,\cdots a_{r-1}\}$ and $\{\beta\}=\{a_{r} \cdots a_n \}$.   The summation is over the ordered permutations as before but factoring out equivalent permutations due to cyclic symmetry (see appendices of ref.~\cite{Dunbar:2023ayw}  for examples of the summations). 
This relation allows the double trace terms to be derived from the leading in color terms only.  This reduces the number of functional forms to be computed in a calculation considerably. 
There are some analogues between the one-loop relation eq.~(\ref{eq:OneLoopSubLeading}) and the tree relation eq.~(\ref{eq:KK}).  
 This relation can be obtained from the decoupling equations for $n\leq 5$ but beyond $n=5$ the decoupling equations are not sufficient: explicitly,  at $n=6$,   decoupling identities determine the combination 
 \begin{equation}
 A_{6:4}^{(1)}  (a_1,a_2,a_3  ;  a_4,a_5,a_6 )+A_{6:4}^{(1)}  (a_1,a_3,a_2  ;  a_4,a_5,a_6 )
 \end{equation}
 but in themselves do not determine the individual terms.

Turning to  two-loop amplitudes,  for $n=5$ there are relations which allow the
$A_{5:1B}^{(2)}$ to be expressed in terms of the $A_{5:1}^{(2)}$
and $A_{5:3}^{(2)}$ thus reducing the number of independent functional forms from three to two~\cite{Edison:2011ta}.  
This relation is outside of decoupling identities:  the $A_{n:1B}^{(2)}$  only mixes with itself in decoupling identities.      In fact the 
$A_{n:1B}^{(2)}$ satisfies the identical decoupling relations as the tree amplitude $A_{n}^{(0)}$ and so satisfy (\ref{eq:KK}) for $n\leq 6$ at least. 

Examining the all-plus  amplitudes split into irreducible representations at specific kinematic points allows us to search for possible identities.    Firstly,   a significant number of the projections vanish: these are indicated as ``\YZ''  on the table.  
Secondly,  where an amplitude has multiple versions of a representation, then these may not be independent but proportional to each other. This proportionality is indicated as equality on the table. 
The numerical number of relations of a single vanishing or equality 
 is given by the dimension of the representation.   
We will first present the results for the six point amplitudes for comparison which are essentially the results of 
ref.~\cite{Edison:2012fn} together with a single extra identity identified in ref.~\cite{Dunbar:2023ayw}.  The irreducible representation content of the amplitudes is shown on table~\ref{table:sixpt}



\def\YZ{$I\hskip -5 pt=\hskip -3 pt 0$}
\def\YY{$I$}
\def\YE{$I\hskip -4 pt=\hskip -4 ptI$}
\def\YEE{$I\hskip -5 pt =\hskip -5 ptI\hskip -5 pt=\hskip -5 ptI$}
\def\YZZ{$I\hskip -5 pt=\hskip -5 pt I \hskip -5 pt=\hskip -5 pt0$}
\def\YEZ{$I\hskip -5 pt=I\hskip -5 pt,  I\hskip -5 pt=\hskip -5 pt0$}

\begin{table}
\begin{tabular}{|c|c|c|c|c|c|c|c|c|c|c|}
\hline
&  & $|R_i|$   &
$A^{(2)}_{6:1}$  &
$A^{(2)}_{6:2}$ &
$A^{(2)}_{6:3}$  &
$A^{(2)}_{6:4}$  &
$A^{(2)}_{6:1,1}$  &
$A^{(2)}_{6:1,2}$  &
$A^{(2)}_{6:2,2}$  &
$A^{(2)}_{6:1B}$  
\\
\hline
$|A^{(2)}_{6:x}|$ &  &  &   60 & 72 & 45 & 30 & 45 & 60 & 15 &  60 
\\
\hline\hline
$R_1$ &  $6$  &  1  &  \YY & \YY  & \YY & \YY & \YY & \YY &  \YY & \YZ
\\
\hline
$R_2$ &  $5,1$  & 5 & . & \YZ &  \YZ & . & \YZ &  \YZZ & . & . 
\\
\hline
$R_3$ &  $4,2$  &  9  & \YY,\YY &  \YY & \YY,\YY & \YY & \YY,\YY & \YE & \YY & \YE
\\
\hline
$R_4$ &  $4,1^2$  &   10 & . & .& . & .&  .& \YZ  & . & .
\\
\hline
$R_5$ &  $3^2$  &   5 & . & \YZ & . & .&  .& \YZ  & . & .
\\
\hline
$R_6$ &  $3,2,1$  & 16 & \YY & \YZ,\YY  & \YY & . & \YY & \YE & . & \YE 
\\
\hline
$R_7$ &  $3,1^3$  &  10 & \YY & .  & . & .&  .&   & . &  \YY
\\
\hline
$R_8$ &  $2^3$  &  5  & \YY,\YY & \YY & \YY & \YY &\YY & .  & \YZ & \YZ, \YY 
\\
\hline
$R_9$ &  $2^2,1$  &  9 & . & \YZ & . & .&  .& . & . & .
\\
\hline
$R_{10}$ &  $2,1^4$  &  5 & \YY & \YY & . & \YY &  .& .   & . & \YZ
\\
\hline
$R_{11}$ &  $1^6$  & 1  & . & \YZ & . & .&  .& .  & . & .
\\ 
\hline
\end{tabular}
\caption{The irreducible $S_6$ representation content of the six point partial amplitudes.  The number of independent partial amplitudes of each type after applying symmetries is shown in second row.  Each $I$ indicates a potential combination of the partial amplitudes after applying cyclic and flip symmetry.    Using the actual kinematic form of the all-plus amplitude some of these vanish which vanish or are proportional (indicated by equality).  }
\label{table:sixpt}
\end{table}

First note the relations among the $A^{(2)}_{6:1B}$, 
\begin{eqnarray}
A^{(2)}_{6:1B} 
\left(
\hbox{\begin{ytableau}  
\null  & \null   &
\null  & \null &  \null &  \null 
\end{ytableau}}
\right)  
&=&
A^{(2)}_{6:1B} 
\left(
\hbox{\begin{ytableau}  
\null  & \null   &
\null  \\ \null &  \null \\  \null 
\end{ytableau}}
\right)  
=
A^{(2)}_{6:1B} 
\left(
\hbox{\begin{ytableau}  
\null  & \null   \\
\null  & \null \\  \null &  \null 
\end{ytableau}}
\right) 
=
A^{(2)}_{6:1B} 
\left(
\hbox{\begin{ytableau}  
\null  & \null   \\
\null  \\ \null \\  \null \\  \null 
\end{ytableau}}
\right)_1  
=0 
\notag\\
A^{(2)}_{6:1B} 
\left(
\hbox{\begin{ytableau}  
\null  & \null   &
\null  & \null \\  \null &  \null 
\end{ytableau}}
\right)_1  
&=&
A^{(2)}_{6:1B} 
\left(
\hbox{\begin{ytableau}  
\null  & \null   &
\null  &  \null \\  \null &   \null 
\end{ytableau}}
\right)_2    \;  . 
\end{eqnarray}
These constitute 36 identities in total which reduce the 60 $A^{(2)}_{6:1B}(1,\sigma'(2,3,4,5,6))$ to the 
24 $A^{(2)}_{6:1B}(1,\sigma(2,3,4,5),6)$.   These are the Kleiss-Kuijf/decoupling identities.

Secondly, we have a set of relations involving the $A^{(2)}_{6:2,2} $
\begin{eqnarray}
A^{(2)}_{6:2,2} 
\left(
\hbox{\begin{ytableau}  
\null  & \null   \\
\null  & \null \\  \null &  \null 
\end{ytableau}}
\right)  
&=&0
\notag\\
A^{(2)}_{6:2,2} 
\left(
\hbox{\begin{ytableau}  
\null  & \null   &
\null  & \null \\  \null &  \null 
\end{ytableau}}
\right)
&=&
 -8 A^{(2)}_{6:4} 
\left(
\hbox{\begin{ytableau}  
\null  & \null   &
\null  & \null \\  \null &  \null 
\end{ytableau}}
\right)
\notag\\
A^{(2)}_{6:2,2} 
\left(
\hbox{\begin{ytableau}  
\null  & \null   &
\null  & \null &  \null &  \null 
\end{ytableau}}
\right)   &=& 
-24 A^{(2)}_{6:4} 
\left(
\hbox{\begin{ytableau}  
\null  & \null   &
\null  & \null &  \null &  \null 
\end{ytableau}}
\right)  
\; . 
\end{eqnarray}
Since the relations involve all the representations of $A^{(2)}_{6:2,2} $  this indicates that $A^{(2)}_{6:2,2} $ can be solved in terms of the other amplitudes.   This is known from the decoupling identities. 
We have some relations amongst the $A^{(2)}_{6:3} $  and $A^{(2)}_{6:4} $ 
\begin{eqnarray}
A^{(2)}_{6:3} 
\left(
\hbox{\begin{ytableau}  
\null  & \null   &
\null  & \null &  \null \\  \null 
\end{ytableau}}
\right)  
&=&
0
\notag\\
2 A^{(2)}_{6:3} 
\left(
\hbox{\begin{ytableau}  
\null  & \null   &
\null  & \null \\  \null &  \null 
\end{ytableau}}
\right)_1   
&+&
 A^{(2)}_{6:3} 
\left(
\hbox{\begin{ytableau}  
\null  & \null   &
\null  & \null \\  \null &  \null 
\end{ytableau}}
\right)_2 
 +4 A^{(2)}_{6:4} 
\left(
\hbox{\begin{ytableau}  
\null  & \null   &
\null  & \null \\  \null &  \null 
\end{ytableau}}
\right)
=0
\notag\\
A^{(2)}_{6:1} 
\left(
\hbox{\begin{ytableau}  
\null  & \null   &
\null  & \null &  \null &  \null 
\end{ytableau}}
\right)  
&=& \frac{1}{8}A^{(2)}_{6:3} 
\left(
\hbox{\begin{ytableau}  
\null  & \null   &
\null  & \null &  \null &  \null 
\end{ytableau}}
\right)   
=-\frac{1}{3}A^{(2)}_{6:4} 
\left(
\hbox{\begin{ytableau}  
\null  & \null   &
\null  & \null &  \null &  \null 
\end{ytableau}}
\right)   
\end{eqnarray}
which form constraints but are not sufficient to solve for either.      They are sufficient to indicate the the $A^{(2)}_{6:2,2}$ can be expressed either purely in terms of the $A^{(2)}_{6:4}$  or $A^{(2)}_{6:3}$.  

Finally we have  relations involving the $A^{(2)}_{6:1B} $
\begin{eqnarray}
\frac{8}{5} A^{(2)}_{6:1B} 
\left(
\hbox{\begin{ytableau}  
\null  & \null   &
\null  & \null \\  \null &  \null 
\end{ytableau}}
\right)_1    &=&
8 A^{(2)}_{6:1} 
\left(
\hbox{\begin{ytableau}  
\null  & \null   &
\null  & \null \\  \null &  \null 
\end{ytableau}}
\right)_1   
-4 A^{(2)}_{6:1} 
\left(
\hbox{\begin{ytableau}  
\null  & \null   &
\null  & \null \\  \null &  \null 
\end{ytableau}}
\right)_2 
 -2 A^{(2)}_{6:3} 
\left(
\hbox{\begin{ytableau}  
\null  & \null   &
\null  & \null \\  \null &  \null 
\end{ytableau}}
\right)_1
 + A^{(2)}_{6:3} 
\left(
\hbox{\begin{ytableau}  
\null  & \null   &
\null  & \null \\  \null &  \null 
\end{ytableau}}
\right)_2
\notag \\
2 A^{(2)}_{6:1B} 
\left(
\hbox{\begin{ytableau}  
\null  & \null   \\
\null  & \null \\  \null &  \null 
\end{ytableau}}
\right)  &=& 
-10 A^{(2)}_{6:1} 
\left(
\hbox{\begin{ytableau}  
\null  & \null   \\
\null  & \null \\  \null &  \null 
\end{ytableau}}
\right)_1
+
  A^{(2)}_{6:3} 
\left(
\hbox{\begin{ytableau}  
\null  & \null   \\
\null  & \null \\  \null &  \null 
\end{ytableau}}
\right)
+2 A^{(2)}_{6:4} 
\left(
\hbox{\begin{ytableau}  
\null  & \null   \\
\null  & \null \\  \null &  \null 
\end{ytableau}}
\right)
\; . 
\end{eqnarray}
These fourteen relations constrain the 24 $A^{(2)}_{6:1B}$ but are insufficient to solve - unlike the five point situation.  

For the seven-point amplitude we have the decomposition of the partial amplitudes as shown in table~\ref{table:sevenpt} where we also indicate the vanishing and proportionality for the all-plus amplitude of this paper. 

\begin{table}
\begin{tabular}{|c|c|c|c|c|c|c|c|c|c|c|c|}
\hline
&  & $|R_i|$  &
$A^{(2)}_{7:1}$  &
$A^{(2)}_{7:2}$ &
$A^{(2)}_{7:3}$  &
$A^{(2)}_{7:4}$  &
$A^{(2)}_{7:1,1}$  &
$A^{(2)}_{7:1,2}$  &
$A^{(2)}_{7:1,3}$  &
$A^{(2)}_{7:2,2}$  &
$A^{(2)}_{7:1B}$  
\\
\hline
$|A^{(2)}_{7:x}|$&  &  &   360 & 420 & 252 &210 &252 & 315 & 140 & 105 & 360 
\\
\hline\hline
$R_1$ &  $7$  &  1  &  . & . & . & . & . & . & . & . & .
\\
\hline
$R_2$ &  $6,1$  & 6 & . & . & . & . & . & . & . & . & .
\\
\hline
$R_3$ &  $5,2$  &  14  & . & . & . & . & . & . & . & . & .
\\
\hline
$R_4$ &  $5,1^2$  &   15 & \YY,\YY,\YY  & \YY, \YY & \YY,\YY & \YE & \YY,\YY & \YY  & \YY & \YY & \YEE
\\
\hline
$R_5$ &  $4,3$  &   14 & \YY & \YY & .  &  & . & . & . & . &  \YZ
\\
\hline
$R_6$ &  $4,2,1$  & 35 & \YY, \YY & \YEE & \YY , \YY & \YY & \YY,\YY & \YE & \YY & . & \YZ, \YY
\\
\hline
$R_7$ &  $4,1^3$  &  20 & \YY & \YEZ & \YY,\YY  & \YE & \YY,\YY & \YE & \YE  & \YY & \YZ 
\\
\hline
$R_8$ &  $3^2,1$  &  21  & \YY,\YY,\YY & \YY,\YY & \YY,\YY & \YY &\YY,\YY & \YY & . & \YY & \YEE
\\
\hline
$R_9$ &  $3,2^2$  &  21 &  . & \YZ & . & . & . & \YZ & . & . & . 
\\
\hline
$R_{10}$ &  $3,2,1^2$  &  35 & \YY,\YY,\YY &\YE , \YY& \YY,\YY &\YY,\YY & \YY, \YY & \YE,\YZ & \YY & \YY & \YEE
\\
\hline
$R_{11}$ &  $3,1^4$  & 15 &  . & \YZ & . & . & . & \YZ & \YZ & . & .
\\
\hline
$R_{12}$ &  $2^3,1$  &  14 & \YY & \YY & . & .  &. & \YZ & . & . & \YZ
\\
\hline
$R_{13}$ &  $2^2,1^3$  &  14 & \YY, \YY & \YY & . & \YY & . & \YZ & . & \YZ  & \YE
\\
\hline
$R_{14}$ &  $2,1^5$  &  6 &  . & . & . & . & . & . & . & . & .
\\
\hline
$R_{15}$ &  $1^7$  & 1  &\YY  &  . & . & . & . & . & . & . &  \YZ
\\ \hline
\end{tabular}
\caption{The irreducible $S_7$ representation content of the seven point partial amplitudes.   Those combinations which vanish or are equal for the all-plus are indicated.   }
\label{table:sevenpt}
\end{table}

\vfill\eject
 
For the all-plus seven-point amplitude we have relations amongst the $A^{(2)}_{7:1B}$
\begin{eqnarray}
A^{(2)}_{7:1B} 
\left(
\hbox{\begin{ytableau}  
\null  & \null   &
\null  & \null \\  \null &  \null &
\null
\end{ytableau}}
\right) &=&
A^{(2)}_{7:1B} 
\left(
\hbox{\begin{ytableau}  
\null  & \null   &
\null  & \null \\  \null &  \null \\
\null
\end{ytableau}}
\right)_1 =
A^{(2)}_{7:1B} 
\left(
\hbox{\begin{ytableau}  
\null  & \null   &
\null  & \null \\  \null \\  \null \\
\null
\end{ytableau}}
\right) =
A^{(2)}_{7:1B} \left(
\hbox{\begin{ytableau}  
\null  & \null   \\
\null  & \null \\  \null &  \null \\
\null
\end{ytableau}}
\right)
=A^{(2)}_{7:1B} 
\left(\hbox{\begin{ytableau}  
\null  \\ \null   \\
\null  \\ \null \\  \null \\  \null \\
\null
\end{ytableau}}
\right)
=0
\notag
\\
A^{(2)}_{7:1B} 
\left(
\hbox{\begin{ytableau}  
\null  & \null   &
\null  & \null &  \null \\  \null \\
\null
\end{ytableau}}
\right)_1
&=&
A^{(2)}_{7:1B} 
\left(
\hbox{\begin{ytableau}  
\null  & \null   &
\null  & \null &  \null \\  \null \\
\null
\end{ytableau}}
\right)_2
=
A^{(2)}_{7:1B} 
\left(
\hbox{\begin{ytableau}  
\null  & \null   &
\null  & \null &  \null \\  \null \\
\null
\end{ytableau}}
\right)_3
\notag
\\
A^{(2)}_{7:1B} 
\left(
\hbox{\begin{ytableau}  
\null  & \null   &
\null  \\ \null &  \null &  \null \\
\null
\end{ytableau}}
\right)_1
&=&
A^{(2)}_{7:1B} 
\left(
\hbox{\begin{ytableau}  
\null  & \null   &
\null  \\  \null &  \null &  \null \\
\null
\end{ytableau}}
\right)_2
=
A^{(2)}_{7:1B} 
\left(
\hbox{\begin{ytableau}  
\null  & \null   &
\null  \\  \null &  \null &  \null \\
\null
\end{ytableau}}
\right)_3
\notag
\\
A^{(2)}_{7:1B} 
\left(
\hbox{\begin{ytableau}  
\null  & \null   &
\null  \\ \null &  \null \\  \null \\
\null
\end{ytableau}}
\right)_1
&=&
A^{(2)}_{7:1B} 
\left(
\hbox{\begin{ytableau}  
\null  & \null   &
\null  \\  \null &  \null \\  \null \\
\null
\end{ytableau}}
\right)_2
=
A^{(2)}_{7:1B} 
\left(
\hbox{\begin{ytableau}  
\null  & \null   &
\null  \\  \null &  \null \\  \null \\
\null
\end{ytableau}}
\right)_3
\notag
\\
A^{(2)}_{7:1B} 
\left(
\hbox{\begin{ytableau}  
\null  & \null   \\
\null  & \null \\  \null \\  \null \\
\null
\end{ytableau}}
\right)_1
&=&
A^{(2)}_{7:1B} 
\left(
\hbox{\begin{ytableau}  
\null  & \null   \\
\null  &  \null \\  \null \\  \null \\
\null
\end{ytableau}}
\right)_2
\; . 
\end{eqnarray}
In total,   the all-plus $A^{(2)}_{7:1B}$ satisfies 240 relations which allow a Kleiss-Kuijf (\ref{eq:KK})  type relation amongst themselves.   The expected number from decoupling identities is 239. The extra relation required is the vanishing of the $1^7$ representation.  
\begin{equation}
A^{(2)}_{7:1B} 
\left(\hbox{\begin{ytableau}  
\null  \\ \null   \\
\null  \\ \null \\  \null \\  \null \\
\null
\end{ytableau}}
\right)
=0
\end{equation}
which in terms of amplitudes is
 \begin{equation}
   \sum_{\sigma_6}  (-1)^{|\sigma_6|} A_{7:1B}^{(2)} (1, \sigma_6(2,3,4,5,6,7) ) =0
   \; . 
 \end{equation}
This is satisfied for the all plus amplitude.  It provides a future test for other helicity amplitudes as to whether the Kleiss-Kuijf relations persist beyond six point for $A_{n:1B}^{(2)}$ and arbitrary helicity.

We have relations involving $A^{(2)}_{7:2,2}$,
\begin{eqnarray}
A^{(2)}_{7:2,2} 
\left(\hbox{\begin{ytableau}  
\null  & \null   \\
\null  &  \null \\  \null \\  \null \\
\null
\end{ytableau}}
\right)
&=&0
\notag
\\
A^{(2)}_{7:2,2} 
\left(\hbox{\begin{ytableau}  
\null  & \null   &
\null  &  \null &  \null \\  \null \\
\null
\end{ytableau}}
\right) &=& 
6 A^{(2)}_{7:3} 
\left(\hbox{\begin{ytableau}  
\null  & \null   &
\null  &  \null &  \null \\  \null \\
\null
\end{ytableau}} \right)_1
+6 A^{(2)}_{7:3} 
\left(\hbox{\begin{ytableau}  
\null  & \null   &
\null  &  \null &  \null \\  \null \\
\null
\end{ytableau}}
\right)_2
-8 A^{(2)}_{7:4} 
\left(\hbox{\begin{ytableau}  
\null  & \null   &
\null  &  \null &  \null \\  \null \\
\null
\end{ytableau}}
\right)_2
\notag
\\
A^{(2)}_{7:2,2} 
\left(\hbox{\begin{ytableau}  
\null  & \null   &
\null  &  \null \\  \null \\  \null \\
\null
\end{ytableau}}
\right)
&=&
12 A^{(2)}_{7:3} 
\left(\hbox{\begin{ytableau}  
\null  & \null   &
\null  &  \null \\  \null \\  \null \\
\null
\end{ytableau}}
\right)_1
+12 A^{(2)}_{7:3} 
\left(\hbox{\begin{ytableau}  
\null  & \null   &
\null  &  \null \\  \null \\  \null \\
\null
\end{ytableau}}
\right)_2
\notag
\\
A^{(2)}_{7:2,2} 
\left(\hbox{\begin{ytableau}  
\null  & \null   &
\null  \\  \null &  \null &  \null \\
\null
\end{ytableau}}
\right)
&=&
2 A^{(2)}_{7:3} 
\left(\hbox{\begin{ytableau}  
\null  & \null   &
\null  \\  \null &  \null &  \null \\
\null
\end{ytableau}}
\right)_1
\notag
\\
A^{(2)}_{7:2,2} 
\left(\hbox{\begin{ytableau}  
\null  & \null   &
\null  \\  \null &  \null \\  \null \\
\null
\end{ytableau}}
\right)
&=&
A^{(2)}_{7:3} 
\left(\hbox{\begin{ytableau}  
\null  & \null   &
\null  \\  \null &  \null \\  \null \\
\null
\end{ytableau}}
\right)_1
+
A^{(2)}_{7:3} 
\left(\hbox{\begin{ytableau}  
\null  & \null   &
\null  \\  \null &  \null \\  \null \\
\null
\end{ytableau}}
\right)_2  \; . 
\end{eqnarray}
Since these determine all the irreducible representations of $A^{(2)}_{7:2,2} $ this indicates that $A^{(2)}_{7:2,2}$ can be expressed in terms of the sub-leading partial amplitudes $A^{(2)}_{7:3}$ and $A^{(2)}_{7:4}$.  This can be seen from decoupling identities. 

There are two further identities relating the remaining 
$A^{(2)}_{7:1B}$
\begin{eqnarray}
18A^{(2)}_{7:1B} 
\left(\hbox{\begin{ytableau}  
\null  & \null   &
\null  &  \null &  \null \\  \null \\
\null
\end{ytableau}}
\right)
 +
126A^{(2)}_{7:1}
\left(\hbox{\begin{ytableau}  
\null  & \null   &
\null  &  \null &  \null \\  \null \\
\null
\end{ytableau}}
\right)_1
 & & -42A^{(2)}_{7:1} 
\left(\hbox{\begin{ytableau}  
\null  & \null   &
\null  &  \null &  \null \\  \null \\
\null
\end{ytableau}}
\right)_3
+21A^{(2)}_{7:3} 
\left(\hbox{\begin{ytableau}  
\null  & \null   &
\null  &  \null &  \null \\  \null \\
\null
\end{ytableau}}
\right)_2
+70A^{(2)}_{7:4} 
\left(\hbox{\begin{ytableau}  
\null  & \null   &
\null  &  \null &  \null \\  \null \\
\null
\end{ytableau}}
\right)_1
=  0
\notag
\\
20A^{(2)}_{7:1B} 
\left(\hbox{\begin{ytableau}  
\null  & \null   &
\null  \\  \null &  \null &  \null \\
\null
\end{ytableau}}
\right)
  +210A^{(2)}_{7:1B} 
\left(\hbox{\begin{ytableau}  
\null  & \null   &
\null  \\  \null &  \null &  \null \\
\null
\end{ytableau}}
\right)_1
& & +270A^{(2)}_{7:1B} 
\left(\hbox{\begin{ytableau}  
\null  & \null   &
\null  \\  \null &  \null &  \null \\
\null
\end{ytableau}}
\right)_2
-450A^{(2)}_{7:1B} 
\left(\hbox{\begin{ytableau}  
\null  & \null   &
\null  \\  \null &  \null &  \null \\
\null
\end{ytableau}}
\right)_3
+
\notag
\\
& & 
-21A^{(2)}_{7:3} 
\left(\hbox{\begin{ytableau}  
\null  & \null   &
\null  \\  \null &  \null &  \null \\
\null
\end{ytableau}}
\right)_1
-12A^{(2)}_{7:3} 
\left(\hbox{\begin{ytableau}  
\null  & \null   &
\null  \\  \null &  \null &  \null \\
\null
\end{ytableau}}
\right)_2
-30A^{(2)}_{7:4} 
\left(\hbox{\begin{ytableau}  
\null  & \null   &
\null  \\  \null &  \null &  \null \\
\null
\end{ytableau}}
\right) = 0
\; . 
\end{eqnarray}
Although these form constraints (36 in number) upon the $A^{(2)}_{7:1B}$ these is not enough to determine  since the
\begin{equation}
A^{(2)}_{7:1B} 
\left(\hbox{\begin{ytableau}  
\null  & \null   &
\null  &  \null \\  \null &  \null \\
\null
\end{ytableau}}
\right)_2
 \; , 
 A^{(2)}_{7:1B} 
\left(\hbox{\begin{ytableau}  
\null  & \null   &
\null  \\  \null &  \null \\  \null \\
\null
\end{ytableau}}
\right)
\; , 
A^{(2)}_{7:1B} 
\left(\hbox{\begin{ytableau}  
\null  & \null   \\
\null  & \null \\ \null \\  \null \\
\null
\end{ytableau}}
\right)
\end{equation}
are involved in no  relations.

We also a small number of constraints

\begin{eqnarray}
A^{(2)}_{7:4} 
\left(\hbox{\begin{ytableau}  
\null  & \null   &
\null  & \null \\ \null &  \null \\
\null
\end{ytableau}}
\right)
+
A^{(2)}_{7:3} 
\left(\hbox{\begin{ytableau}  
\null  & \null   &
\null  & \null \\ \null &  \null \\
\null
\end{ytableau}}
\right)_1
+
A^{(2)}_{7:3} 
\left(\hbox{\begin{ytableau}  
\null  & \null   &
\null  & \null \\ \null &  \null \\
\null
\end{ytableau}}
\right)_2
&=& 0
\notag
\\
A^{(2)}_{7:4} 
\left(\hbox{\begin{ytableau}  
\null  & \null   &
\null  & \null \\ \null \\  \null \\
\null
\end{ytableau}}
\right)_1
+
A^{(2)}_{7:3} 
\left(\hbox{\begin{ytableau}  
\null  & \null   &
\null  & \null \\ \null \\  \null \\
\null
\end{ytableau}}
\right)_1
+
A^{(2)}_{7:3} 
\left(\hbox{\begin{ytableau}  
\null  & \null   &
\null  & \null \\ \null \\  \null \\
\null
\end{ytableau}}
\right)_2
&=& 0
\notag
\\
2 A^{(2)}_{7:4} 
\left(\hbox{\begin{ytableau}  
\null  & \null   &
\null  \\ \null & \null \\  \null \\
\null
\end{ytableau}}
\right)_1
+
A^{(2)}_{7:3} 
\left(\hbox{\begin{ytableau}  
\null  & \null   &
\null \\  \null & \null \\  \null \\
\null
\end{ytableau}}
\right)_1
+
A^{(2)}_{7:3} 
\left(\hbox{\begin{ytableau}  
\null  & \null   &
\null  \\ \null & \null \\  \null \\
\null
\end{ytableau}}
\right)_2
&=& 0
\notag
\\
-4A^{(2)}_{7:4} 
\left(\hbox{\begin{ytableau}  
\null  & \null   &
\null  & \null & \null \\  \null \\
\null
\end{ytableau}}
\right)_1
+36
A^{(2)}_{7:1} 
\left(\hbox{\begin{ytableau}  
\null  & \null   &
\null  & \null & \null \\  \null \\
\null
\end{ytableau}}
\right)_2+
24 A^{(2)}_{7:1} 
\left(\hbox{\begin{ytableau}  
\null  & \null   &
\null  & \null & \null \\  \null \\
\null
\end{ytableau}}
\right)_3  +& & 
\notag
\\
-9 A^{(2)}_{7:3} 
\left(\hbox{\begin{ytableau}  
\null  & \null   &
\null  & \null & \null \\  \null \\
\null
\end{ytableau}}
\right)_1-
3 A^{(2)}_{7:3} 
\left(\hbox{\begin{ytableau}  
\null  & \null   &
\null  & \null & \null \\  \null \\
\null
\end{ytableau}}
\right)_2  &=& 0
\; . 
\end{eqnarray}

These finalise all possible linear relationships between the partial amplitudes.   Whether they all extend to all helicities remains open 

\vfill\eject

\section{Eight and Nine Point Relations}

   \label{section:eight}
We can use the results for the polylogarithmic terms $P^{(2)}_{n:\lambda}$ of section~\ref{section:four}     as  experimental data to search for linear identities amongst the partial amplitudes.    The results of this study are displayed, in fairly compact form,  in tables~\ref{tables:eightpt}  
and~\ref{table:ninept}.    As before we emphasis these are purely for the all plus configuration and are only indicative for other helicities.

As can be seen the complexity of the decomposition increases with number of legs and in particular there are significant numbers of independent combinations which are in the same representation of $S_n$. 
Consequently,  we have chosen to organise this more compactly by listing the number of relations a particular set of combinations satisfy.  For example the $R_3$ content of the $A^{(2)}_{8:1B}$ representation is given as $''3(3)[2]''$.    The first number is the number of independent copies of the representation.  The second number $(3)$ is the number of independent relations satisfied by these combinations. The third number $[2]$ is the number of independent relations {\it amongst themselves.}. These may be equating a combination to zero or proportionality between terms as in the six and seven point amplitudes.    If there are enough relations involving the terms for all representations.   i.e. the  second number equals the first for all $R_i$,  then that partial amplitude has a solution in terms of the others.

We can summarise some of the content of these (and the six and seven point amplitudes always with the caveats

$\bullet$  There are no relations amongst the leading-in-color $A^{(2)}_{8:1}$ or $A^{(2)}_{9:1}$.  Hence these are all independent
.

$\bullet$ There are some number of relations amongst the amplitudes of the other  partial amplitudes : the total number of which indicated as $[\#]$ at bottom of table.
These are quite numerous for the triple trace terms but much more limited for the double trace. 

$\bullet$ There are 
significant relations amongst the $A^{(2)}_{8:1B}$ and $A^{(2)}_{9:1B}$ with the number precisely that needed 
{\it with no more} to reduce the $(n-1)!/2$ to $(n-2)!$ independent amplitudes.

$\bullet$ The $A^{(2)}_{n:1B}$ satisfy significant further identities. 
These relations are beyond those due to decoupling identities. 
However,   there are not enough relations to completely specify this term.  For $n=8$ there is {\it just} not enough with only a single identity for $R_7$ lacking.   For $n=9$ there are multiple identities lacking.   

$\bullet$  The sub-sub-leading  triple trace terms can be expressed in terms of the double trace partial amplitudes.  
This is in addition to the information present in the table.   
For $A^{(2)}_{n:2,r}$ this can be shown true for all helicities using decoupling identities, that is 
\begin{eqnarray}
\biggl\{A^{(2)}_{n:2,r}  \biggr\}  &\in & \biggl\{  \sum_{s > 2} A^{(2)}_{n:s} \biggr\} 
\end{eqnarray} 
however 
we find this  is also true for $A^{(2)}_{9:3,3}$.   Futhermore there are  simplifications. For $n=9$,  
\begin{eqnarray}
\biggl\{A^{(2)}_{9:2,2}  \biggr\}  &\in & \biggl\{A^{(2)}_{9:3}  , A^{(2)}_{9:5}  \biggr\}  
\notag\\
\biggl\{A^{(2)}_{9:2,3}  \biggr\}  &\in & \biggl\{A^{(2)}_{9:3},   A^{(2)}_{9:5}  \biggr\}   \hbox{ or  } \biggl\{A^{(2)}_{9:4},   A^{(2)}_{9:5} \biggr\} 
\notag\\
\biggl\{A^{(2)}_{9:3,3}  \biggr\}  &\in &\biggl\{A^{(2)}_{9:3}  , A^{(2)}_{9:4}  \biggr\}  \hbox{ or  }
\biggl\{A^{(2)}_{9:4}  , A^{(2)}_{9:5}  \biggr\} 
\; . 
\end{eqnarray} 
 In ref~\cite{Dunbar:2023ayw} it was speculated that  $A^{(2)}_{9:3,3}$ could be expressed a  sum of $A^{(2)}_{9:4}$ 
 however this is not the case.

\begin{table}
\begin{tabular}{|c|c|c||c|c|c|c|c|c|c|c|}
\hline
&  & $|R_i|$  &
$A^{(2)}_{8:1}$  &
$A^{(2)}_{8:3}$  &
$A^{(2)}_{8:4}$  &
$A^{(2)}_{8:5}$  &
$A^{(2)}_{8:2,2}$  &
$A^{(2)}_{8:2,3}$  &
$A^{(2)}_{8:1B}$   &\# 
\\
\hline
$|A_{8:x}^{(2)}|$&  &  &   2520  & 1680 &  1344
&  630  & 630  &560  & 2520  & 
\\
\hline\hline
$R_1$ &  $8$  &  1  &  1(1)[0]   & 1(1)[0]  &1(1)[0] &  1(1)[0] & 1(1)[0] & 1(1)[0]  & 1(1)[1] &  4
\\
\hline
$R_2$ &  $7,1$  & 7   & 0 & 1(1)[0]   & 1(1)[0]  & 0 & 1(1)[0] & 1(1)[0] & 0& 1 
\\
\hline
$R_3$ &  $6,2$  &  20 & 3(3)[0]   & 3(3)[0]   & 2(2)[0]  & 2(2)[0]  & 3(3)[1] & 2(2)[0]  & 3(3)[2]   & 7
 \\
\hline
$R_4$ &  $6,1^2$  &   21 & 0 & 0  & 0   & 0&0 & 0 & 0    & 
\\
\hline
$R_5$ &  $5,3$  &   28 & 1(0)[0]    &  2(1)[0]  & 2(2)[1]  &0  & 1(1)[1] & 1(1)[0] & 1(1)[1]   & 3  
\\
\hline
$R_6$ &  $5,2,1$  & 64 & 4(2)[0]   &  3(2)[0]   & 2(2)[0]  &1(1)[0] & 2(2)[1] & 1(1)[0]  & 4(4)[3]   & 7 
\\
\hline
$R_7$ &  $5,1^3$  &  35 &  2(0)[0]   &  1(0)[0]  & 0  &0 & 0 & 0 & 2(1)[1]    & 1
\\
\hline
$R_8$ &  $4^2$  &  14  & 3(2)[0]   &  2(2)[0]   & 0  &2(2)[1]  & 2(2)[1] & 1(1)[0] &  3(3)[2]   & 5 
\\
\hline
$R_9$ &  $4,3,1$  &  70 & 3(1)[0]   & 3(2)[0]   & 1(1)[0]  &0  & 1(1)[0] & 1(1)[0]  & 3(3)[2]   & 4
\\
\hline
$R_{10}$ &  $4,2^2$  &  56 & 7(4)[0]   &  5(4)[0]   &4(4)[0] &3(3)[0]  & 3(3)[1] & 2(2)[0]  & 7(7)[5]   & 12
\\
\hline
$R_{11}$ &  $4,2,1^2$  & 90 & 4(1)[0]   &  2(1)[0]   & 3(2)[0]  &0 &0 & 0 & 4(4)[3]   & 5
\\
\hline
$R_{12}$ &  $4,1^4$  &  35 &  4(2)[0]   &  2(1)[0]   & 3(2)[0]  & 1(1)[0] & 0& 1(1)[0]   & 4(4)[3]   & 6 
\\
\hline
$R_{13}$ &  $3^2,2$  &  42 &  1(0)[0]   &  1(1)[0]   & 1(1)[0] &0 & 0 & 0   & 1(1)[1]   & 2
\\
\hline
$R_{14}$ &  $3^2,1^2$  &  56 & 5(2)[0]   &  2(1)[0]   & 0   &2(1)[0]  & 1(1)[0] & 0 &5(5)[3]     & 5 
\\
\hline
$R_{15}$ &  $3,2^2,1$  & 70 &  4(1)[0]   &  3(2)[0]   & 3(2)[0]  &1(1)[0] & 1(1)[0] & 1(1)[0]   & 4(4)[3]   & 6
\\
\hline
$R_{16}$ &  $3,2,1^3$  &  64 & 4(1)[0]   & 2(1)[0]   & 2(1)[0] &1(1)[0]  & 0& 1(1)[0]  & 4(4)[3]   & 5
\\
\hline
$R_{17}$ &  $3,1^5$  &  21 & 1(0)[0]   &  1(1)[0]  & 3(2)[1]  &0  & 0& 1(1)[0]  & 1(1)[1] & 3
\\
\hline
$R_{18}$ &  $2^4$  & 14 &  3(1)[0]   & 2(1)[0]   & 0  &2(1)[0]   & 1(1)[0] & 1(1)[0]  & 3(3)[2]   & 4 
\\
\hline
$R_{19}$ &  $2^3,1^2$  &  28 & 0 &  0  &0  &0 &0  & 0 & 0   &.   
\\
\hline
$R_{20}$ &  $2^2,1^4$  &  20 & 2(1)[0]   &  1(1)[0]   & 0  &1(1)[0]  & 0 & 1(1)[0]  & 2(2)[1]   & 3 
\\
\hline
$R_{21}$ &  $2,1^6$  &  7 & 0 &  0  & 0  &0& 0 & 0 &0 &.  
\\
\hline
$R_{22}$ &  $1^7$  & 1  & 0 &  0  & 0  & 0& 0  & 0 & 0 &.  
\\
\hline
$[\#]$ &    &   & 0 &  0  & 49  & 14  & 182   & 0   &  1800 &.  
\\
\hline
\end{tabular}
\caption{Irreducible representation content of the eight point amplitudes and relations.  This table contains various information.  The first number indicates the number of independent copies of the representation in the partial amplitude using only its cyclic and flip symmetry.  The other two numbers refer to the actual amplitudes of the all-plus configuation.  The number in round brackets indicates the number of independent linear relations those combinations are involved in.  The number in square brackets is the number of linear relations those combinations obey amongst themselves.  The bottom row indicates the total number of identities a specific partial amplitude satisfies amongst themselves. The final column is the total number of identities amongst the $A^{(2)}_{8:1}$, $A^{(2)}_{8:3}$,$A^{(2)}_{8:4}$,$A^{(2)}_{8:5}$ and $A^{(2)}_{8:1B}$.    }
\label{tables:eightpt}\end{table}

\vfill\eject

\begin{table}
\hskip -1.0 truecm
\begin{tabular}{|c|c|c|c|c|c|c|c|c|c|c|c|}
\hline
&  & $|R_i|$  &
$A^{(2)}_{9:1}$   &
$A^{(2)}_{9:3}$  &
$A^{(2)}_{9:4}$  &
$A^{(2)}_{9:5}$  &
$A^{(2)}_{9:2,2}$  &
$A^{(2)}_{9:2,3}$  &
$A^{(2)}_{9:3,3}$  &
$A^{(2)}_{9:1B}$  & 
\#  
\\
\hline
$|A_{9:x}| $ &  & &      20160  &   12960 &  10080   & 9072 & 4536 & 7560 & 1120      & 20160
\\
\hline\hline
$R_1$ &  $9$  & $1$  &  0    & 0  & 0  &  0 & 0&  0 & 0 & 0  & .
\\   \hline
$R_2$ &  $8,1$  & $8$  &  0   & 0  & 0  &  0 & 0 &  0 & 0 & 0 & .
\\   \hline
$R_3$ &  $7,2$  & $27$  &  0   &  0 & 0 &  0  &   0  & 0 & 0 & 0  & . 
\\ \hline
$R_4$ &  $7,1^2$  & $28$  &  4(4)[0]   & 3(3)[0]     &  3(3)[0]   &  3(3)[0]   &  2(2)[0]  & 2(2)[0] & 1(1)[0] & 4(4)[3] & 10
\\  \hline
$R_5$ &  $6,3$  & $48$  &  3(2)[0]      & 1(1)[0]    & 1(1)[0]   &  0 &  0 &0   & 0 & 3(3)[2] & 4 
\\  \hline
$R_6$ &  $6,2,1$  & $105$  &  5(1)[0]      & 5(3)[0]     &  3(3)[0]   & 3(3)[0]   &  2(2)[0]  &  3(3)[2]  &0    & 5(4)[4] & 9 
\\  \hline
$R_7$ &  $6,1^3$  & $56$  &  3(2)[0]    & 4(3)[0]  &  4(4)[0]  &  3(3)[0]   &  2(2)[0]  & 4(4)[2]  & 1(1)[0] & 3(3)[2] & 9 
\\  \hline
$R_8$ &  $5,4$  & $42$  &  1(0)[0]     & 1(0)[0]    & 0 &   0  & 0 &  0   &  0 & 1(1)[1] & 1
\\     \hline
$R_9$ &  $5,3,1$  & $162$  &  12(6)[0]     &  9(7)[0]    & 6(6)[0]   &   5(5)[0]  & 4(4)[1]  & 4(4)[2] & 1(1)[1]  & 12(12)[9]  & 20 
\\  \hline
$R_{10}$ &  $5,2^2$  & $120$  &  3(0)[0]     & 2(1)[0]    & 1(1)[0]   &   0  &  0  & 1(1)[1] & 0 & 3(2)[2] & 3 
\\  \hline
$R_{11}$ &  $5,2,1^2$  & $189$  &  12(4)[0]      &  9(6)[0]   &  7(6)[0]   & 6(5)[1]  &  4(4)[1]  & 7(7)[5] &1(1)[1]  & 12(11)[9]  & 20 
\\  \hline
$R_{12}$ &  $5,1^4$  & $70$  &  1(0)[0]      & 1(1)[0]    & 1(1)[0]   &  0 &  0  &  2(2)[2]  & 0 &  1(1)[1]  & 2 
\\    \hline
$R_{13}$ &  $4^2,1$  & $84$  &  3(1)[0]    & 3(2)[0]    & 2(2)[1]  &  1(1)[0]   &   0 &  1(1)[1]  & 0  & 3(3)[2]  & 5 
\\    \hline
$R_{14}$ &  $4,3,2$  & $168$  &  9(3)[0]     & 6(4)[0]    & 3(3)[0]     &   4(3)[0]    &  2(2)[1]  &2(2)[1]   & 0 & 9(9)[7] & 12 
\\    \hline
$R_{15}$ &  $4,3,1^2$  & $216$  &  12(4)[0]      & 9(6)[0]   & 7(6)[1]     & 7(6)[1]   &  4(4)[2]   & 6(6)[4]  & 1(1)[1] & 12(12)[9] & 20 
\\    \hline
$R_{16}$ &  $4,2^2,1$  & $216$  &  12(3)[0]    & 6(4)[0]    &  4(3)[0]   & 4(3)[0]    &  2(2)[1] & 3(3)[2] & 0  & 12(11)[9]  & 14 
\\    \hline
$R_{17}$ &  $4,2,1^3$  & $189$  &  12(3)[0]      & 6(4)[0]     & 6(4)[0]     &   4(3)[0]    & 2(2)[1] &  5(5)[3]  & 1(1)[1] &   12(11)[9] & 15 
\\    \hline
$R_{18}$ &  $4,1^5$  & $56$  &  3(0)[0]     & 0    & 1(0)[0]    &   0  & 0 &  0  & 0 &   3(2)[2]  & 2 
\\    \hline
$R_{19}$ &  $3^3$  & $42$  &  6(4)[0]      & 3(3)[0]    &  3(3)[0]   &   3(3)[0]    & 2(2)[1] &  1(1)[0] & 1(1)[0] &   6(6)[4] & 9 
\\ \hline
$R_{20}$ &  $3^2,2,1$  & $168$  &  9(2)[0]      & 6(4)[0]    & 4(3)[0]    &   5(4)[0]    &  2(2)[1] &   3(3)[1]& 0 & 9(9)[7]  & 12
\\ \hline
$R_{21}$ &  $3^2,1^3$  & $120$  & 3(1)[0]      & 3(2)[0]    & 2(2)[0]    &  2(2)[0]   & 0  &  2(2)[1]  & 0 & 3(3)[2]  & 5 
\\ \hline
$R_{22}$ &  $3,2^3$  & $84$  &  3(1)[0]     & 1(1)[0]    & 1(0)[0]    &  1(1)[0]     & 0 &  0  & 0 & 3(3)[2]  & 3 
\\ \hline
$R_{23}$ &  $3,2^2,1^2$  & $162$  &  12(4)[0]       & 6(4)[0]    & 6(4)[0]    &  6(5)[0]   &  2(2)[1]  & 3(3)[1]  & 1(1)[0] & 12(12)[9]  & 16 
\\ \hline
$R_{24}$ &  $3,2,1^4$  & $105$  &  5(1)[0]     & 2(1)[0]    &  3(2)[0]    &  2(2)[0]     &  0  & 1(1)[0]  &  0 & 5(5)[4] & 6 
\\ \hline
$R_{25}$ &  $3,1^6$  & $28$  &  4(2)[0]      & 1(1)[0]   & 2(1)[0]   &  1(1)[0]    & 0 &0  & 1(1)[0] & 4(4)[3]   & 5 
\\ \hline
$R_{26}$ &  $2^4,1$  & $42$  &  1(0)[0]      & 1(1)[0]    & 1(1)[0]    &  1(1)[0]    &  0   &0  & 0& 1(1)[1]  & 2 
\\ \hline
$R_{27}$ &  $2^3,1^3$  & $48$  &  3(1)[0]      & 2(1)[0]    &  3(2)[0]    & 2(2)[0]    &  0  & 1(1)[0]&  1(1)[0] & 3(3)[2]  & 5 
\\ \hline
$R_{28}$ &  $2^2,1^5$  & $27$  &  0    & 0  & 1(1)[0]    &  1(1)[0]    & 0 &  0  &0  &  0  & 1 
\\ \hline
$R_{29}$ &  $2,1^7$  & $8$  &  0    & 1(1)[0]    & 1(1)[0]    &  1(1)[0]    &  0 &  0 &0  &   0  & 2 
\\ \hline
$R_{30}$ &  $1^9$  & $1$  &  0   & 0   & 0 &  0&  0 &  0   & 0  &   0  & . 
\\
\hline
$[\#]$  &  &  & 0 & 0 & 300 & 405 & 1728 & 4416 & 756 & 15120  & 
\\\hline\end{tabular}
\caption{Irreducible representation content of the nine point amplitudes and relations using the same format as for the eight point 
amplitude.}
\label{table:ninept}
\end{table}

 \vfill\eject  
   
\section{Conclusions}

Using the techniques of four dimensional unitarity cuts and augmented recursion, we have obtained the two-loop seven-gluon all-plus helicity Yang-Mills amplitude in a compact, analytic form.
By separating the procedure into these two parts, we have avoided the need for a more difficult D-dimensional unitarity approach.
Our method has only required evaluation of one-loop integrals, to obtain a two-loop result.

The technique of four dimensional unitarity is manifestly gauge invariant throughout, involving on-shell amplitudes which are themselves gauge invariant as well as box integrals.
Working in four dimensions allows the spinor helicity formalism to be used straightforwardly.
Helicity considerations greatly constrain the number of diagrams that contribute to the unitarity.

The BCFW recursion is also manifestly gauge invariant.
The only ingredients are gauge invariant lower-point amplitudes.
Although the process introduces a new reference momentum to allow the amplitude to be treated as a complex function, the result is independent of this reference and it is not a gauge choice.

The only gauge dependence of our process takes place in the augmented recursion portion.
To ensure that both leading and sub-leading poles are accounted for, we carry out some explicit loop integrals involving currents.
Internal off-shell legs are treated in the spinor helicity setup with an axial gauge formalism, which introduces a reference momentum gauge choice.
Despite the individual contributions being gauge dependent, the overall result is independent of the reference momentum, which is a powerful consistency check.
The particular choice of axial gauge for the gauge dependent step is also convenient because ghosts decouple from gluons in this gauge, meaning they did not appear in our diagrams.  The final result is gauge invariant, as would be expected for an observable such as an amplitude.

Our calculation presents a new $n$-point expression  for the two-loop all-plus polylogarithmic piece $P_{n:\lambda}^{(2)}(1^+,2^+,\cdots,n^+)$.
It also finds agreement with a previous n-point conjecture, for the $N_c$ independent single-trace rational piece $R_{n:1B}^{(2)}(1^+,2^+,\cdots,n^+)$~\cite{Dunbar:2020wdh}.
Both were found in large part by identifying patterns in compact analytic amplitude expressions, demonstrating the value of calculating such objects.

Although for a specific helicity configuration,  this amplitude can be used to explore the properties of amplitudes and hopefully lead to further insights.    We have used the full seven-point amplitude together with the $n$-point polylogarithmic amplitude for $n=8,9$ to test for linear relations between the partial amplitudes akin to the Kleiss-Kuijf relations of tree  amplitudes. The results are limited and in line with previous studies but we provide evidence that the most sub-leading in color can be derived from the other partial amplitudes.

This work was partially supported by the UKRI Science and Technology Facilities Council (STFC) Consolidated Grant No. ST/T000813/1.    For the purpose of open access, the authors have applied a Creative Commons Attribution (CC BY) licence.

\appendix

\section{Spinors Helicity Conventions}
\label{app:spin}

Amplitudes can be represented more compactly if spinors are used to represent the momenta and 
polarisations of massless particles \cite{DeCausmaecker:1981jtq,Berends:1981uq,Kleiss:1985yh,Xu:1986xb}, 
in what is known as the ``spinor-helicity formalism''.
Each four-momentum $p_i^\mu$ is replaced by  a pair of two-component spinors $\lambda_i^\alpha$ and $\Tilde{\lambda}_i^{\dot{\alpha}}$, according to
\begin{align}
    p^{\dot{\alpha}\alpha} =
    p^\mu \bar{\sigma}_\mu^{\dot{\alpha}\alpha} =
    \Tilde{\lambda}^{\dot{\alpha}} \lambda^\alpha.
\end{align}
The factor $\bar{\sigma}_\mu^{\dot{\alpha}\alpha} = (\mathbb{I},\vec{\sigma})$ contains the Pauli matrices.
Expressions are then  built out of Lorentz-invariant spinor products, defined by
\begin{align}
    \la ij \ra \equiv & \lambda_i^\alpha \lambda_{j\alpha} = \eps_{\alpha\beta} \lambda_i^\alpha \lambda_j^\beta = -\la ji \ra,
    \notag\\
    [ij] \equiv &
    \Tilde{\lambda}_{i\dot{\alpha}} \Tilde{\lambda}_j^{\dot{\alpha}} = 
    -\eps_{\dot{\alpha}\dot{\beta}} \Tilde{\lambda}_i^{\dot{\alpha}}  \Tilde{\lambda}_j^{\dot{\beta}} = -[ji],
\end{align}
where the Levi-Civita antisymmetric tensor raises and lowers spinor indices.

The common Mandelstam variable $s_{ij}$ is expressed in terms of spinor products as
\begin{align}
    s_{ij} \equiv (p_i+p_j)^2 = 2 p_i \cdot p_j = \spa{i}.{j} \spb{j}.{i},
\end{align}
and we also define a three-particle momentum factor,
\begin{align}
    t_{ijk} \equiv (p_i+p_j+p_k)^2 = s_{ij} + s_{jk} + s_{ki},
\end{align}
for convenience.
A compact notation can be used to join two spinor products of opposite type, in a way that illustrates the equivalence of full four-momentum factors and their spinors,
\begin{align}
    [i|j|k\ra \equiv \spb{i}.{j} \spa{j}.{k} = \Tilde{\lambda}_i^{\dot{\alpha}} p_{j\mu} \bar{\sigma}^\mu_{\dot{\alpha}\alpha} \lambda_k^\alpha .
\end{align}
We also use notation
\begin{align}
    [i|ab|k] \equiv \spb{i}.{a} \spa{a}.{b} \spb{b}.{k}
\end{align}
and
\begin{align}
  [i|k_{ab}|k\ra  \equiv \spb{i}.{a} \spa{a}.{j}+ \spb{i}.{b} \spa{b}.{j} 
\end{align}
together with obvious generalisations. 

\section{Currents}

\label{app:cur}

The augmented recursion procedure requires the input of currents, which are objects with two off-shell legs.
We accommodate these legs in terms of spinors, which individually only represent massless particles, by introducing the axial gauge formalism \cite{Kosower:1989xy,Kosower:2004yz,Schwinn:2005pi}.
A null reference momentum $q$ is required, so that any non-null momentum $K$ can be represented as a sum of a null momentum $K^\flat$ and a piece proportional to $q$.
Momentum $K$ is said to be ``nullified'' according to
\begin{align}
    K^\flat = K - \frac{K^2}{[q|K|q\ra} q,
\end{align}
where $(K^\flat)^2=0$, $q^2=0$ and $K^2 \neq 0$.
We note that a superscript ($^\flat$) will be used to denote the nullified form of off-shell momenta when they appear in our calculations.
However, spinor labels will not use these superscripts as there is no ambiguity -- spinors are on-shell objects and any spinor we use could be considered to be nullified.

We do not need to derive the entire current, because only those terms with poles in relevant momenta contribute to the recursion process.
Therefore we can derive a simpler ``good enough'' current approximation that contains only necessary structures.
Two conditions on these structures have been identified previously, which allow them to be generated \cite{Dunbar:2016aux,Dunbar:2016dgg}.
For a current with legs $\{\alpha, \beta\}$ that are in general off-shell, we have the rules:
\begin{align}
    \text{(C1) } &\text{The current must reproduce the leading $s_{\alpha\beta}$ singularities as $s_{\alpha\beta}\rightarrow 0$, for any }
    \notag\\
    &\text{choices of momenta $\{\alpha, \beta\}$ with $\alpha^2,\beta^2\neq0$.}
    \notag\\
    \text{(C2) } &\text{The current must reproduce the appropriate amplitude when $\alpha^2,\beta^2\rightarrow0$ and }
    \notag\\
    &\text{$s_{\alpha\beta}$ takes a general value (which can be $s_{\alpha\beta}\neq 0$).}
    \notag
\end{align}
By following these rules, the terms involving poles in $s_{\alpha\beta}$ are specified, but various other contributions can be omitted.

To obtain the new currents required by the seven-point two-loop calculation, we follow a new, systematic derivation method.
This procedure guarantees a ``good enough'' current that fulfils the rules C1 and C2:
\begin{enumerate}
    \item Start by writing the ``good enough'' current as
    \begin{align}\label{eq:good_enough_current}
        \tau_n^{(1)}(\alpha,\beta,\cdots)=A_n^{(1)}\vert_{\alpha^2,\beta^2=0}(\alpha,\beta,\cdots)+\Ord(\alpha^2,\beta^2), \notag
    \end{align}
    where $A_n^{(1)}\vert_{\alpha^2,\beta^2=0}$ represents the amplitude $A_n^{(1)}$, with nullified $\alpha^\flat$ and $\beta^\flat$ as arguments.
    Any instances of $s_{\alpha\beta}$ have been replaced by $\spa{\alpha}.\beta\spb{\beta}.\alpha$. 
    Condition C2 has been satisfied.

    \item Add the leading singularities to the current. These are given by all the lower-point amplitude factorisations where the propagator is $s_{\alpha\beta}$. 
    In this case, it is the full off-shell $s_{\alpha\beta}$ that should be written.
    We denote the contribution $\tau_{LS}^{(1)}$.
    This satisfies condition C1, but introduces extra structures that break C2.

    \item Subtract the piece $\tau_{LS}^{(1)}\vert_{\alpha^2,\beta^2=0}$, which represents the leading singularity terms with off-shell momenta replaced by the nullified forms $\alpha \rightarrow \alpha^\flat$ and $\beta \rightarrow \beta^\flat$.
    (In particular, $s_{\alpha\beta} \rightarrow \spa{\alpha}.\beta\spb{\beta}.\alpha$.)
    This restores condition C2, without affecting C1.

    \item Identifying the $\Ord(\alpha^2,\beta^2)$ piece of step 1 with $\tau_{LS}^{(1)}-\tau_{LS}^{(1)}\vert_{\alpha^2,\beta^2=0}$, the entire ``good enough'' current, satisfying both conditions C1 and C2, can be written as
    \begin{align}
        \tau_n^{(1)}(\alpha,\beta,\cdots)=A_n^{(1)}\vert_{\alpha^2,\beta^2=0}(\alpha,\beta,\cdots)+\tau_{LS}^{(1)}-\tau_{LS}^{(1)}\vert_{\alpha^2,\beta^2=0}.
    \end{align}
\end{enumerate}

After reducing the number of distinct currents required by using symmetry properties and decoupling identities, the two new currents to be derived are $\tau^{(1)}_7(\alpha^-,c^+,\beta^+,d^+,e^+,f^+,g^+)$ and $\tau^{(1)}_7(\alpha^-,c^+,d^+,\beta^+,e^+,f^+,g^+)$.
The difference between these structures is the degree of separation between the two off-shell legs.
Referencing this fact, we refer to the currents as the ``singly non-adjacent'' and ``doubly non-adjacent'' currents, respectively.
Also required is the ``adjacent'' current, $\tau^{(1)}_{7:1}(\alpha^-,\beta^+,c^+,d^+,e^+,f^+,g^+)$, which has been calculated previously \cite{Dunbar:2017nfy}.

The singly non-adjacent current can be written
\begin{align}
    \tau^{(1)}_7(\alpha^-,c^+,\beta^+,d^+,e^+,f^+,g^+) =&
    \hat{\tau}^{basis}_7(\alpha^-,c^+,\beta^+,d^+,e^+,f^+,g^+) 
    \notag\\
    &-\hat{\tau}^{basis}_7(\alpha^-,g^+,f^+,e^+,d^+,\beta^+,c^+),
\end{align}
where we define
\begin{align} \label{eq:current_basis_defn}
    \hat{\tau}^{basis}_7 & (\alpha^-,c^+,\beta^+,d^+,e^+,f^+,g^+) =
    \notag\\
    \frac{i}{3} \Bigg( &
    - \frac{\la \alpha|k_{c \beta}|d]  \spa{ \beta}.{e}  \spa{\alpha}.{d} ^3}{\spa{\beta}.{d} ^2 \spa{c}.{\beta}  \spa{d}.{e} ^2 \spa{e}.{f}  \spa{f}.{g}  \spa{g}.{\alpha}  \spa{\alpha}.{c} }
    \notag\\
    &+ \frac{\la \alpha|k_{ef}|g] ^3}{\la d|k_{ef}|g]  \spa{\beta}.{d}  \spa{c}.{\beta}  \spa{e}.{f} ^2 \spa{\alpha}.{c}  t_{efg} }
    \notag\\
    &- \frac{1}{2} \frac{\la \alpha|k_{c\beta}k_{de}|\alpha\ra ^3}{ \la \alpha|k_{c\beta}k_{de}|f\ra  \la \alpha|k_{fg}k_{de}|\beta\ra  \spa{c}.{\beta}  \spa{d}.{e} ^2 \spa{f}.{g}  \spa{g}.{\alpha}  \spa{\alpha}.{c} }
    \notag\\
    &- \frac{\spa{c}.{d}  \spa{\alpha}.{\beta} ^3 \spb{c}.{\beta} }{\spa{\beta}.{d} ^2 \spa{c}.{\beta} ^2 \spa{d}.{e}  \spa{e}.{f}  \spa{f}.{g}  \spa{g}.{\alpha} }
    \notag\\
    &+ \frac{\spb{c}.{g} ^3}{\spa{\beta}.{d}  \spa{d}.{e}  \spa{e}.{f}  \spb{g}.{\alpha}  \spb{\alpha}.{c}  t_{g\alpha c} } 
    \left( \frac{1}{2}\spb{\beta}.{f} - \frac{ [c|k_{\beta d}k_{ef}|g]  s_{de} }{2 \la \beta|k_{\alpha c}|g]  \la f|k_{g\alpha}|c] } +  \frac{ [f|k_{de}k_{ef}|g] }{\la \beta|k_{\alpha c}|g] }  \right)
    \notag\\
    &+ \frac{\la \alpha|k_{c\beta}|g] ^3}{\la \beta|k_{\alpha c}|g]  \spa{c}.{\beta}  \spa{d}.{e}  \spa{e}.{f}  \spa{\alpha}.{c}  t_{def}  t_{\alpha c\beta} }
    \left( \frac{\la \alpha|k_{c\beta}|d]  s_{de} }{\la \alpha|k_{c\beta}k_{de}|f\ra } + \frac{ [f|k_{de}k_{ef}|g] }{\la d|k_{ef}|g] } \right)
    \Bigg).
\end{align}

The doubly non-adjacent current takes a similar form, and can be written in terms of the same basis function as
\begin{align}
    \tau^{(1)}_7(\alpha^-,c^+,d^+,\beta^+,e^+,f^+,g^+) =&
    \hat{\tau}^{basis}_7(\alpha^-,c^+,\beta^+,d^+,e^+,f^+,g^+) 
    \notag\\
    &-\hat{\tau}^{basis}_7(\alpha^-,g^+,f^+,e^+,\beta^+,d^+,c^+).
\end{align}

These currents are then integrated as part of augmented recursion diagrams.

\section{Group Theory Conventions}

In this appendix we define the conventions which specify the translation between the group theory representations and specific combinations of amplitudes.  

Irreducible representations for the symmetric group.are specified by Young diagrams  These consist of diagrams of exactly $n$ boxes.    For the symmetric group,  the diagram determines the dimension of the corresponding irreducible representation.   Furthermore the defining representation contains the same number of copies of an irreducible representation as its dimension.   This can be seen in the relation
\begin{equation}
n! = \sum_{\tau}  d_{\tau}^2
\end{equation}
where $d_{\tau}$ is the dimension of representation $\tau$.   The $d_\tau$ representations are given by the individual proper tableaux for that diagram.  

The partial  amplitudes, in general, form a reducible representation of the symmetry group but due to cyclic and reflection symmetry not the full representation.    The content of the various partial amplitudes is given in the tables.   For a given irreducible representation there can be multiple copies appearing.  

What this means for amplitudes is that there are specific linear combinations of the amplitudes which act as the representations under permutations.  These combinations with be of the form
\begin{equation}
T_i^\tau =\sum_{\sigma}  c^\tau_{\sigma,i}  A_{n:x}(\sigma(1,2,\cdots n)
\end{equation}
and these transform under permutations as 
\begin{equation}
g: T^\tau_i   \longleftarrow   M^\tau(g)_{ij}   T^\tau_j
\end{equation}
where $M^\tau(g)$ is the matrix of permutation $g$ in representation $\tau$.

\ytableausetup{centertableaux, boxsize=0.9em}

We now describe a process to create these combinations. (following ref.~\cite{Edison:2012fn} and references therein)     The starting point is a Young diagram and its proper Young tableaux.    Each proper Young tableaux corresponds to a potential copy of the representation.    For example, 
{\begin{ytableau}  
1  & 2  & 3  & 4 \\  5 & 6 \\  7 
\end{ytableau}}
is a proper Young tableaux of the Young diagram 
{\begin{ytableau}  
\null  & \null   & \null  & \null  \\  \null & \null \\  \null   
\end{ytableau}}.  
From the  tableaux,   projection operators  can be formed
\begin{equation}
e^\tau_{ij}   =  \frac{d_\tau}{n!} Q_i  S_{ij}  P_j  M_j
\end{equation}
Where $P_j$ is the subset of the permutations which permutes elements of the rows within the row of tableaux $j$.   $Q_i$ is the subset of the permutations which permutes elements of the columns within the column of tableaux $i$ but signed by the order of the permutation, $S_{ij}$ is the permutation which converts tableaux $j$ to $i$ and 
\begin{equation}
M_i  = \sum_{j=1}^{i-1}  e^\tau_{jj}  \; . 
\end{equation}
The $ e^\tau_{ij}$ satisfy
\begin{equation}
e^{\tau}_{ij} e^{\tau'}_{kl}  =\delta^{\tau,\tau'}  \delta_{jk}  e^\tau_{il}  \; . 
\end{equation} 
The `diagonal' $e^\tau_{ii}$ form projection operators i.e.
\begin{eqnarray}
e^{\tau}_{ii} e^{\tau'}_{jj}  &=&0  \;\;   i \neq j \hbox{ or } \tau \neq \tau'
\notag \\
e^{\tau}_{ii} e^{\tau}_{ii}  &=& e^{\tau}_{ii}  
\notag \\
\sum_{\tau}\sum_{i}   e^{\tau}_{ii} &=& 1  \; . 
\end{eqnarray}
For a given $\tau$ and $i$ the $e^\tau_{ij}$ project onto the elements of the representation.  
So taking
\begin{equation}
e^\tau_{ii}  A_{n:r}( 1,2,\cdots n)
\end{equation}
gives us the first combination of the amplitudes in the representation.  The other combinations in the representation  are obtained as 
$e^\tau_{ij}  A_{n:r}( 1,2,\cdots n)$ or by action of $S_n$.     Although we can form $d_\tau$ combinations many of these will be zero or not independent due to cyclic and flip symmetry.    The precise combinations we choose for $n=7$ are included in the attached 
mathematica file  {\it Amp7reps.m}.   This defines our conventions which then in turns specifies the  normalisations of the specific relations
in sections \ref{section:group}   and \ref{section:eight}.    

As a consistency check,  the number of independent combinations can be computed from the subgroup $G$ of $S_n$ which leaves a partial amplitude invariant (up to minus sign).  For the leading in color $A^{(2)}_{n:1}$,  this is just the dihedral group $D_{n}$.  
The number of independent combinations for a given representation $R_i$ is given using the characters of $S_n$
\begin{equation}
n_i  = \frac{1}{|G|} \left(  \sum_j  \epsilon_j  m_j  \chi_j(R_i)   \right)  
\end{equation}
where $m_j$ is the number of elements of conjugacy class $j$ within $G$ and $\epsilon_j$ is plus one if the partial amplitude is invariant and $-1$ is anti-invariant.    The $n_i$ evaluated in this way have been used as a check  for $n=7,8,9$.




\bibliographystyle{JHEP}

\end{document}